\documentclass[format=acmsmall, review=false, screen=true]{acmart}

\usepackage{booktabs}
\usepackage{amsmath}
\usepackage{bm}
\usepackage{longtable}
\usepackage{multirow}
\usepackage{makecell}
\usepackage{amssymb}
\usepackage{color}
\usepackage{subcaption}
\allowdisplaybreaks

\usepackage[ruled]{algorithm2e}

\SetAlFnt{\small}
\SetAlCapFnt{\small}
\SetAlCapNameFnt{\small}
\SetAlCapHSkip{0pt}
\IncMargin{-\parindent}

\DeclareMathOperator{\E}{\mathbb{E}}
\DeclareMathOperator{\sig}{\text{sig}}
\DeclareMathOperator*{\argmin}{arg\!\min}
\DeclareMathOperator*{\argmax}{arg\!\max}

\acmPrice{}

\setcopyright{none}
\renewcommand\footnotetextcopyrightpermission[1]{} 
\makeatletter
\renewcommand\@formatdoi[1]{\ignorespaces}
\makeatother

\begin{document}
\title{Attribute-aware Collaborative Filtering: Survey and Classification}  
\author{Wen-Hao Chen}
\affiliation{%
  \institution{National Taiwan University}
}
\author{Chin-Chi Hsu}
\authornote{This author has equal contributions as first author to this paper.}
\affiliation{%
  \institution{Academia Sinica}
}
\author{Yi-An Lai}
\affiliation{%
  \institution{National Taiwan University}
}
\author{Vincent Liu}
\authornote{This author has equal contributions as third author to this paper.}
\affiliation{%
  \institution{National Taiwan University}
}
\author{Mi-Yen Yeh}
\affiliation{%
  \institution{Academia Sinica}
}
\author{Shou-De Lin}
\affiliation{%
  \institution{National Taiwan University}
}


\begin{abstract}


Attribute-aware CF models aims at rating prediction given not only the historical rating from users to items, but also the information associated with users (e.g. age), items (e.g. price), or even ratings (e.g. rating time).
This paper surveys works in the past decade developing attribute-aware CF systems, and discovered that mathematically they can be classified into four different categories. We provide the readers not only the high level mathematical interpretation of the existing works in this area but also the mathematical insight for each category of models. Finally we provide in-depth experiment results comparing the effectiveness of the major works in each category.

\end{abstract}

%
%


%
%




\maketitle
\thispagestyle{empty}

\renewcommand{\shortauthors}{C-C Hsu, W-H Chen, Y-A Lai, V. Liu, M-Y Yeh, S-D Lin.}

\section{Introduction}
\label{section:introduction}

Collaborative filtering is arguably the most effective idea in building a recommender system.
It assumes that a user's preferences on items can be inferred collaboratively from other users' preferences.
In practice, users' past records toward items, such as explicit ratings or implicit feedback (e.g. binary access records), are typically used to infer similarity of taste among users for recommendation.
In the past decade, \emph{matrix factorization} (MF) has become a widely adopted realization of collaborative filtering.
Specifically, MF learns a latent representation vector for a user and an item, and compute their inner products as the predicted rating. The learned latent user/item factors are supposed to embed the specific information about the user/item accordingly. That is, two users with similar latent representation shall have similar taste to items with similar latent vectors.

\begin{figure}
    \centering
    \includegraphics[width=0.5\linewidth]{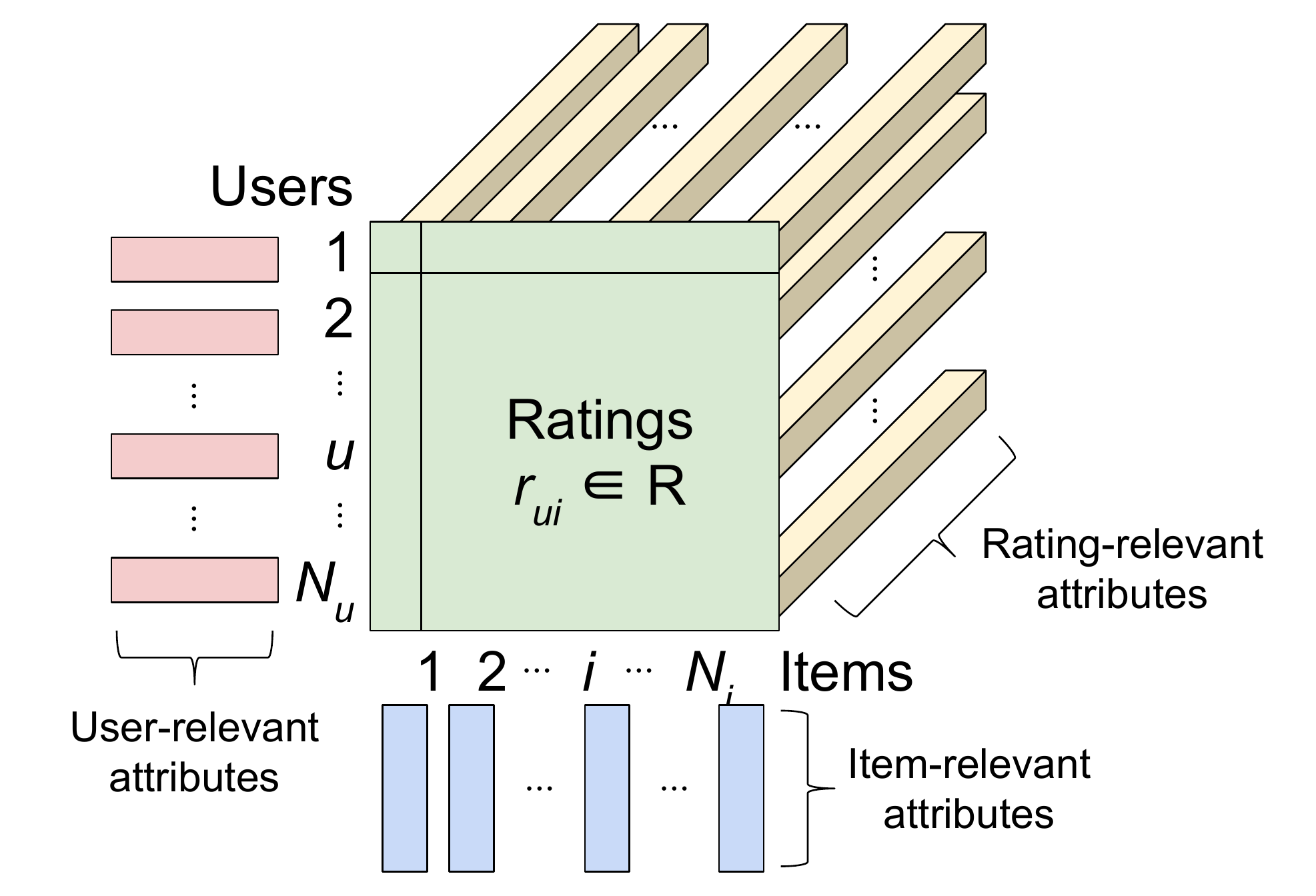}
    \caption{Interpretation of inputs, including ratings and attributes, in attribute-aware collaborative filtering based recommender systems.}
    \label{figure:attribute_aware_recommender_system}
\end{figure}

In big data era, classical MF using only ratings suffer a serious drawback for not being able to exploit other accessible information such as the attributes of users/items/ratings. 
For instance, data could contain the location and time about where and when a user rated an item. These rating-relevant attributes, or \emph{contexts}, could be useful in determining the scale of a user liking an item. The \emph{side information} or attributes relevant to users or items (e.g. the demographic information of users or the item genera) can also reveal useful information.  Such side information is particularly useful for situation when the ratings about a user or an item is sparse, which is known as the \emph{cold-start} problem for recommender systems.
Therefore, researchers have formulated the \emph{attribute-aware recommender systems} (see Figure \ref{figure:attribute_aware_recommender_system}) aiming at leverage not only the rating information but also the attributes associated with ratings/users/items to improve the quality of recommendation.


Researchers have proposed different methods to extend existing collaborative filtering models in recent years, such as factorization machines, probabilistic graphical models, kernel tricks and models based on deep neural networks.
We notice that those papers can also be categorized according to what kinds of attributes are incorporated into models.
If attributes are relevant to users (e.g. age, gender, occupation) or items (e.g. expiration, price), then the class of \emph{recommender systems with side information} (e.g., \cite{Adams10DPMF, Fang11MCRI, Guo17CoEmbed, Kim14VBMFSICA, Lu16SIMMCSI, Ning12SSLIM, Park13HBMFSI, Porteous10BMFSI, Xu13Maxide, Yu17SQ, Zhao16PCFSI, Zhao17LDRSSI, Zhou12KPMF, Zhou17CGSI}) consider such attributes when predicting ratings.
On the other hand, \emph{context-aware recommender systems} (e.g., \cite{Baltrunas11CAMF, Chen14CCTRSMF2, Hidasi12iTALS, Hidasi15GFF, Hidasi16GFF, Karatzoglou10TF, Li10PLRM, Liu13SoCo, Liu15CALR, Nguyen14GPFM, Rendle11FM, Shi12TFMAP, Shi14CARS2, Shin09CARAUC}) enhances themselves by considering the attributes appended to each rating (e.g. rating time, rating location).
Other terms may be used to indicate attributes interchangably such as \emph{metadata} \cite{Kula15LightFM}, \emph{features} \cite{Chen12SVDFeature} , \emph{taxonomy} \cite{Koenigstein11YMR}, \emph{entities} \cite{Yu14PHeteRec}, \emph{demographical data} \cite{Safoury13DACR}, \emph{categories} \cite{Chen16LRPPMCF}, 
\emph{contexture information} \cite{Weng09MultiRecom}, etc.
The above setups all share the same mathematical representation; thus technically we do not distinguish them in this paper. That is, we regard whichever information associated with user/item/rating as user/item/rating attributes, regardless whether they are location, time, or demographical features.
Therefore, a CF model that take advantage of not only ratings but also associated attributes are called \emph{attribute-aware recommender} in this paper.

\begin{table}
    \centering
    \begin{tabular}{c c c}
        \toprule
        Difference & Previous Works \cite{Adomavicius11CAR, Verbert12CAR, Bobadilla13RSS, Shi14CFB} & Our Work \\
        \midrule
        Attribute discussions & \makecell{Categories and definitions \\ of diversified attributes} & \makecell{Mathematical formulations \\ of the most general attribute vectors} \\
        Model introduction & \makecell{High-level summary \\ of text descriptions} & \makecell{Mathematical interpretation \\ of model design criteria} \\
        Comparison Experiments & \makecell{For memory-based models in \cite{Bobadilla13RSS}; \\ no experiments in others} & \makecell{For seven model-based models \\ on seven benchmark datasets} \\
        \bottomrule
    \end{tabular}
    \caption{Presentation differences between previous works and our work.}
    \label{table:survey_paper_comparison}
\end{table}

Note that the attribute-aware recommender systems discussed in this paper is not equivalent to hybrid recommender systems.
The former treats addtional information as attributes while the latter emphasizes the combination of collaborative filtering based methods and content based methods.
To be more precise, this survey covers only works that assume unstructured and independent attributes, either in binary or numerical format, for each user, item or rating.
The reviewed models do not have prior knowledge of the dependency between attributes, such as the adjancent terms in a document or user relationships in a social network.

This survey covers more than one hundred papers in this area in the past decade. We found that the majority of the works propose an extension of matrix factorization to incorporate attribute information in collaborative filtering.
The main contribution in this paper is to not only provide the review report, but rather a means to classify these works into  four categories: (I) \emph{discriminative matrix factorization}, (II) \emph{generative matrix factorization}, (III) \emph{generalized factorization}, and (IV) \emph{heterogeneous graphs}. Inside each category, we provide the probabilistic interpretation of the models.
The major distinction of these four categories lies in the representation of the interactions of users, items and attributes.
The discriminative matrix factorization models extend the traditional MF by treating the attributes as prior knowledge to learn the latent representation of users or items.
Generative matrix factorization further considers the distributions of attributes, and learn such together with the rating distributions.
Generalized factorization models view the user/item identity simply as a kind of attribute, and various models are designed for learning the low-dimensional representation vectors for rating prediction.
The last category of models propose to represent the users, items and attributes using a heterogeneous graph, where a recommendation task can be cast into a link prediction task on the heterogeneous graph.
In the following sections, we will elaborate the general mathematical explanations of the four types of model designs, and discuss the similarity/difference among models.

There have been four prior survey works \citep{Adomavicius11CAR, Verbert12CAR, Bobadilla13RSS, Shi14CFB} introducing attribute-aware recommender systems.
We claim three major differences between our work and the existing papers. First, previous survey mainly focuses on grouping different types of attributes, and discussing the distinctions of memory-based collaborative filtering and model-based collaborative filtering. In contrast, we are the first that aims at classifying the existing works based on the methodology proposed, instead of the type of data used. We further provide mathematical connections for different types of models so the readers can better understand the spirit of the design of different models as well as their technical differences. 
Second, we are the first to provide thorough experiment results (7 different models on 8 benchmark datasets) to compare different types of attribute-award recommendation systems. Note that \cite{Bobadilla13RSS} is the only previous survey work with experiment results. However, it performed experiments to compare different similarity measures in collaborative filtering algorithms, instead of directly verifying the effectiveness of different attribute-aware recommender systems.
Finally, we cover the latest works on attribute-aware recommender systems. We have realized that the existing survey papers do not include about \emph{forty} papers after $2015$. Especially in recent years several deep neural network based solutions have provided the state-of-the-art performance for this task.


Table \ref{table:survey_paper_comparison} shows the comparisons between our work and previous surveys.

We will introduce basic ideas about recommender systems in Section \ref{section:preliminary}, followed by the formal analyses on attribute-aware recommender systems in Section \ref{section:survey} and \ref{section:common_integration}.
A series of experiments in Section \ref{section:experiment} are conducted to compare the accuracy and parameter sensitivity of six widely adopted models.
Finally Section \ref{section:conclusion} concludes this review work and some tasks to be done in the future.
\section{Preliminaries}
\label{section:preliminary}

\subsection{Problem Definition of Recommender Systems}
\label{section:problem_definition}

\emph{Recommender systems} act as skilled agents to assist \emph{users} in conquering information overload while making selection decisions over \emph{items} by providing customized recommendations.
\emph{Users} and \emph{items} are general phrases denoting entities actively browsing and making choices and entities being selected such as goods and services, respectively.

Formally, recommender systems leverage one or more of three information sources to discover user preferences and generate recommendations: \emph{user-item interactions}, \emph{side information}, and  \emph{contexts}.
\emph{User-item interactions}, or \emph{ratings}, are collected explicitly by prompting users to provide numerical feedbacks towards items and acquired implicitly by tracking user behaviors such as clicks, browsing time, or purchase history.
These information are commonly represented as a matrix that encodes preferences of users and is naturally sparse since users normally interact with a limited fraction of items.
\emph{Side information} are rich information attached to individual user or item that depict user  characteristics such as educations and jobs or item properties such as descriptions and product categories.
Side information can span over diverse structures with rich meaning ranging from numerical status, texts, images to videos, locations, or networks.
On the other hand, \emph{contexts} refer to all the information collected when a user interacts with an item such as timestamps, locations, or textual reviews.
These contextual information usually serve as an additional information source appended to the user-item interaction matrix.

The goal of recommender systems is to disclose unknown user preferences over items that users never interact with and recommend the most preferred items to them.
In practice, recommender systems learn to generate recommendations based on three types of approaches: \emph{pointwise}, \emph{pairwise}, and \emph{listwise}.
\emph{Pointwise} approach is the most common approach and demands recommendation systems to provide accurate numerical predictions on observed ratings.
Items that a user never interacts with are then sorted by their rating predictions and a number of items with the highest ratings are recommended to the user.
On the other hand, \emph{pairwise} approach seeks to preserve the ordering of any pair of items based on ratings, while in the \emph{listwise} approach recommender systems aim to preserve the relative order of all rated items as a list for each user.
Pairwise approach and listwise approach are together considered as \emph{item ranking} that only requires recommender systems to output ordering of items but not ratings for individual items.


The problem definition of recommender systems can be defined as follows:
Given $N_{u}$ users, $N_{i}$ items, and information sources \emph{user-item ratings} $\bm{R} \in \mathbb{R}^{N_{u} \times N_{i}}$ with $N_Z$ known entries, \emph{side information of users} $\bm{X} \in \mathbb{R}^{K_{X} \times N_{u}}$, \emph{side information of items} $\bm{Y} \in \mathbb{R}^{K_{Y} \times N_{i}}$, \emph{contexts} $\bm{Z} \in \mathbb{R}^{K_{Z} \times N_{r}}$,
and under the assumption that ratings $r_{ui} > r_{uj} \Leftrightarrow$ an item preference relation $i \succ_{u} j$ for user $u$, 
a recommender system is a function $f$ that outputs a permutation of items for each user with more preferred items in front:
\begin{align}
    f: \bm{R} \times \bm{X} \times \bm{Y} \times \bm{Z} \rightarrow 
    \begin{bmatrix}
        \pi_{u}^{-1}(1) & \pi_{u}^{-1}(2) & \ldots & \pi_{u}^{-1}(N_{i})
    \end{bmatrix}
    \label{equation:problem_definition}
\end{align}
such that
\begin{align}
    \pi_{u}(i) < \pi_{u}(j) \ \Rightarrow 
    \ i \succ_{u} j
    \quad \forall \ u, i, j \ ,
\end{align}
where function $\pi_{u}(\cdot)$ moves item $i$ from index $i$ to index $\pi_{u}(i)$ in the list, with respect to user $u$, and $\pi_{u}^{-1}(\cdot)$ is its inverse function.
Note that the dimension $K_{X}, K_{Y}$ of side information attribute matrix $\bm{X}, \bm{Y}$ might be zero denoting that there is no side information about users or items. Likewise, if there is no contextual information about user-item interactions, $K_{Z}$ will be zero.

The core techniques or algorithms to realize recommender systems are generally classified into three categories: \emph{content-based filtering}, \emph{collaborative filtering}, and \emph{hybrid filtering} \cite{Bobadilla13RSS, Shi14CFB, ISINKAYE2015261}.
\emph{Content-based filtering} generates recommendations based on properties of items and user-item interactions.
Content-based techniques exploit domain knowledge and seek to transform item properties in raw attribute structures such as texts, images, or locations into numerical item profiles.
Each item is represented as a vector and the matrix of side information of items $\bm{Y}$ is constructed.
A representation of each user is then created by aggregating profiles of items that this user interacted with and a similarity measure is leveraged to retrieve a number of the most similar items as recommendations.
Note that content-based filtering doesn't require information from any other user to make recommendations.
\emph{Collaborative filtering} strives to identify a group of users with similar preferences for each user based on the past user-item interactions and items preferred by these users are recommended.
Since discovering users with common preferences is generally based on user-item ratings $\bm{R}$, collaborative filtering becomes the first choice when item properties are inadequate in describing their content such as movies or songs.
\emph{Hybrid filtering} is the extension or combination of content-based and collaborative filtering. 
Examples are building an ensemble from both techniques, using item rating history of collaborative filtering as part of item profiles for content-based filtering, or extending collaborative filtering to incorporate user characteristics $\bm{X}$ or item properties $\bm{Y}$.
This survey focuses on \emph{attribute-aware recommender systems} that shed light on not only user-item interactions $\bm{R}$ but also side information of users or items $\bm{X}, \bm{Y}$, and contexts $\bm{Z}$ which is a subset of hybrid filtering.

\subsection{Collaborative Filtering and Matrix Factorization}
\label{section:collaborative_filtering_matrix_factorization}

Collaborative filtering (CF) has become the most prevailing technique to realize recommender systems in recent years \cite{1423975, Shi14CFB, Adomavicius11CAR, ISINKAYE2015261}.
It assumes preferences that users exhibit towards interacted items can be generalized and used to infer their preferences towards items they have never interacted with through leveraging records of other users with similar preferences.
This section briefly introduces conventional CF techniques that assumes the availability of only user-item interactions, or the rating matrix $\bm{R}$.
In practice, they are commonly categorized into \emph{memory-based} CF and \emph{model-based} CF \cite{Shi14CFB, ISINKAYE2015261, 1423975}.

\emph{Memory-based} CF directly exploits rows or columns in the rating matrix $\bm{R}$ as representations of users or items and identifies a group of similar users or items by a pre-defined similarity measure.
Commonly used similarity metrics include the Pearson correlation, the Jaccard similarity coefficient, the cosine similarity, or their variants.
Memory-based CF techniques can be divided into \emph{user-based} or \emph{item-based} approaches indicating that a technique tries to identify a group of either similar users or similar items. 
For user-based approaches, $K$ nearest neighbors --- or the $K$ most similar users --- are extracted, and their preferences or ratings towards a target item are aggregated into a rating prediction using similarities between users as weights.
The rating prediction of user $u$ to item $i$, $\hat{r}_{ui}$, can be formulated as:
\begin{align}
    \hat{r}_{ui} = \frac{1}{Z} \sum_{v \in U_u}{\text{sim}(u, v) r_{vi}},
\end{align}
where function $\text{sim}(\cdot)$ is a similarity measure, $Z$ is the normalization constant and $U_u$ is the set of similar users to user $u$ \cite{Shi14CFB}.
Rating predictions of item-based approaches can be formulated in a similar way.
The calculated pairwise similarities between users or items act as the \emph{memory} of the recommender system since they can be saved for  generating later recommendations.

\emph{Model-based} CF, on the other hand, takes the rating matrix $\bm{R}$ to train a predictive model with a set of parameters $\bm{\theta}$ to make recommendations \cite{1423975, Shi14CFB}. 
Predictive models can be formulated as a function that output ratings for \emph{rating predictions} or numerical preference scores for \emph{item ranking} given a user-item pair $(u, i)$:
\begin{align}
    \hat{r}_{ui} = f_{\bm{\theta}}(u, i) .
\end{align}
Model-based CF then ranks and selects $K$ items with the highest ratings or scores $r_{ui}$ as recommendations.
Common core algorithms for model-based CF involve Bayesian classifiers, clustering techniques, graph-based approaches, genetic algorithms, and dimension reduction methods such as Singular Value Decomposition (SVD) \cite{Bobadilla13RSS, Shi14CFB, Adomavicius11CAR, ISINKAYE2015261, 1423975}.
Over the last decade, a class of \emph{latent factor models}, called \emph{matrix factorization}, has been popularized and is commonly adopted as the basis of advanced techniques because of its success in the development of algorithms for the Netflix competition \cite{Koren:2009:MFT:1608565.1608614, Koren2011}.
In general, latent factor models aim to learn a low-dimensional representation, or latent factor, for each entity and combine latent factors of different entities using specific methods such as inner product, bilinear map, or neural networks to make predictions.
As a member of latent factor models, matrix factorization for recommender systems characterizes each user and item by a low-dimensional vector and predicts ratings based on inner product.

\emph{Matrix factorization} (MF) \cite{Shi14CFB, Koren:2009:MFT:1608565.1608614, Paterek2007ImprovingRS, Koren2011}, in the basic form, represents each user $u$ as a parameter vector $\bm{w}_{u} \in \mathbb{R}^K$ and each item $i$ as $\bm{h}_{i} \in \mathbb{R}^K$, where $K$ is the dimension of latent factors.
The prediction of user $u$'s rating or preference towards item $i$, denoted as $\hat{r}_{ui}$, can be computed using inner product:
\begin{align}
    \hat{r}_{ui} = \bm{w}_{u}^{\top} \bm{h}_{i},
\end{align}
which captures the interaction between them.
MF seeks to generate rating predictions as close as possible to those recorded ratings.
In matrix form, it can be written as finding $\bm{W}, \bm{H}$ such that $\bm{R} \approx \bm{W}^{\top} \bm{H}$ where $\bm{R} \in \mathbb{R}^{N_u \times N_i}$.
MF is essentially learning a low-rank approximation of the rating matrix since the dimension of representations $K$ is usually much smaller than the number of users $N_{u}$ and items $N_{i}$.
To learn the latent factors of users and items, the system tries to find $\bm{W}, \bm{H}$ that minimize the regularized square error on the set of known ratings $\delta ( \bm{R} )$:
\begin{align}
    \bm{W}^{\bm{*}},\ \bm{H}^{\bm{*}} = 
    \argmin_{\bm{W}, \bm{H}} \sum_{(u, i) \in \delta ( \bm{R} )} \frac{1}{2} \left ( r_{ui} - \bm{w}_{u}^{\top} \bm{h}_{i} \right ) ^2
    + \frac{\lambda_{W}}{2} \sum_{u = 1}^{N_{u}} \| \bm{w}_{u} \|_{2}^{2}
    + \frac{\lambda_{H}}{2} \sum_{i = 1}^{N_{i}} \| \bm{h}_{i} \|_{2}^{2} ,
\end{align}
where $\lambda_{W}$ and $\lambda_{H}$ are regularization parameters.
MF tends to cluster users or items with similar rating configuration into groups in the latent factor space which implies that similar users or items will be close to each other. 
Furthermore, MF assumes the rank of rating matrix $\bm{R}$ or the dimension of the vector space generated by rating configuration of users is far smaller than the number of users $N_{u}$.
This implies that each user's rating configuration can be obtained by a linear combination of ratings from a group of other users since they are all generated by $K$ principle vectors.
Thus MF entails the spirit of collaborative filtering, which is to infer a user's unknown ratings by ratings of several other users.

\emph{Biased matrix factorization} \cite{Koren:2009:MFT:1608565.1608614, Paterek2007ImprovingRS, Koren2011}, as an improvement of MF, models characteristics of each user and each item and the global tendency that are independent of user-item interactions.
The obvious drawback of MF is that only user-item interactions $\bm{w}_{u}^{\top} \bm{h}_{i}$ are considered in rating predictions.
However, ratings usually contain universal shifts or exhibit systematic tendencies with respect to users and items.
For instance, there might be a group of users inclined to give significant higher ratings than others or a group of items widely considered as high-quality ones and receiving higher ratings.
Besides, it is common that all ratings are non-negative which implies the overall average might not be close to zero and causes a difficulty for training of small-value-initialized representations.
With issues mentioned above, biased MF augments MF rating predictions with linear biases that account for user-related, item-related, and global effects.
The rating prediction is extended as follows:
\begin{align}
    \hat{r}_{ui} = \mu + c_{u} + d_{i} + \bm{w}_{u}^{\top} \bm{h}_{i},
\end{align}
where $\mu, c_i, d_j$ are global bias, bias of user $i$, and bias of item $j$, respectively.
Biased MF then finds the optimal $\bm{W}, \bm{H}, \bm{c}, \bm{d}, \mu$ that minimize the regularized square error as follows:
\begin{align}
    \bm{W}^{\bm{*}}, \bm{H}^{\bm{*}}, \bm{c}^{\bm{*}}, \bm{d}^{\bm{*}}, \mu^{\bm{*}} = 
    \argmin_{\bm{W}, \bm{H}, \bm{c}, \bm{d}, \mu}
    \sum_{(u, i) \in \delta ( \bm{R} )}
    (r_{ui} - \mu - c_{u} - d_{i} - \bm{w}_{u}^{\top} \bm{h}_{i})^2
    + \lambda \left (  \|\bm{W}\|_F^2 + \|\bm{H}\|_F^2 
    + \| \bm{c} \|_2^2 + \| \bm{d} \|_2^2 \right ) ,
\end{align}
where $\| \bm{W} \|_{F}^{2} = \sum_{u = 1}^{N_{u}} \| \bm{w}_{u} \|_{2}^{2} $ denotes the squared Frobenius norm.
The regularization parameter $\lambda$ is tuned by cross-validation.

\begin{figure}
    \centering
    \begin{subfigure}{0.5\linewidth}
        \centering
        \includegraphics[width=0.8\linewidth]{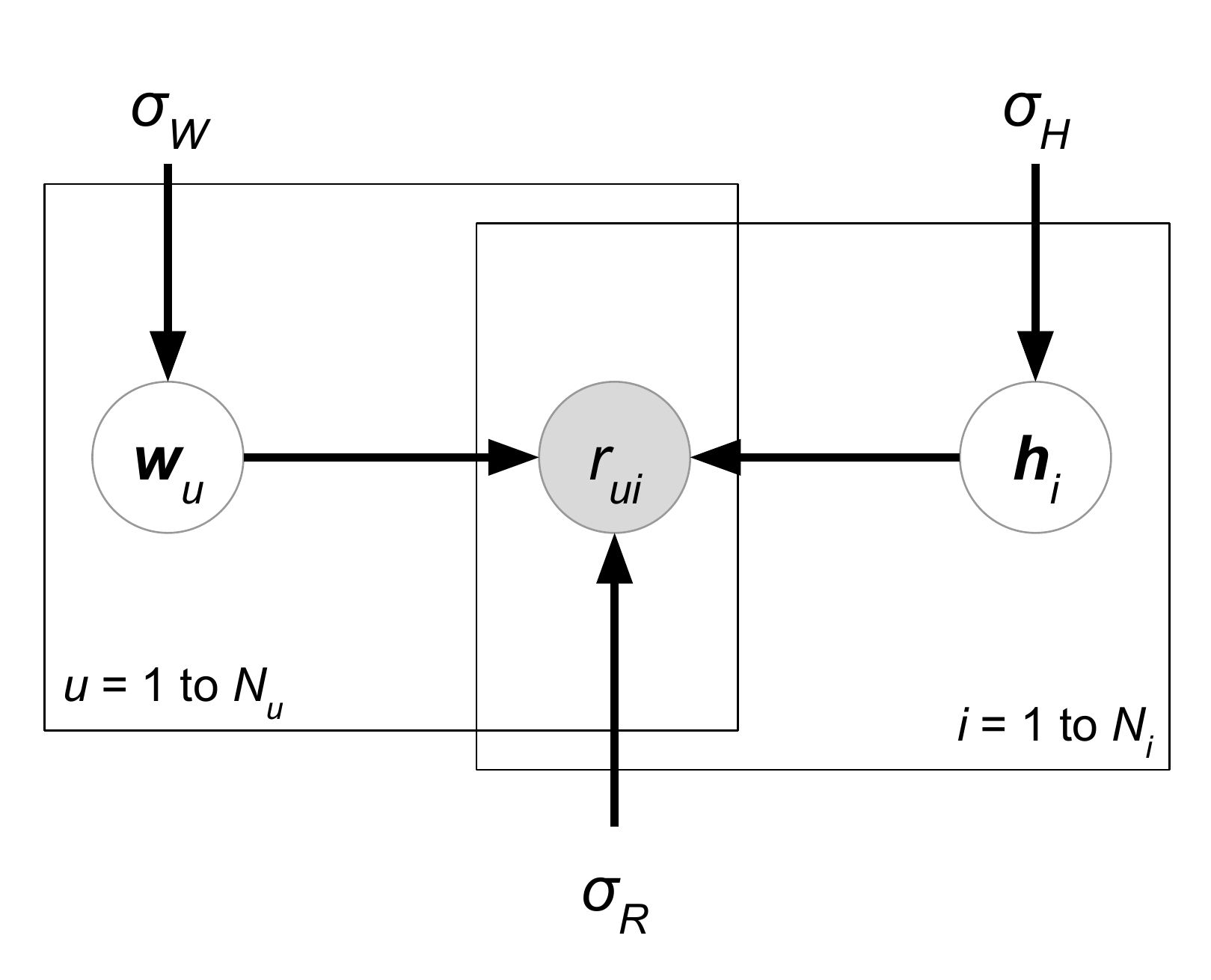}
        \caption{PMF}
    \end{subfigure}%
    \begin{subfigure}{0.5\linewidth}
        \centering
        \includegraphics[width=0.8\linewidth]{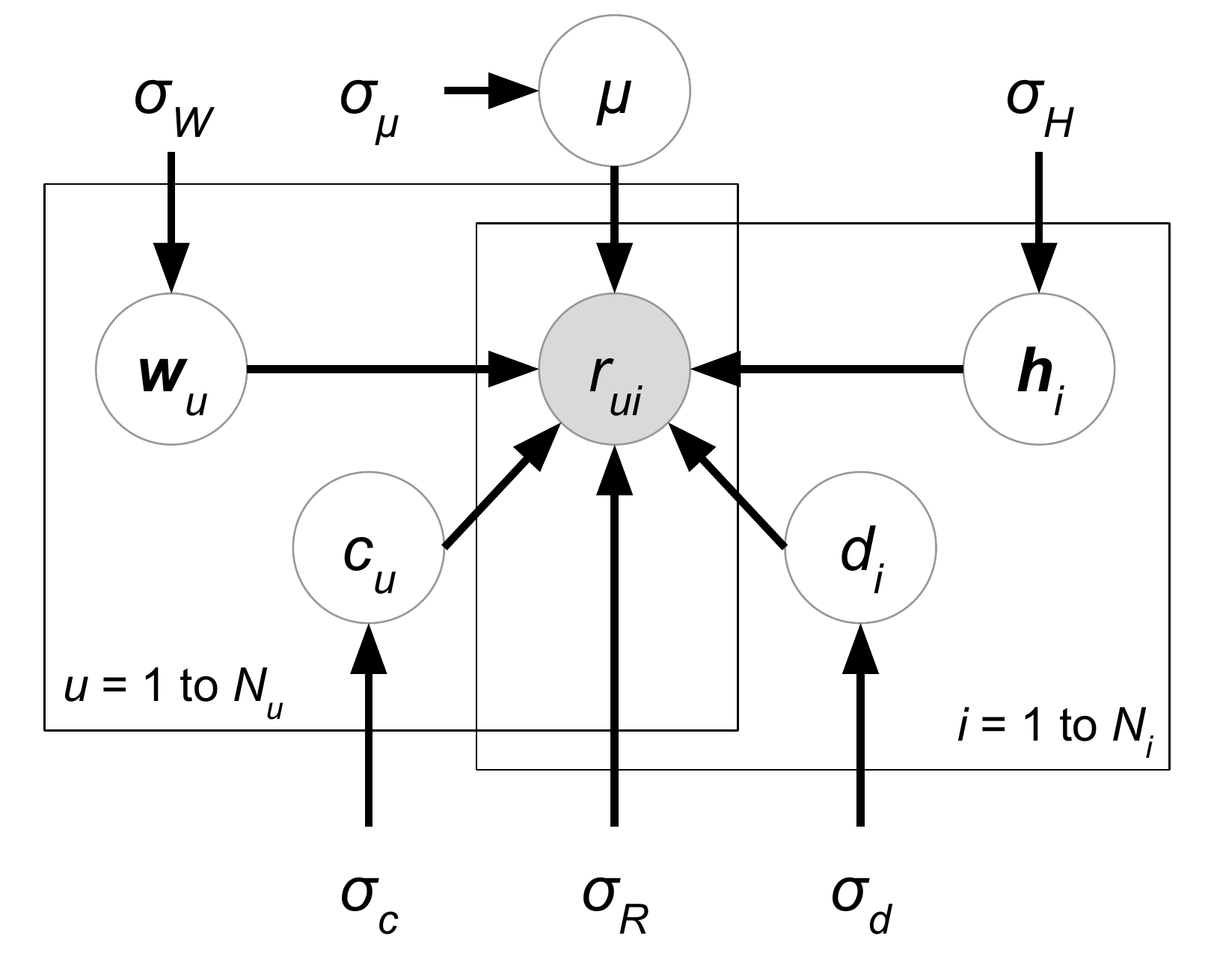}
        \caption{Biased PMF}
    \end{subfigure}
    \caption{Graphical interpretation of Probabilistic Matrix Factorization (PMF). User or item latent factors $\bm{W}, \bm{H}$ are put to generate observed ratings $\bm{R}$. We can put biase terms $\bm{c}, \bm{d}, \mu$ to learn the latent shifts between $\bm{R}$ and $\bm{W}^{\top} \bm{H}$. Parameters $\sigma_{W}, \sigma_{H}, \sigma_{R}, \sigma_{c}, \sigma_{d}, \sigma_{\mu}$ control the certainty in the generation process.}
    \label{figure:probabilistic_matrix_factorization}
\end{figure}

\emph{Probabilistic matrix factorization} (PMF, Figure \ref{figure:probabilistic_matrix_factorization})  \cite{Salakhutdinov:2007:PMF:2981562.2981720, Salakhutdinov:2008:BPM:1390156.1390267} is a probabilistic linear model with observed Gaussian noise and can be viewed as a probabilistic extension of MF.
PMF adopts the assumption that users and items are  independent and represents each user or each item with a zero-mean spherical multivariate Gaussian distribution as follows:
\begin{align}
    p \left ( \bm{W} \mid \sigma_{W}^2 \right ) = 
    \prod_{u=1}^{N_{u}}{ \mathcal{N} \left (
    \bm{w}_u \mid \bm{0}, \sigma_{W}^2 \bm{I} \right ) }, \qquad
    p \left ( \bm{H} \mid \sigma_{H}^2 \right ) = 
    \prod_{i=1}^{N_{i}}{ \mathcal{N} \left (
    \bm{h}_i \mid \bm{0}, \sigma_{H}^2 \bm{I} \right ) },
\end{align}
where $\sigma_{W}^2$ and $\sigma_{H}^2$ are observed user-specific and item-specific noise.
PMF then formulates the conditional probability over the observed ratings as
\begin{align}
    p \left (\bm{R} \mid \bm{W}, \bm{H}, \sigma^2 \right ) = 
    \prod_{(i, j) \in \delta ( \bm{R} )}{\mathcal{N} \left ( 
    r_{ui} \mid \bm{w}_{u}^{\top} \bm{h}_{i}, \sigma_{R}^2 \right )},
\end{align}
where $\delta ( \bm{R} )$ is the set of known ratings and $\mathcal{N} ( x \mid \mu, \sigma^2 )$ denotes the Gaussian distribution with mean $\mu$ and variance $\sigma^2$.
Learning of PMF is conducted by maximum a posteriori (MAP) estimation, which is equivalent to maximize the log of the posterior distribution of $\bm{W}, \bm{H}$:
\begin{align}
    \log p \left ( \bm{W}, \bm{H} \mid \bm{R}, \sigma_R^2,
    \sigma_W^2, \sigma_H^2 \right ) = &
    \log p \left ( \bm{R} \mid \bm{W}, \bm{H}, \sigma_{R}^2 \right )
    + \log p \left ( \bm{W} \mid \sigma_W^2 \right )
    + \log p \left ( \bm{H} \mid \sigma_H^2 \right ) + C \nonumber \\
    =& - \frac{1}{2 \sigma_R^2} \sum_{(u, i) \in \delta ( \bm{R} )}
    {\left ( r_{ui} - \bm{w}_{u}^{\top} \bm{h}_{i} \right )^2}
    - \frac{1}{2 \sigma_W^2} \sum_{u = 1}^{N_{u}}{ \bm{w}_{u}^{\top} \bm{w}_u }
    - \frac{1}{2 \sigma_H^2} \sum_{i = 1}^{N_{i}}{ \bm{h}_{i}^{\top} \bm{h}_i } \nonumber \\
    & - \frac{1}{2} \Big ( | \delta ( \bm{R} ) | \log \sigma_{R}^2 + N_{u} K \log \sigma_W^2 + N_{i} K \log \sigma_H^2 \Big ) + C
\end{align}
where $C$ is a constant independent of all parameters and $K$ is the dimension of user or item representations.
With Gaussian noise $\sigma_R^2, \sigma_W^2, \sigma_H^2$ observed, maximizing the log-posterior is identical to minimize the objective function with the form:
\begin{align}
    \sum_{(u, i) \in \delta ( \bm{R} )}
    { \frac{1}{2} (r_{ui} - \bm{w}_{u}^{\top} \bm{h}_{i})^2
    + \frac{\lambda_{W}}{2} \sum_{u = 1}^{N_{u}}{ \|\bm{w}_u\|_2^2} 
    + \frac{\lambda_{H}}{2} \sum_{i = 1}^{N_{i}}{ \|\bm{h}_i\|_2^2 }},
    \label{equation:PMF_objective}
\end{align}
where $\lambda_W= \sigma_R^2 / \sigma_W^2, \lambda_H = \sigma_R^2 / \sigma_H^2$.
Note that (\ref{equation:PMF_objective}) has exactly the same form as the regularized square error of MF and gradient descent or its extensions can then be applied in training PMF.

Since collaborative filtering techniques only consider rating matrix $\bm{R}$ in making recommendations, they cannot discover preferences of users or items with scant user-item interactions.
This problem is referred as the \emph{cold-start issue}.
In Section \ref{section:survey}, we will review recommendation systems that extend CF to incorporate contexts or rich side information regarding users and items to alleviate the cold-start problem.

\section{Attribute-Aware Recommender Systems}
\label{section:survey}

\subsection{Overview}
\label{section:survey_overview}

\begin{figure}
    \centering
    \includegraphics[width=0.7\linewidth]{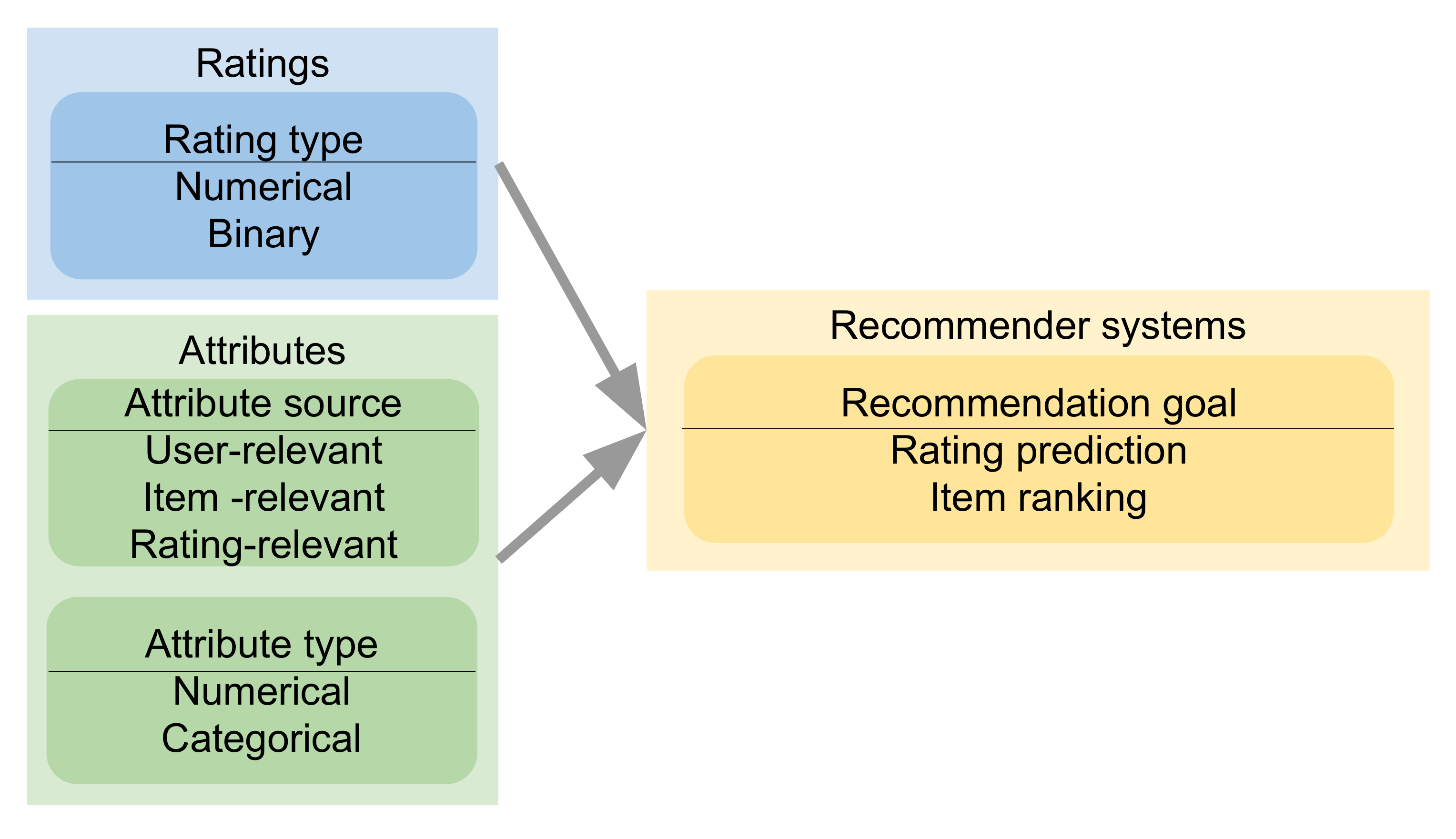}
    \caption{Model design flow of attribute-aware collaborative filtering based recommender system. When reading ratings and attributes for a proposed approach, we have to consider the sources and the types of attributes or ratings, which could affect the recommendation goals and currently common model designs. The evaluation of a proposed recommender system much depends on chosen recommendation goals.}
    \label{figure:design_flow}
\end{figure}

Attribute-aware recommendation models are proposed to tackle the challenges of integrating additional information from user/item/rating.
There are two strategies to design attribute-aware collaborative filtering-based systems.
One direction is to combine content-based recommendation models with CF models, which can directly accept attributes as content to perform recommendation.
On the other hand, researchers also try to extend an existing collaborative filtering algorithm such that it leverages attribute information.

Rather, we will focus on four important factors of designing a attribute-aware recommender system in current works, as shown in Figure \ref{figure:design_flow}.
They are specifically discussed from Section \ref{section:attribute_source} to \ref{section:recommendation_goal}.
With respect to input data, \emph{attribute sources} determine whether a attribute vector is relevant to users, items or ratings.
For example, attribute \emph{age} describes a user instead of item; \emph{rating time} must be appended to ratings, representing when the rating event occurred. 
Different models impose distinct strategies to integrate attributes of specific sources.
Additionally, a model may constrain \emph{attribute types} that can be used.
For instance, graph-based collaborative filtering realizations define attributes as node types, which is not appropriate for numerical attributes.
\emph{Rating types} are even the factor that is emphasized by most model designers.
Beside usual numerical ratings, many recommendation models concentrate on binary rating data, where the ratings represent whether users interact with items.
Finally, different recommender systems emphasize on different \emph{recommendation goals}.
One is to predict the ratings from users to items through minimizing the error between the predicted and real ratings.
Another is to produce the ranking among items given a user, instead of caring about the real rating value of a single item.
We then give a table to summarize the design categories of all the surveyed papers in Section \ref{section:summary}.

Throughout this paper, we will use $\bm{X} = \begin{bmatrix} \bm{x}_{1} \bm{x}_{2} \ldots \bm{x}_{N} \end{bmatrix} \in \mathbb{R}^{K \times N}$ to denote the attribute matrix, where each column $\bm{x}_{i}$ represents a $K$-dimensional attribute vector of entity $i$.
Here an entity can refer to a user, an item or a rating, determined by attribute sources (discussed in Section \ref{section:attribute_source}).
If attributes are limited categorical, then $\bm{X} \in \{ 0, 1 \} ^{K \times N}$ can be represented by one-hot encoding (discussed in Section \ref{section:attribute_type}).
Note that our survey does not include models designed specifically for a certain type of attributes, rather covers models that are general enough to accept different types of attributes. For example, Collaborative Topic Regression (CTR) \cite{Wang11CTM} extends matrix factorization with Latent Dirichlet Allocation (LDA) to import text attributes. 
Social Regularization \cite{Ma11RSS} specifically utilizes user social networks to regularize the learning of matrix factorization. Both models are not included since they are not generally enough to deal with general attributes. 

\subsection{Sources of Attributes}
\label{section:attribute_source}

Attributes usually come from a variety of sources. 
Typically, \emph{side information} refers to the attributes appended to users or items.
In contrast, keyword \emph{contexts} indicate the attributes relevant to ratings.
Ratings from the same user can be attached to different contexts, such as "locations where users rate items".
The recommendation models considering rating-relevant attributes are usually called \emph{context-aware recommender systems}.
Although contexts in some papers could include user-relevant or item-relevant ones, in this paper we tend to be precise and use the term \emph{contexts} for only rating-relevant attributes. 

Sections \ref{section:side_information} and \ref{section:context} respectively introduce different attribute sources.
It is worth mentioning our observation as follows.
Even though some of the models we surveyed demand side information, while others require context information, we discover that the two sets of attributes can be represented in a unified manner and thus both types of models can be applied.
We will discuss such unified representation in Section \ref{section:side_information_context} and \ref{section:context_side_information}.

\subsubsection{Side Information: User-relevant or Item-relevant Attributes}
\label{section:side_information}

In the surveyed papers, \emph{side information} could refer to user-relevant attributes, item-relevant attributes or both.
User-relevant attributes determine the characteristics of a user, such as "age", "gender", "education", etc.
In contrast, item-relevant attributes describe the properties of an item, like "movie running time", "product expiration data", etc.
Below we discuss user-relevant attributes, but all the statements can be applied to item-relevant attributes.
Given user-relevant attributes, we can express them with matrix $\bm{X} \in \mathbb{R}^{K \times N_{u}}$ where $N_{u}$ is the number of users.
Each column of $\bm{X}$ is corresponding to $K$ attribute values of a specific user.
The most important characteristic of user-relevant attributes is that they are \emph{assumed unchanged} with the rating process of a user.
For example, every rating from the same user share the identical user-relevant attribute "age".
In other words, even without any of a user's ratings in collaborative filtering, the user's rating behaviors on items could be still extracted from other users that have similar user-relevant attribute values.
Attribute-aware recommender systems that address the cold-start user problems (i.e., there are few ratings of a user) typically adopt user-relevant attributes as their auxiliary information under collaborative filtering.
The attribute leverage methods are presented in Section \ref{section:common_integration}.

Readers may ask why not distinguish user-relevant attributes and item-relevant attributes.
By our observations during survey, most of the recommendation approaches have symmetric model designs for users and items.
In matrix factorization-based methods, rating matrix $\bm{R}$ is factorized into two matrices $\bm{W}$ and $\bm{H}$, respectively referring to user and item latent factors.
However matrix factorization does not change its learning results if we exchange the rows and columns of $\bm{R}$.
Despite the exchange of rows and columns, $\bm{W}$ and $\bm{H}$ just exchange what they learn from ratings: $\bm{W}$ for items but $\bm{H}$ for users.

Following the above conclusions,some of the related work could be further extended in our opinions.
If one attribute-aware recommender system claims to be designed only for user-relevant attributes, then readers could put a symmetric model design for item-relevant attributes, to obtain a more general model.

\subsubsection{Contexts: Rating-relevant Attributes}
\label{section:context}

Collaborative filtering-based recommender systems usually define ratings as \emph{the} interaction between users and items, though it is likely to have more than one interactions.
Since ratings are still the focus of recommender systems, other types of interactions, or rating-relevant attributes, are called \emph{contexts} in related work.
For example, the "time" and the "location" that a user rates an item are recorded with the occurrence of the rating behavior.
Rating-relevant attributes change with rating behaviors, and thus they could offer auxiliary data about why a user determines to give a rating to an item.
Moreever, rating-relevant attributes could capture rating preference change of a user.
If we have time information appended to ratings, then attribute-aware recommender systems could discover users' preferences at different time.

The format of rating-relevant attributes is potentially more flexible than that of user-relevant or item-relevant ones.
In Section \ref{section:generalized_factorization}, we will introduce a factorization-based generalization of matrix factorization.
In this class of attribute-aware recommender systems, even the user and item latent factors are not required to predict ratings; mere rating-relevant attributes can do it using their corresponding latent factor vectors.

\subsubsection{Converting Side Information to Contexts}
\label{section:side_information_context}

Most attribute-aware recommender systems choose to leverage one of the attribute sources.
Some proposed approaches specifically incorporate user or item-relevant attributes, while others are designed for rating-relevant attributes only.
It seems that existing works should be applied according to which attribute sources they use.
However we argue that the usage of attribute-aware recommender systems could be independent of attribute sources, if we convert them to each other using a simple way.

Let $\bm{X} \in \mathbb{R}^{K_{X} \times N_{u}}$ be the user-relevant attribute matrix, where each column $\bm{x}_{u} \in \mathbb{R}^{K_{X}}$ is the attribute set of user $u$.
Similarly, let $Y \in \mathbb{R}^{K_{Y} \times N_{i}}, Z \in \mathbb{R}^{K_{Z} \times N_{r}}$ be respectively the matrices of item-relevant attributes and rating-relevant attributes.
Note that a column index of matrix $\bm{Z}$ is denoted by $\pi (u, i)$ which is associated with user $u$ and item $i$.
To express $\bm{X}$ or $\bm{Y}$ as $\bm{Z}$, a simple concatenation with respect to users and items can achieve the goal, as shown below:
\begin{align}
    \bm{z}_{\pi (u, i)}' = \begin{bmatrix}
        \bm{z}_{\pi (u, i)} \\ \bm{x}_{u} \\ \bm{y}_{i}
    \end{bmatrix} \in \mathbb{R}^{K_{Z} + K_{X} + K_{Y}} .
    \label{equation:side_information_context}
\end{align}
(\ref{equation:side_information_context}) implies that we just extend current rating-revelant attributes $\bm{z}_{\pi (u, i)}$ to $\bm{z}_{\pi (u, i)}'$, using the attributes $\bm{x}_{u}, \bm{y}_{i}$ from corresponding users or items.
If training data do not consist of $\bm{z}_{\pi (u, i)}, \bm{x}_{u}$ or $\bm{y}_{i}$, we can eliminate the notations on the right-hand side of (\ref{equation:side_information_context}).
Advanced attribute selection or dimensionality reduction methods could extract effective dimensions in $\bm{z}_{\pi (u, i)}'$, but the further improvement is beyond our scope.
If missing attribute values exist in $\bm{z}_{\pi (u, i)}'$, then we suggest directly filling $0$ in these attributes.
Please refer to to Section \ref{section:context_side_information} for our reasons.

\subsubsection{Converting Contexts to Side Information}
\label{section:context_side_information}

Following the topic in Section \ref{section:side_information_context}, reader may be curious of how to reversely convert rating-relevant attributes as user or item-relevant ones.
In the following paragraphs, we adopt the same notations in (\ref{section:side_information_context}).
Due to symmetric designs for $\bm{X}$ and $\bm{Y}$, we demonstrate only the conversion from $\bm{Z}$ to $\bm{X}$.
The concatenation is still the simplest way to express $\bm{Z}$ as one part of $\bm{X}$:
\begin{align}
    \bm{x}_{u}' = \begin{bmatrix}
        \bm{x}_{u}^{\top} & \bm{z}_{\pi (u, 1)}^{\top} & \bm{z}_{\pi (u, 2)}^{\top} & \ldots & \bm{z}_{\pi (u, i)}^{\top} & \ldots & \bm{z}_{\pi (u ,N_{i})}^{\top}
    \end{bmatrix} ^{\top} \in \mathbb{R}^{K_{X} + K_{Z} N_{i}} .
    \label{equation:context_side_information}
\end{align}
All the rating-relevant attributes $\bm{z}_{(u, 1)}, \bm{z}_{(u, 2)}, \ldots, \bm{z}_{(u, N_{i})}$ from $N_{i}$ items must be associated with user $u$.
$\bm{x}_{u}$ is thus extended to $\bm{x}_{u}'$ by appending these attributes.
Note that there exist a large number of missing attributes on the right-hand side of (\ref{equation:context_side_information}), since most items were never rated by user $u$ in real-world data.
Eliminating missing $\bm{z}_{\pi (u, i)}$, as what we do in Section \ref{section:side_information_context}, turns out different dimensions between two user-relevant attributes $\bm{x}_{u}', \bm{x}_{v}'$.
To our knowledge, there is no user-relevant attribute-aware recommender system allowing individual dimensions of user-relevant attributes.

Readers can run attribute imputation approaches to remove missing values in $\bm{x}_{u}'$.
However in our opinions, simply filling $0$ in missing elements could be satisfactory for attribute-aware recommender systems.
We explain our reasons by the observations in Section \ref{section:attribute_type}.
For numerical attributes, (\ref{equation:attribute_function_latent_factor}) (\ref{equation:latent_factor_function_attribute}) (\ref{equation:attribute_function_regression}) show the various attribute modeling methods.
If attributes $\bm{X}$ are mapped through function $f$ like (\ref{equation:attribute_function_latent_factor}) or (\ref{equation:attribute_function_regression}), then zero attributes in $f$ will cause no mapping effect (except constant intercept of $f$).
If attributes $\bm{X}$ are fitted by latent factors onto function $f$ such as (\ref{equation:latent_factor_function_attribute}), then typically in the objective design, we can skip the objective computation of missing attributes.
As for categorical attributes, we exploit one-hot encoding to represent them with numerical values.
Then categorical attributes can be handled as numerical attributes.

\subsection{Attribute Types}
\label{section:attribute_type}

In most cases, attribute-aware recommender systems accept a real-valued attribute matrix $\bm{X}$.
However we notice that some attribute-aware recommender systems require attributes to be categorical, which is typically represented by binary encoding.
Specifically, these approaches have to demand a binary attribute matrix where attributes of value $1$ can be modeled as discrete latent information someway.
The summary of both types of attributes are introduced in Section \ref{section:numerical_attribute} and \ref{section:categorical_attribute}.

It is trivial to put one-hot categorical attributes into numerical attribute-aware recommender systems, since binary values $\{ 0, 1 \} \subset \mathbb{R}$.
Nonetheless putting numerical attributes into categorical attribute-aware recommendation approaches has to take a risk of losing attribute information (e.g., quantization processing).

\subsubsection{Numerical Attributes}
\label{section:numerical_attribute}

In our paper, numerical attributes refer to the set of real-valued attributes, i.e., attribute matrix $\bm{X} \in \mathbb{R}^{K \times N}$.
We also classify integer attributes (like movie ratings $\{ 1, 2, 3, 4, 5 \}$) to numerical attributes.
Most of the relevant papers model numerical attributes as their default inputs in recommender systems, as common machine learning approaches.

There are three common model designs for numerical attributes $\bm{X}$ to affect recommender systems.
First, we can map $\bm{X}$ to latent factor space by function $f_{\bm{\theta}}$ with parameters $\bm{\theta}$, and then fit the corresponding user or item latent factor vectors:
\begin{align}
    \argmin_{\bm{\theta}, \bm{W}} & \left \| f_{\bm{\theta}} ( \bm{X} ) - \bm{W} \right \| \text{ or } \bm{W} = f_{\bm{\theta}} ( \bm{X} )  \text{ for user-relevant attributes} , \nonumber \\
    \argmin_{\bm{\theta}, \bm{H}} & \left \| f_{\bm{\theta}} ( \bm{X} ) - \bm{H} \right \| \text{ or } \bm{H} = f_{\bm{\theta}} ( \bm{X} ) \text{ for item-relevant attributes} .
    \label{equation:attribute_function_latent_factor}
\end{align}
Second, like the reverse of (\ref{equation:attribute_function_latent_factor}), we define a mapping function $f_{\bm{\theta}}$ such that mapped values from user or item latent factors can be close to observed attributes:
\begin{align}
    \argmin_{\bm{\theta}, \bm{W}} & \left \| f_{\bm{\theta}} ( \bm{W} ) - \bm{X} \right \| \text{ or } \bm{X} = f_{\bm{\theta}} ( \bm{W} ) \text{ for user-relevant attributes} , \nonumber \\
    \argmin_{\bm{\theta}, \bm{H}} & \left \| f_{\bm{\theta}} ( \bm{H} ) - \bm{X} \right \| \text{ or } \bm{X} = f_{\bm{\theta}} ( \bm{H} ) \text{ for item-relevant attributes} .
    \label{equation:latent_factor_function_attribute}
\end{align}
Finally, numerical attributes can be put into function $f_{\bm{\theta}}$ that is independent of existing user or item latent factors in matrix factorization:
\begin{align}
    \argmin_{\bm{\theta}, \bm{W}, \bm{H}} \left \| f_{\bm{\theta}} ( \bm{X} ) + \bm{W}^{\top} \bm{H}  - \bm{R} \right \| .
    \label{equation:attribute_function_regression}
\end{align}
(\ref{equation:attribute_function_latent_factor}) and (\ref{equation:latent_factor_function_attribute}) are typically seen in user-relevant or item-relevant attributes, while rating-relevant attributes are often put into (\ref{equation:attribute_function_regression})-like formats.
However we emphasize that attribute-aware recommender systems are not restricted to these three model designs.

\subsubsection{Categorical Attributes}
\label{section:categorical_attribute}

The values of a numerical attribute are ordered, though the values of a categorical attribute show no ordered relations of each other.
Given a categorical attribute $\text{Food} \in \{ \text{Rice}, \text{Noodles}, \text{Other} \}$, the meanings of the attribute values do not imply which one is larger than the other.
Thus, it is improper to give categorical attributes ordered dummy variables, like 
$\text{Rice} = 0, \text{Noodles} = 1, \text{Other} = 2$ that could incorrectly imply $\text{Rice} < \text{Noodles} < \text{Other}$, which makes machine learning models misunderstand attribute information.
The most common solution to categorical attribute transformation is \emph{one-hot encoding}.
We generate $d$-dimensional binary attributes that correspond to the $d$ values of a categorical attribute.
Each of the $d$ binary attributes indicate the current value of a categorical attribute.
For example, we express attribute $\text{Food} \in \{ \{ 1, 0, 0 \}, \{ 0, 1, 0 \}, \{ 0, 0, 1 \} \}$.
They are corresponding to the original values $\{ \text{Rice}, \text{Noodles}, \text{Other} \}$.
Since a categorical attribute exactly equals to one value, the mapped binary attributes contain only a $1$ and others $0$.
Once all the categorical attributes are converted to one-hot encoding expressions, we are allowed to apply them to existing numerical attribute-aware recommender systems.

However certain relevant papers are suitable for, or even limited to, categorical attributes.
Heterogeneous graph-based methods (Section \ref{section:heteregeneous_graph}) add new nodes (e.g., three nodes named $\text{Rice}, \text{Noodles}, \text{Other}$) to represent the values of categorical attributes.
Following the latent factor ideas in matrix factorization, some methods propose to assign each categorical attribute value a low-dimensional latent factor vector (e.g., each of $\text{Rice}, \text{Noodles}, \text{Other}$ has a latent factor vector $\bm{w} \in \mathbb{R}^{K}$).
Then these vectors are jointly learned with classical user or item latent factors in attribute-aware recommender systems.

\subsection{Rating Types}
\label{section:rating_type}

Although we always define term \emph{ratings} as the interactions between users and items in this paper, some existing works claim the difference between explicit opinions and implicit feedback.
Taking dataset MovieLens for example, a user gives a rating value in $\{ 1, 2, 3, 4, 5 \}$ toward an item.
The value denotes the \emph{explicit opinion}, which quantifies the preference of the user to that item.
How recommendation methods handling such type of ratings will be introduced in Section \ref{section:numerical_rating}.

Even though modeling explicit opinions is more beneficial for future recommendation, such data is more difficult to gather from users.
Users may hesitate to show their preferences due to privacy consideration, or they are not willing to spend time labeling explicit ratings.
Instead, recommender system developers are more likely to collect \emph{implicit feedback}, like user browsing logs.
Such datasets record a series of binary values, each of which imply whether a user ever saw an item.
User preferences behind implicit feedback assume that all the items seen by a user must be more preferred by the user, than those items having never seen.
We deeply discuss the type of ratings in Section \ref{section:binary_rating}.

There exist controversial numerical rating data, like "the number that a user ever clicked the hyperlink toward the page of an item".
Some of the related work may define such data as implicit feedback, because the number of clicks is not equivalent to explicit user preferences.
However in this paper, we still identify them as explicit opinions.
With respect to model designs, related recommendation approaches take no difference between such data and explicit opinions.

\subsubsection{Explicit Opinions: Numerical Ratings}
\label{section:numerical_rating}

A numerical rating matrix $r \in \mathbb{R}$ expresses users' opinions on items.
Actually numerical ratings in real-world scenarios are often represented by positive integers, such as MovieLens ratings $r \in \{ 1, 2, 3, 4, 5 \}$.
Despite no explicit statements in related work, typically we suppose that a higher rating implies a more positive opinion.

Since in most datasets the gathered rating values are positive, it could incur an unbiased learning problem.
Matrix factorization could not learn the rating bias due to the non-zero mean of ratings $\E ( r ) \neq 0$.
Specifically, in vanilla matrix factorization, we have regularization terms $\| \bm{W} \|_{F}^{2}$ and $\| \bm{H} \|_{F}^{2}$ for user and item latent factor matrix $\bm{W}, \bm{H}$.
That is, we require the expected value $\E ( \bm{W} ) = \E ( \bm{H} ) = \bm{0}$ in the viewpoint of corresponding normal distributions.
Given rating $r_{ui}$ of user $u$ to item $i$, and assuming the independence of $\bm{W}, \bm{H}$ as probabilistic matrix factorization does, we obtain the expected value of rating estimate $\E ( \hat{r}_{ui} ) = \E ( \bm{w}_{u}^{\top} \bm{h}_{i} ) = 0 \ \forall \ (u, i)$, which cannot closely fits true ratings if $\E ( r_{ui} ) \neq 0$.
Biased matrix factorization can alleviate the problem by absorbing the non-zero mean with additional bias terms.
Besides, we are allowed to normalize all the ratings (subtract the rating mean from every rating) to make matrix factorization prediction unbiased.
Real-world numerical ratings also have finite maximum and minimum values.
Some recommendation models choose to normalize the ratings to range $r \in [0, 1]$, and then constrain the range of rating estimate $\sig ( \hat{r} ) \in ( 0, 1 )$ using the sigmoid function $\sig ( x ) = \frac{1}{1 + \exp ( - x ) }$.

\subsubsection{Implicit feedback: Binary Ratings}
\label{section:binary_rating}

Today there are more and more researches that are interested in the scenario of binary ratigns $r \in \{ 0, 1 \}$ (i.e., implicit feedback), since such rating data are more accessible, like "whether a user browsed the information about an item".
Online services do not have to require users to give an explicit numerical ratings, which are often gathered less than binary ones.

Nevertheless, we observe only positive ratings $r = 1$; negative ratings $r = 0$ do not exist in training data.
Taking browsing logs as example, the data collect the items that are browsed by a user (i.e., positive examples).
The items not in the browsing data could imply either absolutely unattractive ($r = 0$) or just unknown ($r \in \{ 0, 1 \}$) to the user.
\emph{One-class collaborative filtering} methods are proposed to address the problem.
Such methods often claim two assumptions:
\begin{itemize}
    \item An item must be attractive to a user ($r = 1$), as long as the user ever saw the item.
    
    \item Since we cannot distinguish the two reasons (absolutely unattractive or just unknown) why an item is unseen, such methods suppose that all the unseen items are less attractive ($r = 0$).
    However the number of unseen items are practically much more than that of seen items.
    To alleviate the problems learning bias toward $r = 0$ together with learning speed, we exploit \emph{negative sampling} that sub-samples partial unseen ratings for training.
    \end{itemize}
To build an objective function satisfying the above assumptions, we can choose either pointwise learning (Section \ref{section:rating_prediction}) or pairwise learning (Section \ref{section:item_ranking}).
Area Under ROC Curve (AUC), Normalized Discounted Cumulative Gain (NDCG), Mean Average Precision (MAP), precision and recall are often used to justfy the quality of recommender systems for binary ratings.

\subsection{Recommendation Goals}
\label{section:recommendation_goal}

Any recommender system needs human developers to offer a training goal of recommendation.
Since collaborative filtering-based recommender systems rely on ratings, the most straightforward goal is to infer what rating will be given by a user for an unseen item, named \emph{rating prediction}.
If the ratings of every item can be accurately predicted, then for any user, a recommender system just sorts predicted ratings and recommends the items of the highest predicted ratings.
In machine learning, such goal for model-based recommender systems can be described as a \emph{pointwise learning}.
That is, given a pair of user and item, a pointwise learning recommendation model directly minimize the error of predicted ratings and true ones.
The related mathematical details is put in Section \ref{section:rating_prediction}.

However in general, our ultimate goal is to recommend unseen items to users without concerning about how these items are rated.
All unseen items in pointwise learning are finally ranked in descent order of their ratings.
In other words, what we truly care about is the order of ratings, but not the true rating values.
Also, some research papers figure out that low error of rating prediction is not always equivalent to high quality of recommended item lists.
Recent model-based collaborative filtering models begin to set optimization goals of \emph{item ranking}.
That is, for the same user, such models maximize the differences between high-rated items and low-rated ones in training data.
The implementation of item ranking includes \emph{pairwise learning and listwise learning} in machine learning domains.
Both learning ideas try to compare the potentially related ranks between at least two items for the same user.
Section \ref{section:item_ranking} will present how to define optimization criteria for item ranking.

\subsubsection{Rating Prediction: Pointwise Learning}
\label{section:rating_prediction}

In the training stage, given a ground-truth rating $r$, a recommender system needs to make a rating estimate $\hat{r}$ that is expected to predict $r$.
Model-based collaborative filtering methods (e.g., matrix factorization) build an objective function to be optimized (either maximization or minimization) for recommendation goals.
For numerical ratings $r \in \mathbb{R}$ (Section \ref{section:numerical_rating}) of users $u$ to items $i$, we can minimize the error between the ground truth and the estimate as follows:
\begin{align}
    \argmin_{\hat{r}} & \sum_{(u, i) \mid r_{ui} \in \delta ( \bm{R} )} \left ( \hat{r}_{ui} - r_{ui} \right ) ^{2} , \nonumber \\
    \argmin_{\hat{r}} & \sum_{(u, i) \mid r_{ui} \in \delta ( \bm{R} )} \left ( \sig ( \hat{r}_{ui} ) - r_{ui} \right ) ^{2} .
    \label{equation:rating_prediction_rmse_minimization}
\end{align}
$\delta ( \bm{R} )$ is the set of training ratings, which are the non-missing entries in rating matrix $\bm{R}$.
As Section \ref{section:numerical_rating} mentioned, if ground-truth ratings $r$ are normalized to $[0, 1]$ in data pre-processing, then in (\ref{equation:rating_prediction_rmse_minimization}) we can put sigmoid function $\sig ( x ) = \frac{1}{1 + \exp ( - x ) } \in ( 0, 1 )$ onto rating estimate $\hat{r}$ that could more fit $r$.
With respective to probability, (\ref{equation:rating_prediction_rmse_minimization}) is equivalent to maximizing normal likelihood:
\begin{align}
    \argmax_{\hat{r}} & \prod_{(u, i) \mid r_{ui} \in \delta ( \bm{R} )} \mathcal{N} \left ( r_{ui} \mid \mu = \hat{r}_{ui}, \sigma^{2} \right )  \nonumber \\
    \argmax_{\hat{r}} & \prod_{(u, i) \mid r_{ui} \in \delta ( \bm{R} )} \mathcal{N} \left ( r_{ui} \mid \mu = \sig ( \hat{r}_{ui} ), \sigma^{2} \right )
    \label{equation:rating_prediction_normal_likelihood}
\end{align}
where $\mathcal{N}$ means the probability density function of a normal distribution with mean $\mu = \hat{r}$ and variance $\sigma^{2}$ being a predefined uncertainty between $r$ and $\hat{r}$.
Taking $( - \log )$ on (\ref{equation:rating_prediction_normal_likelihood}) will obtain (\ref{equation:rating_prediction_rmse_minimization}).
Evidently both (\ref{equation:rating_prediction_rmse_minimization}) and (\ref{equation:rating_prediction_normal_likelihood}) make the rating prediction problem be addressed by regression models over ratings $\bm{R}$.

For binary ratings $r \in \{ 0, 1 \}$ (Section \ref{section:binary_rating}), beside (\ref{equation:rating_prediction_rmse_minimization}) with the sigmoid function, such data can be modeled as a binary classification problem.
Specifically we model $r = 1$ as the positive set, $r = 0$ as the negative set.
Then logistic regression (or Bernoulli likelihood) is built for rating prediction:
\begin{align}
    \argmax_{\hat{r}} & \prod_{(u, i) \mid 1 = r_{ui} \in \delta ( \bm{R} )} \Pr \left ( \hat{r}_{ui} = 1 \right ) \prod_{(u, i) \mid 0 = r_{ui} \in \delta ( \bm{R} )} \Pr \left ( \hat{r}_{ui} = 0 \right ) \nonumber \\
    = & \underbrace{ \prod_{(u, i) \mid 1 = r_{ui} \in \delta ( \bm{R} )} \sig \left ( \hat{r}_{ui} \right ) }_{\text{Positive set}} \underbrace{ \prod_{(u, i) \mid 0 = r_{ui} \in \delta ( \bm{R} )} \left ( 1 - \sig \left ( \hat{r}_{ui} \right ) \right ) }_{\text{Negative set}} .
    \label{equation:rating_prediction_bernoulli_likelihood}
\end{align}

The optimization of (\ref{equation:rating_prediction_rmse_minimization}) (\ref{equation:rating_prediction_normal_likelihood}) corresponds to an evaluation metric: Root Mean Squared Error (RMSE), whose formal definition is shown as follows:
\begin{align}
    \text{RMSE} = \sqrt{ \frac{1}{| \delta ( \bm{R} ) |} \sum_{(u, i) \mid r_{ui} \in \delta ( \bm{R} )} \left (\hat{r}_{ui} - r_{ui} \right ) ^{2} } .
    \label{equation:rmse_definition}
\end{align}
For the convenience of optimization, the regression models eliminate the root function from RMSE, i.e., they optimizes MSE in fact.
Since the root function is monotonically increasing, minimizing MSE is equivalent to minimizing RMSE (\ref{equation:rmse_definition}).

Even though a recommender system selects to optimize (\ref{equation:rating_prediction_bernoulli_likelihood}), the binary classification also corresponds to minimizing RMSE, except that rating estimate $\hat{r}$ is replaced with sigmoid-applied version $\sig ( \hat{r} )$.
Observing the maximization of (\ref{equation:rating_prediction_bernoulli_likelihood}), we obtain a conclusion: $\sig ( \hat{r} ) \to 1$ as $r = 1$, or $\sig ( \hat{r} ) \to 0$ as $r = 0$.
In other words, (\ref{equation:rating_prediction_bernoulli_likelihood}) tries to minimize the error between $\sig ( \hat{r} ) \in (0, 1)$ and $r \in \{ 0, 1 \}$, which has the same optimization goal as RMSE (\ref{equation:rmse_definition}).

\subsubsection{Item Ranking: Pairwise Learning and Listwise Learning}
\label{section:item_ranking}

This class of recommendation goal requires a model to correctly rank two items in the training data, even though the model could inaccurately predict the value of a single rating.
Since recommender systems concern about item ranking for the same user $u$ more than ranking for different users, existing works sample item pairs $(i, j)$ where $r_{ui} > r_{uj}$ given fixed user $u$ (i.e., item $i$ is ranked higher than item $j$ for user $u$), and then let rating estimate pair $(\hat{r}_{ui}, \hat{r}_{uj})$ learn to rank the two items with $\hat{r}_{ui} > \hat{r}_{uj}$.
In particular, we can use the sigmoid function $\sig ( x ) = \frac{1}{1 + \exp ( - x )}$ to model the probabilities in the pairwise comparison likelihood:
\begin{align}
    \argmax_{\hat{r}} & \prod_{ (u, i, j) \mid \substack{ \{ r_{ui}, r_{uj} \} \subseteq \delta ( \bm{R} ) , \\ r_{ui} > r_{uj} } } \Pr \left ( \hat{r}_{ui} > \hat{r}_{uj} \right ) \nonumber \\
    = & \prod_{(u, i, j) \mid \substack{ \{ r_{ui}, r_{uj} \} \subseteq \delta ( \bm{R} ) , \\ r_{ui} > r_{uj} } } \sig \left ( \hat{r}_{ui} - \hat{r}_{uj} \right ) .
    \label{equation:item_ranking_bernoulli_likelihood}
\end{align}
Taking $( - \log )$ on objective function (\ref{equation:item_ranking_bernoulli_likelihood}) will become the log-loss function.
Bayesian Personalized Ranking (BPR) \cite{Rendle09BPR} first investigates the usage and the optimization of  (\ref{equation:item_ranking_bernoulli_likelihood}) for recommender systems.
BPR shows that (\ref{equation:item_ranking_bernoulli_likelihood}) maximizes a differentiable smoothness of evaluate metric Area Under ROC Curve (AUC), one of whose definitions is:
\begin{align}
    \text{AUC} = \frac{1}{T} \sum_{ (u, i, j) \mid \substack{ \{ r_{ui}, r_{uj} \} \subseteq \delta ( \bm{R} ) , \\ r_{ui} > r_{uj} } } \mathbb{I} \left ( \hat{r}_{ui} > \hat{r}_{uj} \right ) ,
    \label{equation:auc_definition}
\end{align}
where $T$ is the number of training instances $\{ (u, i, j) \mid \{ r_{ui}, r_{uj} \} \subseteq \delta ( \bm{R} ), r_{ui} > r_{uj} \}$.
$\mathbb{I} ( x ) \in \{ 0, 1 \}$ denote an indicator function whose output is $1$ if and only if condition $x$ is judged true.
We show the connection between (\ref{equation:item_ranking_bernoulli_likelihood}) and (\ref{equation:auc_definition}) below:
\begin{align}
    \argmax_{\hat{r}} \  (\ref{equation:auc_definition}) & = \sum_{ (u, i, j) \mid \substack{ \{ r_{ui}, r_{uj} \} \subseteq \delta ( \bm{R} ) , \\ r_{ui} > r_{uj} } } \mathbb{I} \left ( \hat{r}_{ui} - \hat{r}_{uj} > 0 \right ) \nonumber \\
    & \approx \sum_{ (u, i, j) \mid \substack{ \{ r_{ui}, r_{uj} \} \subseteq \delta ( \bm{R} ) , \\ r_{ui} - r_{uj} > 0 } } \sig \left ( \hat{r}_{ui} - \hat{r}_{uj} \right ) \nonumber \\
    & = \sum_{ (u, i, j) \mid \substack{ \{ r_{ui}, r_{uj} \} \subseteq \delta ( \bm{R} ) , \\ r_{ui} - r_{uj} > 0 } } \log \sig \left ( \hat{r}_{ui} - \hat{r}_{uj} \right ) \nonumber \\
    & = \log \prod_{ (u, i, j) \mid \substack{ \{ r_{ui}, r_{uj} \} \subseteq \delta ( \bm{R} ) , \\ r_{ui} - r_{uj} > 0 } } \sig \left ( \hat{r}_{ui} - \hat{r}_{uj} \right ) .
    \label{equation:auc_bernoulli_connection}
\end{align}
Under the condition of $\argmax$, we make non-differentiable indicator function $\mathbb{I} ( x )$ be approximated by differentiable sigmoid function $\sig ( x )$.
The maximization of (\ref{equation:auc_bernoulli_connection}) is equivalent to optimizing  (\ref{equation:item_ranking_bernoulli_likelihood}) due to the monotonically increasing logarithm function.
AUC evaluates whether all the predicted item pairs follow the ground-truth rating comparisons in the whole item list.
By our observation, most of the reviewed approaches based on item ranking build their objective functions with AUC optimization.
There are other choices of optimization functions to approxmately maximize AUC, like hinge loss:
\begin{align}
    \argmin_{\hat{r}} \sum_{(u, i, j) \mid \substack { \{ r_{ui}, r_{uj} \} \subseteq \delta ( \bm{R} ) , \\ r_{ui} > r_{uj} } } \max \left \{ 0, \hat{r}_{uj} - \hat{r}_{ui} \right \} .
    \label{equation:item_ranking_hinge_loss}
\end{align}

In the domain of top-$N$ recommendation, the item orders outside top-$N$ ranks is unimportant for recommender systems.
Maximizing AUC could fail to recommend items since AUC gives the same penalty to all items.
That is, a recommender system could gain high AUC when it accurately ranks the bottom-$N$ items, but it is not beneficial for real-world recommendation since a user pays attention to the top-$N$ items.
Listwise evaluation metrics like Mean Reciprocal Rank (MRR), Normalized Discounted Cumulative Gain (NDCG) or Mean Average Precision (MAP) are proposed to give different penalty values to item ranking positions.
There have been works to optimize differential versions of the above metrics, such as CliMF \cite{Shi12CLiMF}, SoftRank \cite{Taylor08SoftRank} and TFMAP \cite{Shi12TFMAP}.

As our observations to the surveyed papers, recommender systems reading binary ratings (Section \ref{section:binary_rating}) more prefer to optimize an item-ranking objective function.
Compared with numerical ratings (Section \ref{section:numerical_rating}), a single binary rating reveals less information on a user's absolute preference.
Pairwise learning methods could capture more information by modeling a user's relative preferences, because the number of rating pairs $r_{ui} = 1 > 0 = r_{uj}$ is more than the number of ratings for each user.

\subsection{Summary of Related Work}
\label{section:summary}

After introducing the above categories that we propose for attribute-aware recommender systems, we then demonstrate Table \ref{table:model_category}, listing which categories each paper belongs to.
Here Table \ref{table:model_category} also shows all the publications that we have surveyed.
We trace back to the publications to summarize the recent ten-year trend of attribute-aware recommender systems.

\begin{footnotesize}
\begin{longtable}{c c c c c c c c c c c}
    \caption{List of model categories. The numbers in parentheses refer to the corresponding sections for category elaborations. All the model names come from the proposing publications, except that we use the title abbreviations if the authors do not name their approaches. Long model names are commented in footnotes.} \\
    \toprule
    \multirow{2}{*}{Model} & \multirow{2}{*}{Year} & \multicolumn{3}{c}{Attri. Source (\ref{section:attribute_source})} & \multicolumn{2}{c}{Attri. Type (\ref{section:attribute_type})} & \multicolumn{2}{c}{Rating Type (\ref{section:rating_type})} & \multicolumn{2}{c}{Recom. Goal (\ref{section:recommendation_goal})} \\
    \cline{3-11}
     & & User & Item & Rating & Num. & Cat. & Num. & Bin. & Pred. & Rank. \\
     & & (\ref{section:side_information}) & (\ref{section:side_information}) & (\ref{section:context}) & (\ref{section:numerical_attribute}) & (\ref{section:categorical_attribute}) & (\ref{section:numerical_rating}) & (\ref{section:binary_rating}) &     (\ref{section:rating_prediction}) & (\ref{section:item_ranking}) \\
    \midrule
    \endhead
    CMF \cite{Singh08CMF} & 2008 & \checkmark & \checkmark &  & \checkmark &  & \checkmark &  & \checkmark &  \\
TBM \cite{Gunawardana08TBM} & 2008 &  & \checkmark &  & \checkmark &  &  & \checkmark & \checkmark &  \\
WNMCTF \cite{Yoo09WNMCTF} & 2009 & \checkmark & \checkmark &  &  & \checkmark & \checkmark &  & \checkmark &  \\
CAR-AUC \cite{Shin09CARAUC} & 2009 &  &  & \checkmark &  & \checkmark &  & \checkmark & \checkmark &  \\
Multi. Recom. \footnote{Multidimensional Recommendation} \cite{Weng09MultiRecom} & 2009 &  &  & \checkmark &  & \checkmark & \checkmark &  & \checkmark &  \\
\hline
RLFM \cite{Agarwal09RLFM} & 2009 & \checkmark & \checkmark & \checkmark & \checkmark &  & \checkmark &  & \checkmark &  \\
Unified Boltz \cite{Gunawardana09UnifiedBoltz} & 2009 &  & \checkmark &  & \checkmark &  &  & \checkmark & \checkmark &  \\
Matchbox \cite{Stern09Matchbox} & 2009 & \checkmark & \checkmark & \checkmark & \checkmark &  & \checkmark & \checkmark & \checkmark &  \\
BMFSI \cite{Porteous10BMFSI} & 2010 & \checkmark & \checkmark &  & \checkmark &  & \checkmark &  & \checkmark &  \\
wAMAN. \footnote{wAMANWithSchKW} \cite{Li10wAMANWithSchKW} & 2010 & \checkmark &  &  & \checkmark &  &  & \checkmark & \checkmark &  \\
\hline
CACF \cite{Lee10CACF} & 2010 &  &  & \checkmark & \checkmark &  &  & \checkmark & \checkmark &  \\
PLRM \cite{Li10PLRM} & 2010 & \checkmark & \checkmark &  &  & \checkmark & \checkmark &  & \checkmark &  \\
LAFM \cite{Gantner10LAFM} & 2010 & \checkmark & \checkmark &  & \checkmark &  &  & \checkmark &  & \checkmark \\
GPMF \cite{Shan10GPMF} & 2010 &  & \checkmark &  &  & \checkmark & \checkmark &  & \checkmark &  \\
LFL \cite{Menon10LFL} & 2010 &  &  & \checkmark & \checkmark &  & \checkmark &  & \checkmark &  \\
\hline
TF \cite{Karatzoglou10TF} & 2010 &  &  & \checkmark &  & \checkmark & \checkmark &  & \checkmark &  \\
GWNMTF \cite{Gu10GWNMTF} & 2010 & \checkmark & \checkmark &  & \checkmark &  & \checkmark &  & \checkmark &  \\
DPMF \cite{Adams10DPMF} & 2010 & \checkmark & \checkmark &  & \checkmark &  & \checkmark &  & \checkmark &  \\
SoRec \cite{Ma11SoRec} & 2011 & \checkmark & \checkmark &  & \checkmark &  & \checkmark &  & \checkmark &  \\
UGPMF \cite{Du11UGPMF} & 2011 & \checkmark &  &  & \checkmark &  &  & \checkmark &  & \checkmark \\
\hline
BMCF \cite{Yoo11BMCF} & 2011 & \checkmark & \checkmark &  & \checkmark &  & \checkmark &  & \checkmark &  \\
MCRI \cite{Fang11MCRI} & 2011 & \checkmark & \checkmark &  & \checkmark &  &  & \checkmark & \checkmark &  \\
Hybrid. \footnote{Hybrid+LogReg++} \cite{Menon11HybridLogReg} & 2011 &  &  & \checkmark & \checkmark &  &  & \checkmark & \checkmark &  \\
YMR \cite{Koenigstein11YMR} & 2011 & \checkmark & \checkmark &  &  & \checkmark & \checkmark &  & \checkmark &  \\
CAMF \cite{Baltrunas11CAMF} & 2011 &  &  & \checkmark &  & \checkmark & \checkmark &  & \checkmark &  \\
\hline
GFREC \cite{Lee11GFREC} & 2011 &  &  & \checkmark &  & \checkmark &  & \checkmark &  & \checkmark \\
FM \cite{Rendle11FM} & 2011 &  &  & \checkmark & \checkmark &  & \checkmark &  & \checkmark &  \\
FIP \cite{Yang11FIP} & 2011 & \checkmark & \checkmark &  & \checkmark &  &  & \checkmark & \checkmark &  \\
iTALS \cite{Hidasi12iTALS} & 2012 &  &  & \checkmark &  & \checkmark &  & \checkmark & \checkmark &  \\
HVBMCF \cite{Yoo12HVBMCF} & 2012 & \checkmark & \checkmark &  & \checkmark &  & \checkmark &  & \checkmark &  \\
\hline
LCR \cite{Weston12LCR} & 2012 &  & \checkmark &  & \checkmark &  &  & \checkmark &  & \checkmark \\
HierIntegModel \cite{Lu12HierIntegModel} & 2012 &  & \checkmark &  &  & \checkmark & \checkmark &  & \checkmark &  \\
SVDFeature \cite{Chen12SVDFeature} & 2012 & \checkmark & \checkmark & \checkmark & \checkmark &  & \checkmark &  & \checkmark &  \\
SSLIM \cite{Ning12SSLIM} & 2012 &  & \checkmark &  & \checkmark &  &  & \checkmark & \checkmark &  \\
KPMF \cite{Zhou12KPMF} & 2012 & \checkmark & \checkmark &  & \checkmark &  & \checkmark &  & \checkmark &  \\
\hline
TFMAP \cite{Shi12TFMAP} & 2012 &  &  & \checkmark &  & \checkmark &  & \checkmark &  & \checkmark \\
CCMF \cite{Bouchard13CCMF} & 2013 & \checkmark & \checkmark &  & \checkmark &  & \checkmark &  & \checkmark &  \\
GFMF \cite{Chen13GFMF} & 2013 & \checkmark & \checkmark &  & \checkmark &  & \checkmark &  & \checkmark &  \\
KBMF \cite{Gonen13KBMF} & 2013 & \checkmark & \checkmark &  & \checkmark &  & \checkmark &  & \checkmark &  \\
HBMFSI \cite{Park13HBMFSI} & 2013 & \checkmark & \checkmark &  & \checkmark &  & \checkmark &  & \checkmark &  \\
\hline
DACR \cite{Safoury13DACR} & 2013 & \checkmark &  &  &  & \checkmark & \checkmark &  & \checkmark &  \\
Maxide \cite{Xu13Maxide} & 2013 & \checkmark & \checkmark &  & \checkmark &  & \checkmark &  & \checkmark &  \\
MF-EFS \cite{Koenigstein13MFEFS} & 2013 &  & \checkmark &  & \checkmark &  &  & \checkmark & \checkmark &  \\
HeteroMF \cite{Jamali13HeteroMF} & 2013 & \checkmark & \checkmark &  & \checkmark &  & \checkmark &  & \checkmark &  \\
SoCo \cite{Liu13SoCo} & 2013 &  &  & \checkmark & \checkmark & \checkmark & \checkmark &  & \checkmark &  \\
\hline
C-CTR-SMF2 \cite{Chen14CCTRSMF2} & 2014 & \checkmark & \checkmark & \checkmark & \checkmark &  & \checkmark &  & \checkmark &  \\
VBMFSI-CA \cite{Kim14VBMFSICA} & 2014 & \checkmark & \checkmark &  & \checkmark &  & \checkmark &  & \checkmark &  \\
IMC \cite{Natarajan14IMC} & 2014 & \checkmark & \checkmark &  & \checkmark &  &  & \checkmark & \checkmark &  \\
CARS$^{2}$ \cite{Shi14CARS2} & 2014 & \checkmark & \checkmark &  & \checkmark &  & \checkmark &  & \checkmark & \checkmark \\
LLR \cite{Ji14LLR} & 2014 &  & \checkmark &  &  & \checkmark & \checkmark &  & \checkmark &  \\
\hline
GBFM \cite{Cheng14GBFM} & 2014 &  &  & \checkmark & \checkmark &  & \checkmark &  & \checkmark &  \\
SCF \cite{Sedhain14SCF} & 2014 & \checkmark &  &  &  & \checkmark & \checkmark &  & \checkmark &  \\
LCE \cite{Saveski14LCE} & 2014 &  & \checkmark &  & \checkmark &  & \checkmark &  & \checkmark &  \\
CSEL \cite{Zhang14CSEL} & 2014 & \checkmark & \checkmark &  &  & \checkmark & \checkmark &  & \checkmark &  \\
GPFM \cite{Nguyen14GPFM} & 2014 &  & \checkmark & \checkmark &  & \checkmark & \checkmark &  & \checkmark & \checkmark \\
\hline
NCRPD-MF \cite{Hu14NCRPDMF} & 2014 &  & \checkmark & \checkmark & \checkmark &  & \checkmark &  & \checkmark &  \\
HeteRec \cite{Yu14PHeteRec} & 2014 &  & \checkmark &  &  & \checkmark &  & \checkmark &  & \checkmark \\
CAPRF \cite{Gao15CAPRF} & 2015 & \checkmark & \checkmark & \checkmark &  &  & \checkmark &  &  & \checkmark \\
mSDA-CF \cite{Li15mSDACF} & 2015 & \checkmark & \checkmark &  & \checkmark &  & \checkmark &  & \checkmark &  \\
BIMC \cite{Shin15BIMC} & 2015 & \checkmark & \checkmark &  & \checkmark &  &  & \checkmark & \checkmark &  \\
\hline
Convex FM \cite{Blondel15ConvexFM} & 2015 &  &  & \checkmark & \checkmark &  & \checkmark &  & \checkmark &  \\
CDL \cite{Wang15CDL} & 2015 &  & \checkmark &  & \checkmark &  &  & \checkmark & \checkmark &  \\
LightFM \cite{Kula15LightFM} & 2015 & \checkmark & \checkmark &  &  & \checkmark &  & \checkmark & \checkmark &  \\
DCT \cite{Barjasteh15DCT} & 2015 & \checkmark & \checkmark &  & \checkmark &  & \checkmark &  & \checkmark &  \\
GFF \cite{Hidasi15GFF} & 2015 &  &  & \checkmark &  & \checkmark &  & \checkmark & \checkmark &  \\
\hline
CALR \cite{Liu15CALR} & 2015 & \checkmark & \checkmark &  & \checkmark &  & \checkmark &  & \checkmark &  \\
VBPR \cite{He16VBPR} & 2016 &  & \checkmark &  & \checkmark &  &  & \checkmark &  & \checkmark \\
GFF \cite{Hidasi16GFF} & 2016 &  &  & \checkmark &  & \checkmark &  & \checkmark & \checkmark &  \\
PNFM \cite{Blondel16PNFM} & 2016 &  &  & \checkmark & \checkmark &  & \checkmark &  & \checkmark &  \\
TCRM \cite{Kasai16TCRM} & 2016 &  &  & \checkmark &  & \checkmark & \checkmark &  & \checkmark &  \\
\hline
PCFSI \cite{Zhao16PCFSI} & 2016 &  & \checkmark &  & \checkmark &  & \checkmark &  & \checkmark &  \\
CKE \cite{Zhang16CKE} & 2016 &  & \checkmark &  & \checkmark &  &  & \checkmark &  & \checkmark \\
CRAE \cite{Wang16CRAE} & 2016 &  & \checkmark &  & \checkmark &  & \checkmark &  & \checkmark &  \\
SIMMCSI \cite{Lu16SIMMCSI} & 2016 & \checkmark & \checkmark &  & \checkmark &  & \checkmark &  & \checkmark &  \\
DSR \cite{Zheng16DSR} & 2016 & \checkmark & \checkmark &  &  & \checkmark & \checkmark &  & \checkmark &  \\
\hline
ALMM \cite{Chou16ALMM} & 2016 &  & \checkmark &  & \checkmark &  & \checkmark &  & \checkmark &  \\
FFM \cite{Juan16FFM} & 2016 & \checkmark & \checkmark & \checkmark & \checkmark &  &  & \checkmark & \checkmark &  \\
ReMF \cite{Yang16ReMF} & 2016 & \checkmark &  &  &  & \checkmark & \checkmark &  & \checkmark &  \\
TAPER \cite{Ge16TAPER} & 2016 &  &  & \checkmark &  & \checkmark &  & \checkmark & \checkmark &  \\
LPRRM-CF \cite{Chen16LRPPMCF} & 2016 &  &  & \checkmark &  & \checkmark & \checkmark &  & \checkmark &  \\
\hline
HeteRS \cite{Pham16HeteRS} & 2016 & \checkmark & \checkmark & \checkmark &  & \checkmark &  & \checkmark &  & \checkmark \\
MVM \cite{Cao16MVM} & 2016 &  &  & \checkmark & \checkmark &  & \checkmark &  & \checkmark &  \\
SQ \cite{Yu17SQ} & 2017 & \checkmark & \checkmark &  & \checkmark &  &  & \checkmark & \checkmark &  \\
LoCo \cite{Sedhain17LoCo} & 2017 & \checkmark &  &  & \checkmark &  &  & \checkmark & \checkmark &  \\
aSDAE \cite{Dong17aSDAE} & 2017 & \checkmark & \checkmark &  & \checkmark &  & \checkmark &  & \checkmark &  \\
\hline
CoEmbed \cite{Guo17CoEmbed} & 2017 & \checkmark & \checkmark &  & \checkmark &  &  & \checkmark & \checkmark &  \\
HMF \cite{Brouwer17HMF} & 2017 & \checkmark & \checkmark &  & \checkmark &  & \checkmark &  & \checkmark &  \\
DeepFM \cite{Guo17DeepFM} & 2017 & \checkmark & \checkmark &  & \checkmark &  &  & \checkmark & \checkmark &  \\
LDRSSI \cite{Zhao17LDRSSI} & 2017 &  & \checkmark &  & \checkmark &  &  & \checkmark & \checkmark &  \\
CGSI \cite{Zhou17CGSI} & 2017 & \checkmark & \checkmark & \checkmark & \checkmark & \checkmark & \checkmark &  & \checkmark &  \\
\hline
Func. Embed. \footnote{Functional Embedding} \cite{Chen17FuncEmbed} & 2017 & \checkmark & \checkmark &  & \checkmark &  &  & \checkmark & \checkmark & \checkmark \\
CVAE \cite{Li17CVAE} & 2017 &  & \checkmark &  & \checkmark &  & \checkmark &  & \checkmark &  \\
entity2rec \cite{Palumbo17entity2rec} & 2017 &  & \checkmark &  &  & \checkmark &  & \checkmark &  & \checkmark \\
NFM \cite{He17NFM} & 2017 &  &  & \checkmark & \checkmark &  & \checkmark &  & \checkmark &  \\
MFM \cite{Lu17MFM} & 2017 &  &  & \checkmark & \checkmark &  & \checkmark &  & \checkmark &  \\
\hline
Focused FM \cite{Beutel17FocusedFM} & 2017 &  & \checkmark &  &  & \checkmark & \checkmark &  & \checkmark &  \\
GB-CENT \cite{Zhao17GBCENT} & 2017 &  &  & \checkmark & \checkmark &  & \checkmark &  & \checkmark &  \\
CML \cite{Hsieh17CML} & 2017 &  & \checkmark &  & \checkmark &  &  & \checkmark &  & \checkmark \\
ATRank \cite{Zhou18ATRank} & 2018 &  &  & \checkmark &  & \checkmark & \checkmark &  & \checkmark &  \\
Div-HeteRec \cite{Nandanwar18DivHeteRec} & 2018 & \checkmark & \checkmark & \checkmark &  & \checkmark &  & \checkmark & \checkmark &  \\
\hline
HeteLearn \cite{Jiang18HeteLearn} & 2018 & \checkmark & \checkmark & \checkmark &  & \checkmark &  & \checkmark &  & \checkmark \\
RNNLatentCross \cite{Beutel18RNNLatentCross} & 2018 &  &  & \checkmark & \checkmark &  & \checkmark &  & \checkmark &  \\
DDL \cite{Zhang18DDL} & 2018 &  & \checkmark &  &  & \checkmark & \checkmark &  & \checkmark &  \\

    \bottomrule
    \label{table:model_category}
\end{longtable}
\end{footnotesize}

\section{Common Model Designs of Attribute-Aware Recommender Systems}
\label{section:common_integration}

In this section we formally introduce the common attribute integration methods of existing attribute-aware recommender systems.
If collaborative filtering approaches are modeled by user or item latent factor structures like matrix factorization, then attribute matrice become either the prior knowledge of the latent factors (Section \ref{section:discriminative_matrix_factorization}) or the generation outputs from the latent factors (Section \ref{section:generative_matrix_factorization}).
On the other hand, some of the works are actually the generalization of matrix factorization (Section \ref{section:generalized_factorization}).
Besides, the interactions between users and items can be recorded by a heterogeneous network, which can incorporate attributes by simply adding attribute-representing nodes (Section \ref{section:heteregeneous_graph}).
The major distinction of these four categories lies in the representation of the interactions of users, items and attributes.
The discriminative matrix factorization models extend the traditional MF by making the attributes prior knowledge input to learn the latent representation of users or items.
Generative matrix factorization further considers the distributions of attributes, and learn such together with the rating distributions.
Generalized factorization models view the user/item identity simply as a kind of attribute, and various models are designed for learning the low-dimensional representation vectors for rating prediction.
The last category of models propose to represent the users, items and attributes using a heterogeneous graph, where a recommendation task can be cast into a link prediction task on the heterogeneous graph.

\begin{table}[H]
\caption{Classification of attribute-aware recommender systems.}

\begin{tabular}{|c|c|c|}
\hline
DMF & Similarity & 
\makecell{\cite{Li10wAMANWithSchKW},\cite{Gu10GWNMTF},\cite{Du11UGPMF},\cite{Zhou12KPMF},
\cite{Barjasteh15DCT}, \cite{Yu17SQ},\cite{Adams10DPMF},
\cite{Chen14CCTRSMF2}, \cite{Gonen13KBMF}} \\
\cline{2-3}
 & Linear &
\makecell{\cite{Porteous10BMFSI},\cite{Menon10LFL},\cite{Menon11HybridLogReg},
\cite{He16VBPR}, \cite{Zhao16PCFSI}, \cite{Guo17CoEmbed},\cite{Zhao17LDRSSI}} \\
\cline{2-3}
 & Bilinear &
\makecell{\cite{Stern09Matchbox},\cite{Li10PLRM}, \cite{Agarwal09RLFM},\cite{Shin15BIMC}
\cite{Yang11FIP}, \cite{Chen12SVDFeature},\cite{Park13HBMFSI}, \cite{Xu13Maxide}, \cite{Kim14VBMFSICA}, \cite{Natarajan14IMC},\cite{Lu16SIMMCSI},\cite{Chou16ALMM}} \\
\hline
GMF & \makecell{Multiple \\ Matrix Factorization} &
\makecell{\cite{Sedhain17LoCo},\cite{Singh08CMF},\cite{Shan10GPMF},
\cite{Ma11SoRec},\cite{Yoo11BMCF},\cite{Fang11MCRI},\cite{Bouchard13CCMF},
\cite{Saveski14LCE},\cite{Gao15CAPRF},\cite{Ge16TAPER},\cite{Brouwer17HMF}} \\
\cline{2-3}
 & \makecell{Deep \\ Neural Networks} &
 \makecell{\cite{Li15mSDACF},\cite{Wang15CDL},\cite{Zhang16CKE},
 \cite{Wang16CRAE}, \cite{Dong17aSDAE}, \cite{Li17CVAE}} \\
\hline
GF & TF &
\makecell{\cite{Zhou17CGSI},\cite{Karatzoglou10TF},\cite{Hidasi12iTALS},
\cite{Hidasi15GFF},\cite{Kasai16TCRM}} \\
\cline{2-3}
 & FM &
 \makecell{\cite{He17NFM},\cite{Rendle11FM},\cite{Cheng14GBFM},
 \cite{Nguyen14GPFM},\cite{Blondel15ConvexFM},\cite{Blondel16PNFM},
 \cite{Juan16FFM},\cite{Cao16MVM},\cite{Guo17DeepFM},\cite{Lu17MFM}} \\
\hline
HG & &
\makecell{\cite{Yu14PHeteRec},\cite{Zheng16DSR},\cite{Palumbo17entity2rec}} \\
\hline
\end{tabular}
\end{table}

\begin{figure}[H]
    \centering
    \includegraphics[width=0.4\linewidth]{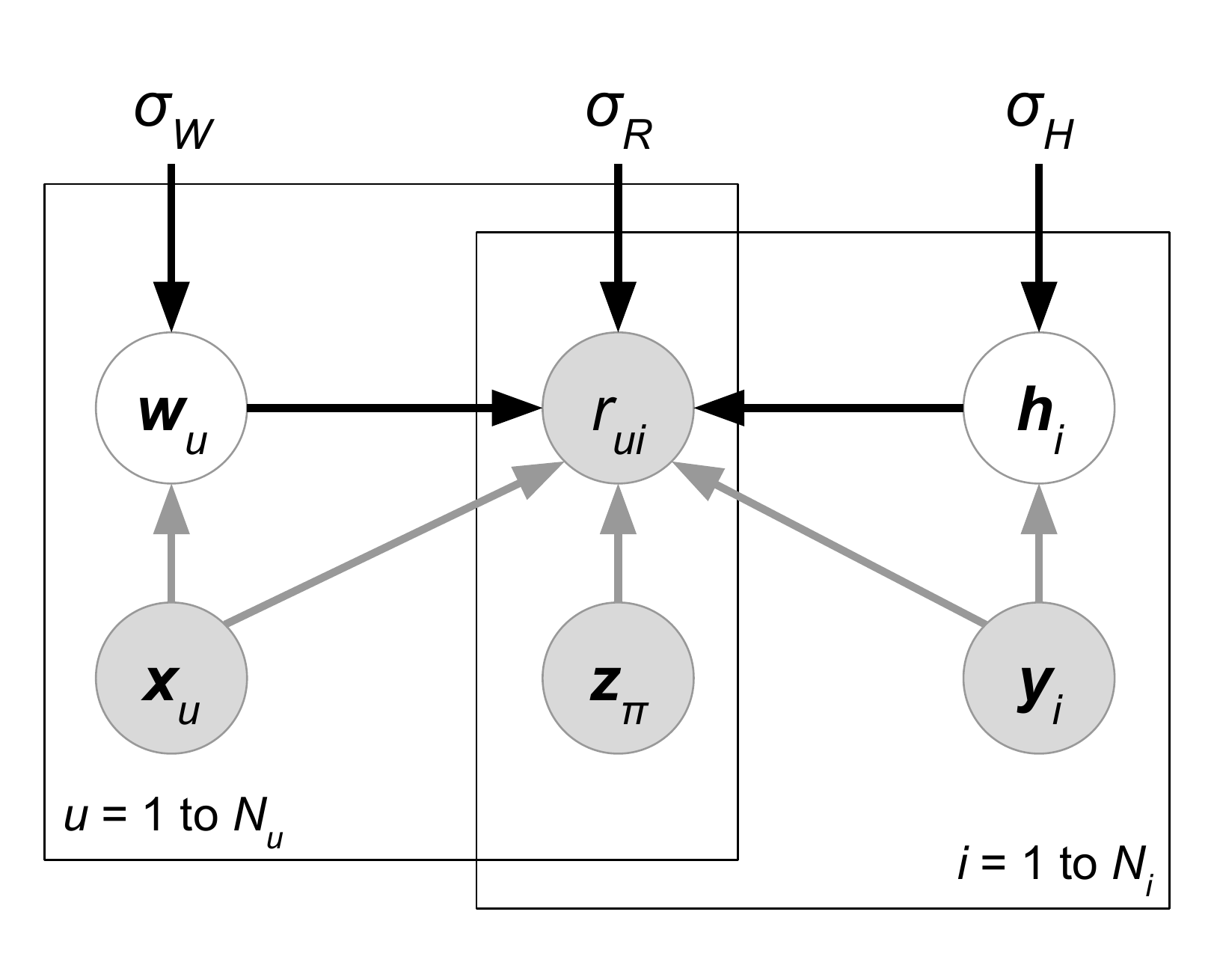}
    \caption{Graphical interpretation of discriminative probabilistic matrix factorization whose attributes $\bm{X}, \bm{Y}, \bm{Z}$ is given for ratings and latent factors. User and item-relevant attributes $\bm{X}, \bm{Y}$ could affect the generation of latent factors $\bm{W}, \bm{H}$ or ratings $\bm{R}$, while rating-relevant attributes $\bm{Z}$ typically determines the rating prediction $\bm{R}$. The models of this class may eliminate some of the gray arrows to imply additional independence assumptions between attributes and other factors.}
    \label{figure:discriminative_matrix_factorization}
\end{figure}

\subsection{Discriminative Matrix Factorization (Figure \ref{figure:discriminative_matrix_factorization})}
\label{section:discriminative_matrix_factorization}

Intuitively, the goal of a attribute-aware recommender system is to import attributes to improve its recommendation performance (either rating prediction or item ranking).
In the framework of matrix factorization, an item is rated or ranked according to the latent factors of the item and its corresponding users.
In order words, the learning of latent factors in classical matrix factorization depend only on ratings.
Thus the learning may fail due to lacks of training ratings.
If we can regularize the latent factors using attributes, or make attribute determine how to rate items, then matrix factorization methods can be more robust to the lacks of rating information in the training data, especially for those users or items that have very few ratings.

Following we choose to describe the attribute participation with probabilistic perspectives.
The learning of Probabilistic Matrix Factorization (PMF) tries to maximize posterior probability $p ( \bm{W}, \bm{H} \mid \bm{R} )$ of two latent factor matrices $\bm{W}$ (for users) and $\bm{H}$ (for items), given observed entries of training rating matrix $\bm{R}$.
Clearly, attribute-aware recommneder systems claim that we are given extra attribute matrix $\bm{X}$.
Then by Bayes' rule, the posterior probability can be shown as follows:
\begin{align}
    \argmax_{\bm{W}, \bm{H}} \underbrace{p \left ( \bm{W}, \bm{H} \mid \bm{R}, \bm{X} \right )}_{\text{Posterior}} & = \frac{p \left ( \bm{R} \mid \bm{W}, \bm{H}, \bm{X} \right ) p \left ( \bm{W}, \bm{H} \mid \bm{X} \right )}{p \left ( \bm{R} \mid \bm{X} \right ) } \nonumber \\
    & = p \left ( \bm{R} \mid \bm{W}, \bm{H}, \bm{X} \right ) p \left ( \bm{W}, \bm{H} \mid \bm{X} \right ) \nonumber \\
    & =  \underbrace{p \left ( \bm{R} \mid \bm{W}, \bm{H}, \bm{X} \right )}_{\text{Likelihood}} \underbrace{p \left ( \bm{W} \mid \bm{X} \right ) p \left ( \bm{H} \mid \bm{X} \right )}_{\text{Prior}} .
    \label{equation:pmf_bayes_rule_attribute_prior}
\end{align}
We eliminate the denominator $p ( \bm{R} \mid \bm{X} )$ since it does not contain variables $\bm{W}, \bm{H}$ for maximization.
At the prior part, we follow the independence assumption $\bm{W} \bot \bm{H}$ of PMF, though here the independence is given attribute matrix $\bm{X}$.
Now compared with classical PMF, both likelihood $p \left ( \bm{R} \mid \bm{W}, \bm{H}, \bm{X} \right )$ and prior $p \left ( \bm{W} \mid \bm{X} \right ) p \left ( \bm{H} \mid \bm{X} \right )$ could be affected by attributes $\bm{X}$.
Attributes in the likelihood can directly help predict or rank ratings, while attributes in the priors regularize the learning directions of latent factors.
Moreover, some current works assumes additional independences between attributes and the matrix factorization formulation.
For ease of explanations, we suppose that all the random variables follow normal distribution $p ( x ) = \mathcal{N} ( x \mid \mu, \sigma^{2} )$ with mean $\mu$ and variance $\sigma^{2}$ or multivariate normal distribution $p ( \bm{x} ) = \mathcal{N} ( \bm{x} \mid \bm{\mu}, \bm{\Sigma} )$ with mean vector $\bm{\mu}$ and covariance matrix $\bm{\Sigma}$.
Theoretically the following models accept other probability distributions.

\begin{figure}
    \centering
    \begin{subfigure}{0.5\linewidth}
        \centering
        \includegraphics[width=0.8\linewidth]{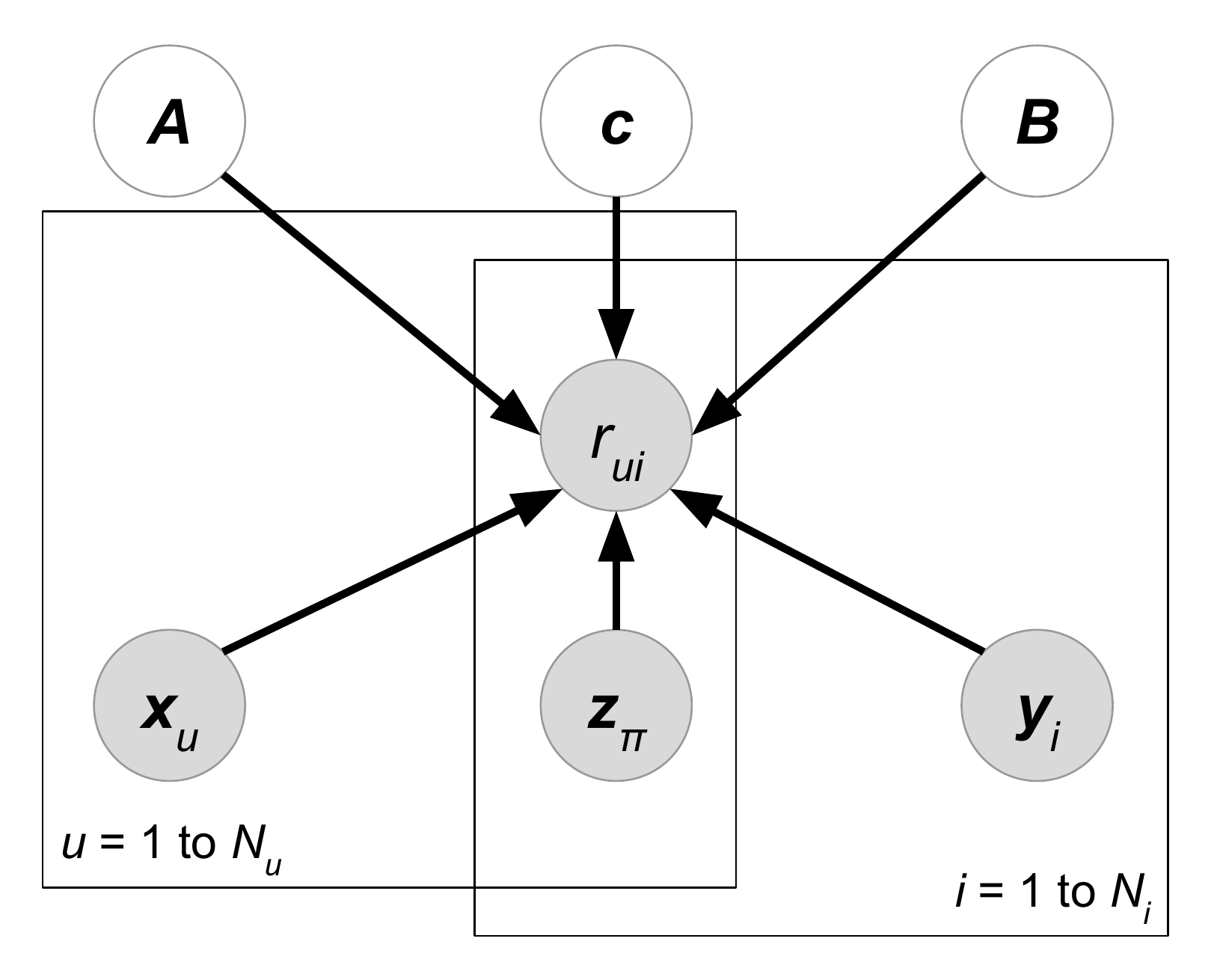}
        \caption{Matchbox}
    \end{subfigure}%
    \begin{subfigure}{0.5\linewidth}
        \centering
        \includegraphics[width=0.8\linewidth]{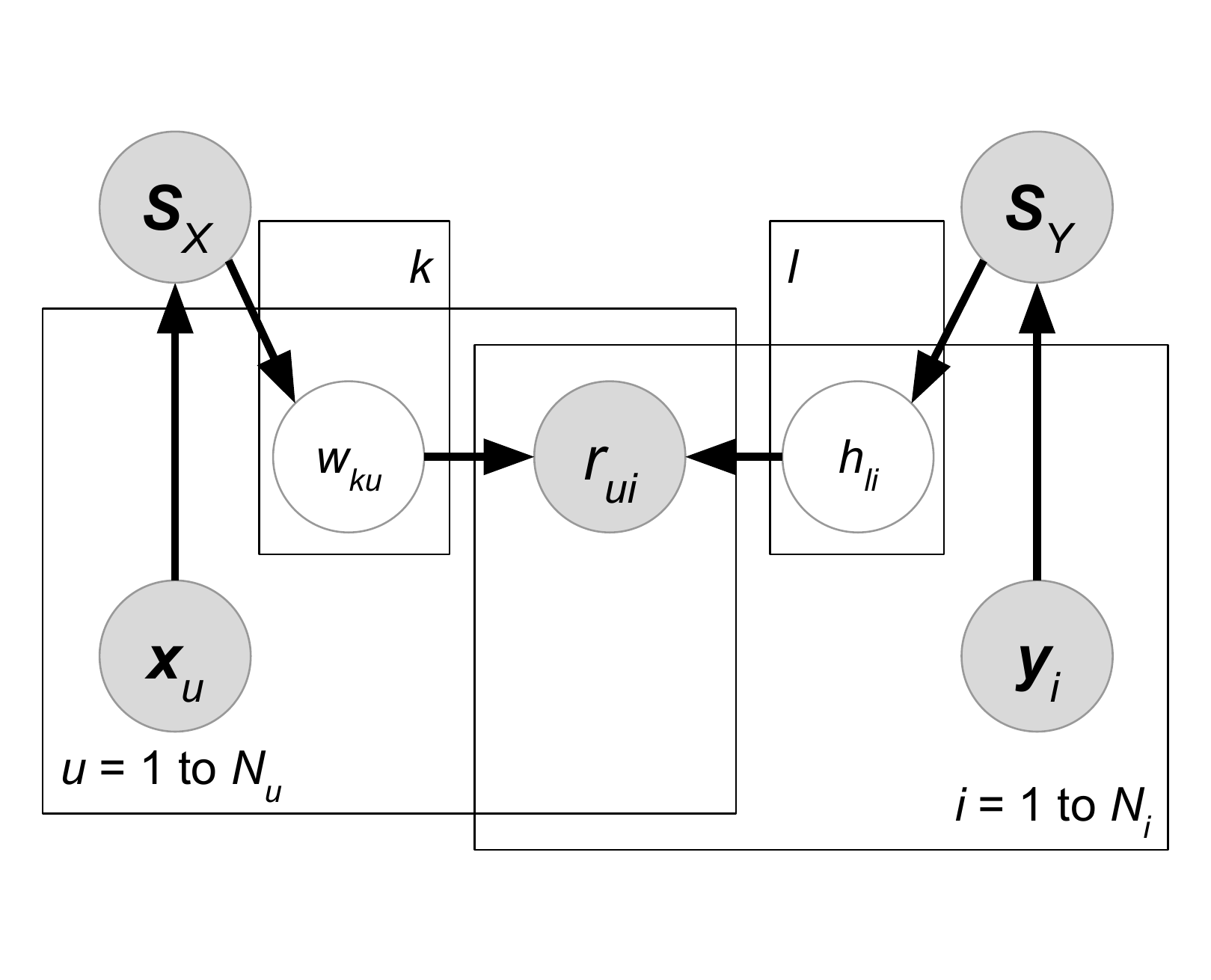}
        \caption{KPMF}
    \end{subfigure}
    \begin{subfigure}{0.5\linewidth}
        \centering
        \includegraphics[width=0.8\linewidth]{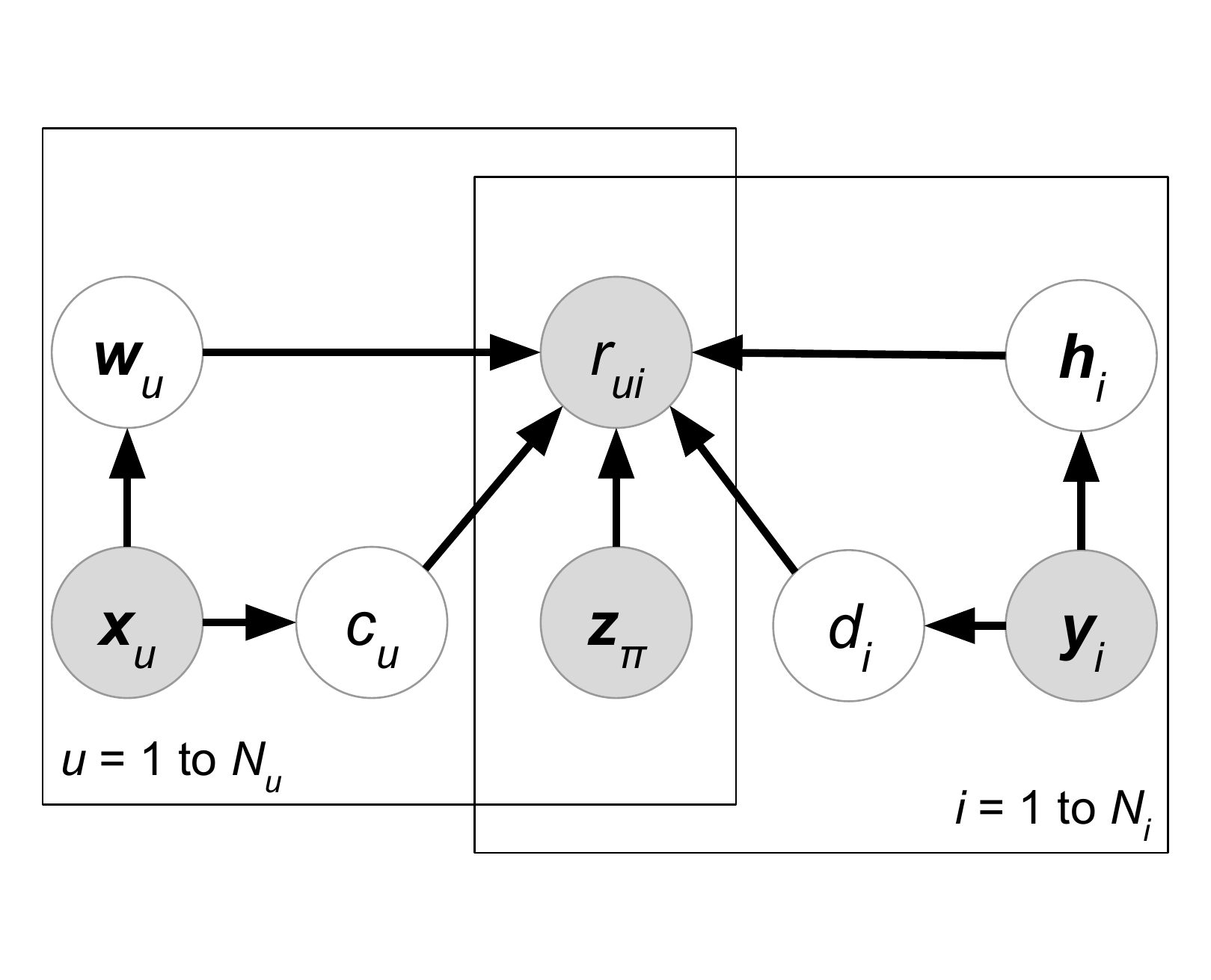}
        \caption{RLFM}
    \end{subfigure}%
    \begin{subfigure}{0.5\linewidth}
        \centering
        \includegraphics[width=0.8\linewidth]{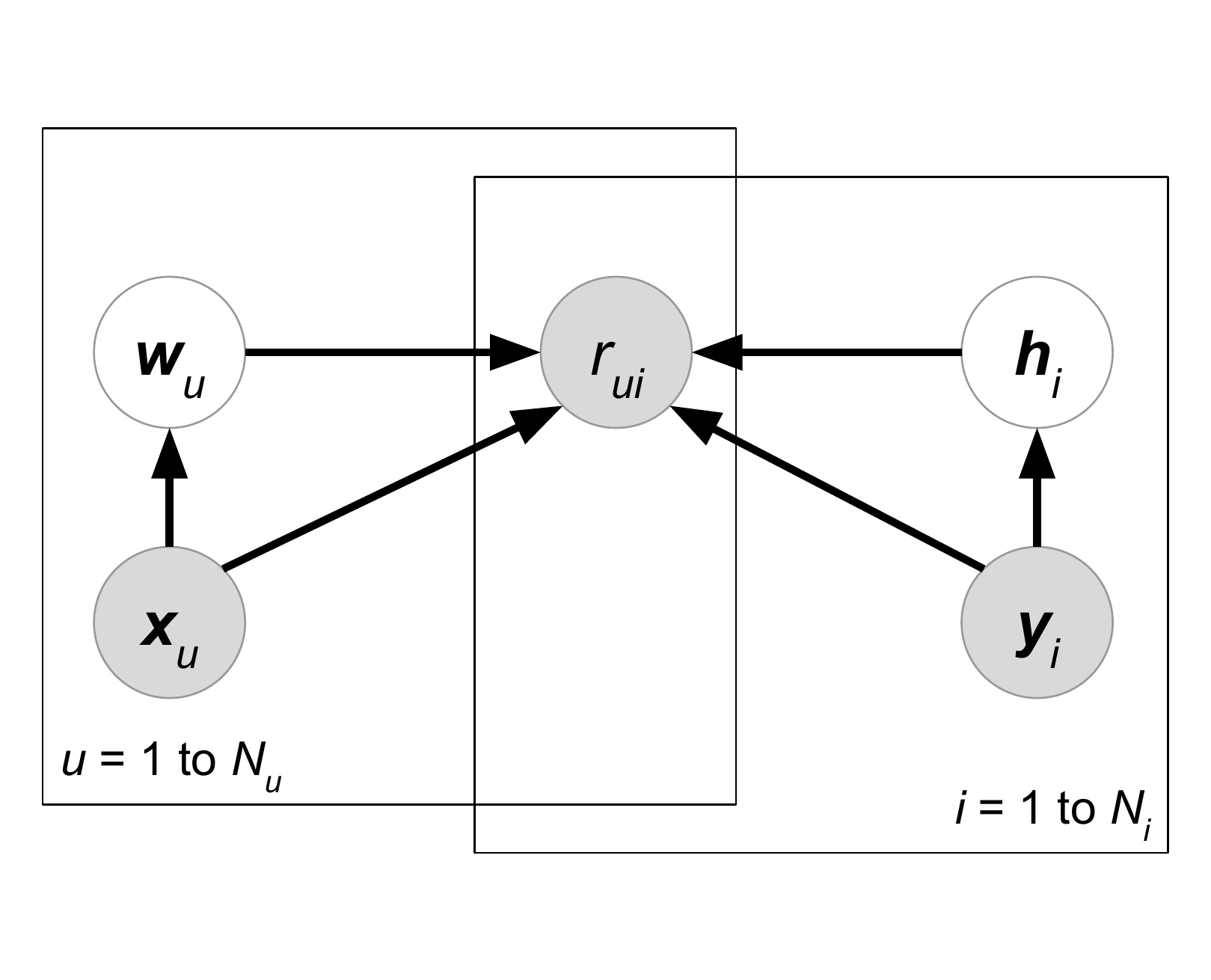}
        \caption{FIP}
    \end{subfigure}
    \caption{Graphical interpretation of the example models whose attributes serve as prior knowledge of latent factors. We eliminate all the hyperparameters for presentation simplicity.}
    \label{figure:discriminative_matrix_factorization_example}
\end{figure}

We further generate the sub-categories as below.

\subsubsection{Attributes in a Linear Model}
\label{section:attribute_linear_model}

This is the generalized form to utilize attributes in this category. Given the attributes, a weight vector is applied to perform linear regression together with classical matrix factorization $\bm{w}_{u}^{\top} \bm{h}_{i}$.
Its characteristic in mathematical form is shown in likelihood functions:
\begin{align}
    \argmax_{\bm{W}, \bm{H}, \theta} & \underbrace{ \prod_{(u, i) \mid r_{ui} \in \delta ( \bm{R} )} \mathcal{N} \left ( r_{ui} \mid \mu_{R} = \bm{w}_{u}^{\top} \bm{h}_{i} + \alpha ( \bm{x}_{u} ) + \beta ( \bm{y}_{i} ) + \gamma ( \bm{z}_{\pi(u, i)} ), \sigma_{R}^{2} \right ) }_{\text{Likelihood}} \underbrace{ p ( \bm{W} \mid \bm{X} ) p ( \bm{H} \mid \bm{Y} ) }_{\text{Prior}};
    ,
    \label{equation:linear_model_form}
\end{align}
where $\theta = \{ \alpha, \beta, \gamma \}$, while $\delta ( \bm{R} )$ denotes the non-missing ratings in the training data, and $\pi ( u, i )$ is the column index corresponding to user $u$ and item $i$.  $\bm{X} \in \mathbb{R}^{K \times N_{u}}, \bm{Y} \in \mathbb{R}^{K \times N_{i}}, \bm{Z} \in \mathbb{R}^{K \times | \delta ( \bm{R} ) |}$ respectively denote attribute matrices relevant to user, item and ratings, while $\alpha, \beta, \gamma$ are their corresponding transformation functions where attribute space is mapped toward the rating space identical with $\bm{w}_{u}^{\top} \bm{h}_{i}$.
Most early models select simple linear transformations, i.e., $\alpha ( \bm{x} ) = \bm{a}^{\top} \bm{x}, \beta ( \bm{y} ) = \bm{b}^{\top} \bm{y}, \gamma ( \bm{z} ) = \bm{c}^{\top} \bm{z}$ which has shown recommendation boosting, but recent works consider neural networks for non-linear $\alpha, \beta, \gamma$ mapping functions.
A simple linear regression model can be expressed as a likelihood function of normal distribution $\mathcal{N} ( r \mid \mu, \sigma^{2} )$ with mean $\mu$ and variance $\sigma^{2}$.
Ideally the distributions of latent factors $\bm{W}, \bm{H}$ shall have prior knowledge from attributes $\bm{X}, \bm{Y}$, but we have not yet observed an approach aiming at designing attribute-aware priors as the last two terms of (\ref{equation:linear_model_form}). 

\begin{itemize}
    \item \textbf{Bayesian Matrix Factorization with Side Information (BMFSI)} \citep{Porteous10BMFSI} is an example case in this sub-category.
    On the basis of Bayesian Probabilistc Matrix Factorization (BPMF) \citep{Salakhutdinov08BPMF}, BMFSI uses a linear combination like (\ref{equation:linear_model_form}) to introduce attribute information to rating prediction.
    It is formulated as:
    \begin{align}
        \argmax_{\bm{W}, \bm{H}, \theta} & \underbrace{p ( \bm{R} \mid \bm{W}, \bm{H}, \theta)}_{\text{Likelihood}} \underbrace{p ( \bm{W} ) p ( \bm{H} )}_{\text{Priors}} \nonumber \\
        = \argmax_{\bm{W}, \bm{H}, \theta} & \underbrace{\prod_{(u, i) \mid r_{ui} \in \delta ( \bm{R} )} \mathcal{N} \left ( r_{ui} \mid \bm{w}_{u}^{\top} \bm{h}_{i} + \bm{a}_{u}^{\top} \bm{x}_{u} + \bm{b}_{i}^{\top} \bm{y}_{i}, \sigma_{R}^{2} \right )}_{\text{Matrix factorization using attributes}} 
        \underbrace{\prod_{u} \mathcal{N} \left ( \bm{w}_{u} \mid \bm{\mu}_{u}, \bm{\Sigma}_{u} \right )
        \prod_{i} \mathcal{N} \left ( \bm{h}_{i} \mid \bm{\mu}_{i}, \bm{\Sigma}_{i} \right ) }_{\text{Regularization}}
        ,
    \end{align}
    where $\theta = \{ \bm{a}, \bm{b} \}$ and $\delta ( \bm{R} )$ is the set of training ratings.
    The difference from (\ref{equation:linear_model_form}) is that rating attributes $\bm{z}$ shall be concatenated with either $\bm{x}_{u}$ or $\bm{y}_{u}$, and thus we drop an independent weight variable $c$ in BMFSI.
    We ignore other attribute-free designs of BMFSI (e.g. Dirichlet process).
\end{itemize}

\subsubsection{Attributes in a Bilinear Model}
\label{section:attribute_bilinear_model}

This a popular method when two kinds of attributes (usually user and item) are provided. Given user attribute matrix $\bm{X}$ and item attribute matrix $\bm{Y}$, a matrix $\bm{A}$ is used to model the relation between them. The mathematical form can be viewed as the following:
\begin{align}
    \argmax_{\bm{W}, \bm{H}, \theta} & \underbrace{ \prod_{(u, i) \mid r_{ui} \in \delta ( \bm{R} )}  \mathcal{N} \left ( r_{ui} \mid \mu_{R} = \alpha ( \bm{x}_{u}, \bm{y}_{i} ) + \beta ( \bm{x}_{u} ) + \gamma ( \bm{y}_{i} ) + \bm{b}, \sigma_{R}^{2} \right ) }_{\text{Likelihood}} \underbrace{ p ( \bm{W} | \bm{X} ) p ( \bm{H} | \bm{Y} ) }_{\text{Prior}}
    ,
    \label{equation:bilinear_model_form}
\end{align}
where $\theta = \{ \alpha, \beta, \gamma \}$ are transformation functions from attribute space to rating space.
In particular, function $\alpha$ learns the interior dependency between user attributes $\bm{x}$ and item attributes $\bm{y}$, while $\beta$ and $\gamma$ find the extra factors that $\bm{x}$ or $\bm{y}$ itself affects the rating result.
Compared with (\ref{equation:linear_model_form}), the advantage of  (\ref{equation:bilinear_model_form}) is further considering a set of rating factors that come from the intersections between user and item attributes.
However, such modeling idea cannot work if either user attributes or item attributes are not provided from training data.
Commonly prior works select a simple linear form, named bilinear regression:
\begin{align}
    \mu_{R} & = \alpha ( \bm{x}_{u}, \bm{y}_{i} ) + \beta ( \bm{x}_{u} ) + \gamma ( \bm{y}_{i} ) + b
    \nonumber \\
    & = \bm{x}_{u}^{\top} \bm{A} \bm{y}_{i} + \bm{c}_{u}^{\top} x_{u} + \bm{d}_{i}^{\top} y_{i} + b
    \nonumber \\
    & = \bm{\widetilde{x}}_{u}^{\top} \widetilde{\bm{A}} \bm{\widetilde{y}}_{i}
    .
    \label{equation:bilinear_regression_case}
\end{align}
In fact, as mentioned in \cite{Lu16SIMMCSI},  $\bm{c}_{u}^{\top} \bm{x}_{u} + \bm{d}_{i}^{\top} \bm{y}_{i} + \bm{b}$  
can be absorbed into $\bm{x}_{u}^{\top} \bm{A} \bm{y}_{i}$ and written as form $\bm{\widetilde{x}}_{u}^{\top} \widetilde{\bm{A}} \bm{\widetilde{y}}_{i}$, by appending a new dimension whose value is fixed to $1$ for each $\bm{x}$ and $\bm{y}$:

Works in this category differ in whether the bilinear term is explicit or implicit.
Also, the latent factor matrices $\bm{W}, \bm{H}$ are inherently included in the bilinear form.
Specifically, (\ref{equation:bilinear_regression_case}) implies that the form of the dot product of two linear-transformed attributes $\bm{w}_{u} = \bm{S} \bm{x}_{u}$ and $\bm{h}_{i} = \bm{T} \bm{y}_{i}$ since it can be reformed as $\bm{w}_{u}^{\top} \bm{h}_{i} = \bm{x}_{u}^{\top} ( \bm{S}^{\top} \bm{T} ) \bm{y}_{i}$ where $\bm{A} = \bm{S}^{\top} \bm{T}$.
Some works such as Regression-based Latent Factor Model (see below) chooses to softly constrain $\bm{w}_{u} \approx \bm{S} \bm{x}_{u}$ and $\bm{h}_{i} \approx \bm{T} \bm{y}_{i}$ using priors $p ( \bm{W} \mid \bm{X} ), p ( \bm{H} \mid \bm{Y} )$.

\begin{itemize}
        \item \textbf{Matchbox} \cite{Stern09Matchbox} . Let $\bm{X}, \bm{Y}, \bm{Z}$ be respectively the attribute matrices with respect to users, items and ratings.
        Matchbox assumes a rating being predicted by the linear combinations of $\bm{X}, \bm{Y}, \bm{Z}$:
        \begin{align}
            \argmax_{\bm{A}, \bm{B}, \bm{c}} & \underbrace{ p ( \bm{R} \mid \bm{A}, \bm{B}, \bm{c}, \bm{X}, \bm{Y}, \bm{Z} ) }_{\text{Likelihood}} \underbrace{ p ( \bm{c} ) p ( \bm{A} ) p ( \bm{B} ) }_{\text{Prior}} \nonumber \\
            = & \underbrace{ \prod_{(u, i) \mid r_{ui} \in \delta ( \bm{R} )} \mathcal{N} \left ( r_{ui} \mid \bm{x}_{u}^{\top} \bm{A}^{\top} \bm{B} \bm{y}_{i} + \bm{c}^{\top} \bm{z}_{\pi (u, i)}, \sigma_{R}^{2} \right ) }_{\text{Matrix factorization using attributes}} \nonumber \\
            & \underbrace{ \prod_{m} \mathcal{N} \left ( c_{m} \mid \mu_{cm}, \sigma_{cm}^{2} \right ) \prod_{(u, k)} \mathcal{N} \left ( a_{uk} \mid \mu_{Auk}, \sigma_{Auk}^{2} \right ) \prod_{(i, l)} \mathcal{N} \left ( b_{il} \mid \mu_{Bil}, \sigma_{Bil}^{2} \right ) }_{\text{Regularization}}
            \label{equation:matchbox_posterior}
        \end{align}
        where $\delta ( \bm{R} )$ is the set of non-missing entries in rating matrix $R$. $\bm{x}_{u}, \bm{y}_{i}$ represents the attribute set of user $u$ or item $i$.
        $\bm{z}_{(u, i)}$ denotes the rating-relevant attributes associated with user $u$ and item $i$.
        Note that (\ref{equation:matchbox_posterior}) defines latent factors $\bm{W} = \bm{A} \bm{X}, \bm{H} = \bm{B} \bm{Y}$ and then we just have to learn shared weight matrices $\bm{A}, \bm{B}$.
        The prior distributions of $\bm{A}, \bm{B}, \bm{c}$ are further factorized, which supposes that all the weight entries in these matrices are independent of each other.
        
        \item \textbf{Friendship-Interest Propagation (FIP)} \cite{Yang11FIP} .
        Following the notations from the previous RLFM introduction, FIP considers two types of attribute matrices: $\bm{X}$ and $\bm{Y}$.
        Based on vanilla matrix factorization, FIP encodes attribute information by modeling the potential correlations between $\bm{X}$ and $\bm{Y}$:
        \begin{align}
            \argmax_{\bm{W}, \bm{H}, \bm{A}, \bm{B}, \bm{C}} & \underbrace{ p ( \bm{R} \mid \bm{W}, \bm{H}, \bm{C}, \bm{X}, \bm{Y} ) }_{\text{Likelihood}} \underbrace{ p ( \bm{W} \mid \bm{A}, \bm{X} ) p ( \bm{H} \mid \bm{B}, \bm{Y} ) }_{\text{Prior}} \nonumber \\
            = & \underbrace{ \prod_{(u, i) \mid r_{ui} \in \delta ( \bm{R} )} \mathcal{N} \left ( r_{ui} \mid \bm{w}_{u}^{\top} \bm{h}_{i} + \bm{x}_{u}^{\top} \bm{C} \bm{y}_{i}, \sigma_{R}^{2} \right ) }_{\text{Matrix factorization using attributes}} \underbrace{ \prod_{u} \mathcal{N} \left ( \bm{w}_{u} \mid \bm{A} \bm{x}_{u}, \bm{\Sigma}_{W} \right ) \prod_{i} \mathcal{N} \left ( \bm{h}_{i} \mid \bm{B} \bm{y}_{i}, \bm{\Sigma}_{H} \right ) }_{\text{Regularization using attributes}}
            \label{equation:fip_posterior}
        \end{align}
        where matrix $\bm{C}$ forms the correlations between attribute matrices $\bm{X}$ and $\bm{Y}$.
        
        \item \textbf{Regression-based Latent Factor Model (RLFM)} \citep{Agarwal09RLFM} .
    Given three types of attribute matrices: user-relevant $\bm{X}$, item-relevant $\bm{Y}$ and rating-relevant $\bm{Z}$, RLFM models them in different parts of biased matrix factorization.
    $\bm{X}, \bm{Y}$ serve as the hyperparameters of latent factors, while $\bm{Z}$ joins the regression framework to predict ratings together with latent factors.
    RLFM can be written as:
    \begin{align}
        \argmax_{\bm{W}, \bm{H}, \theta} & \underbrace{ p ( \bm{R} \mid \bm{W}, \bm{H}, \bm{c}, \bm{d}, \bm{\gamma}, \bm{Z} ) }_{\text{Likelihood}} \underbrace{ p ( \bm{W} \mid \bm{A}, \bm{X} ) p ( \bm{H} \mid \bm{B}, \bm{Y} ) p ( \bm{c} \mid \bm{\alpha}, \bm{X} ) p (\bm{d} \mid \bm{\beta}, \bm{Y} ) }_{\text{Prior}} \nonumber \\
        = \argmax_{\bm{W}, \bm{H}, \theta} & \underbrace{ \prod_{(u, i) \mid r_{ui} \in \delta ( \bm{R} )} \mathcal{N} \left ( r_{ui} \mid \bm{w}_{u}^{\top} \bm{h}_{i} + c_{u} + d_{i} + \bm{\gamma}^{\top} \bm{z}_{\pi (u, i)}, \sigma_{R}^{2} \right ) }_{\text{Matrix factorization using attributes}} \nonumber \\
        & \underbrace{ \prod_{u} \mathcal{N} \left ( \bm{w}_{u} \mid \bm{A} \bm{x}_{u}, \bm{\Sigma}_{W} \right ) \mathcal{N} \left ( c_{u} \mid \bm{\alpha}^{\top} \bm{x}_{u}, \sigma_{c}^{2} \right ) \prod_{i} \mathcal{N} \left ( \bm{h}_{i} \mid \bm{B} \bm{y}_{i}, \bm{\Sigma}_{H} \right ) \mathcal{N} \left ( d_{i} \mid \bm{\beta}^{\top} \bm{y}_{i}, \sigma_{d}^{2} \right ) }_{\text{Regularization using attributes}}
        \label{equation:rlfm_posterior}
    \end{align}
    where $\theta =  \{ \bm{c}, \bm{d}, \bm{A}, \bm{B}, \bm{\alpha}, \bm{\beta}, \bm{\gamma} \}$, and $\delta ( \bm{R} )$ is the set of non-missing ratings for training.
    Biased matrix factorization adds two vectors $\bm{c}, \bm{d}$ to learn the biases for each user or item.
    Parameters $\bm{A}, \bm{B}, \bm{\alpha}, \bm{\beta}, \bm{\gamma}$ map attributes with latent factors (for $\bm{X}, \bm{Y}$) or rating prediction (for $\bm{Z}$).
\end{itemize}

\subsubsection{Attributes in a Similarity Matrix}

In this case, a similarity matrix which measures the closeness of attributes between users or between items is presented. Given the user attribute matrix $\bm{X} \in \mathbb{R}^{D \times N_{u}}$, where $N_{u}$ is the number of users and $D$ is the dimension of user attribute, a similarity matrix $ \bm{S} \in \mathbb{R}^{N_{u} \times N_{u}}$ is computed. There are many metrics to for similarity calculation such as Euclidean distance or kernel functions. The similarity matrix is then used for matrix factorization or other solutions. The speciality of this case is that human knowledge is involved in determining how the interactions between attributes should be modeled.
Kernelized Probabilistic Matrix Factorization is an example which utilizes both user similarity matrix and item similarity matrix.
    
\begin{itemize}
    \item \textbf{Kernelized Probabilistic Matrix Factorization (KPMF)} \citep{Zhou12KPMF} .
    Let $K, N_{u}, N_{i}$ be the number of latent factors, users and items.
    Given user-relevant attribute matrix $\bm{X} \in \mathbb{R}^{K \times N_{u}}$ or item-relvant attribute matrix $\bm{Y} \in \mathbb{R}^{K \times N_{i}}$, we can always obtain a similarity matrix $\bm{S}_{\bm{X}} \in \mathbb{R}^{N_{u} \times N_{u}}$ or $\bm{S}_{\bm{Y}} \in \mathbb{R}^{N_{i} \times N_{i}}$ where each entry stores a pre-defined similarity between a pair of users or items.
    Then KPMF formulates the similarty matrix as the prior of its corresponding latent factor matrix:
    \begin{align}
        \argmax_{\bm{W}, \bm{H}} & \underbrace{ p ( \bm{R} \mid \bm{W}, \bm{H} ) }_{\text{Likelihood}} \underbrace{ p ( \bm{W} \mid \bm{X} ) p ( \bm{H} \mid \bm{Y} ) }_{\text{Prior}} \nonumber \\
        = \argmax_{\bm{W}, \bm{H}} & \underbrace{ \prod_{(u, i) \mid r_{ui} \in \delta ( \bm{R} )} \mathcal{N} \left ( r_{ui} \mid \bm{w}_{u}^{\top} \bm{h}_{i}, \sigma_{R}^{2} \right ) }_{\text{Matrix factorization}} \underbrace{ \prod_{k} \mathcal{N} \left ( \bm{w}^{k} \mid \bm{0}, \bm{S}_{\bm{X}} \right ) \prod_{l} \mathcal{N} \left ( \bm{h}^{l} \mid \bm{0}, \bm{S}_{\bm{Y}} \right ) . }_{\text{Regularization using attributes}}
        \label{equation:kpmf_posterior}
    \end{align}
    Here we use subscripts $\bm{w}_{u}$ to denote the $u$-th column vector of a matrix $\bm{W}$, while superscripts $\bm{w}^{k}$ imply the $k$-th row vector of $\bm{W}$.
    Intuitively, the similarity matrices control the learning preferences of user or item latent factors.
    If two users have similar user-relevant attributes (i.e., they have a higher similarity measure in $\bm{S}_{\bm{X}}$), then their latent factors are forced to be closer during the matrix factorization learning.
\end{itemize}

\begin{figure}[H]
    \centering
    \includegraphics[width=0.4\linewidth]{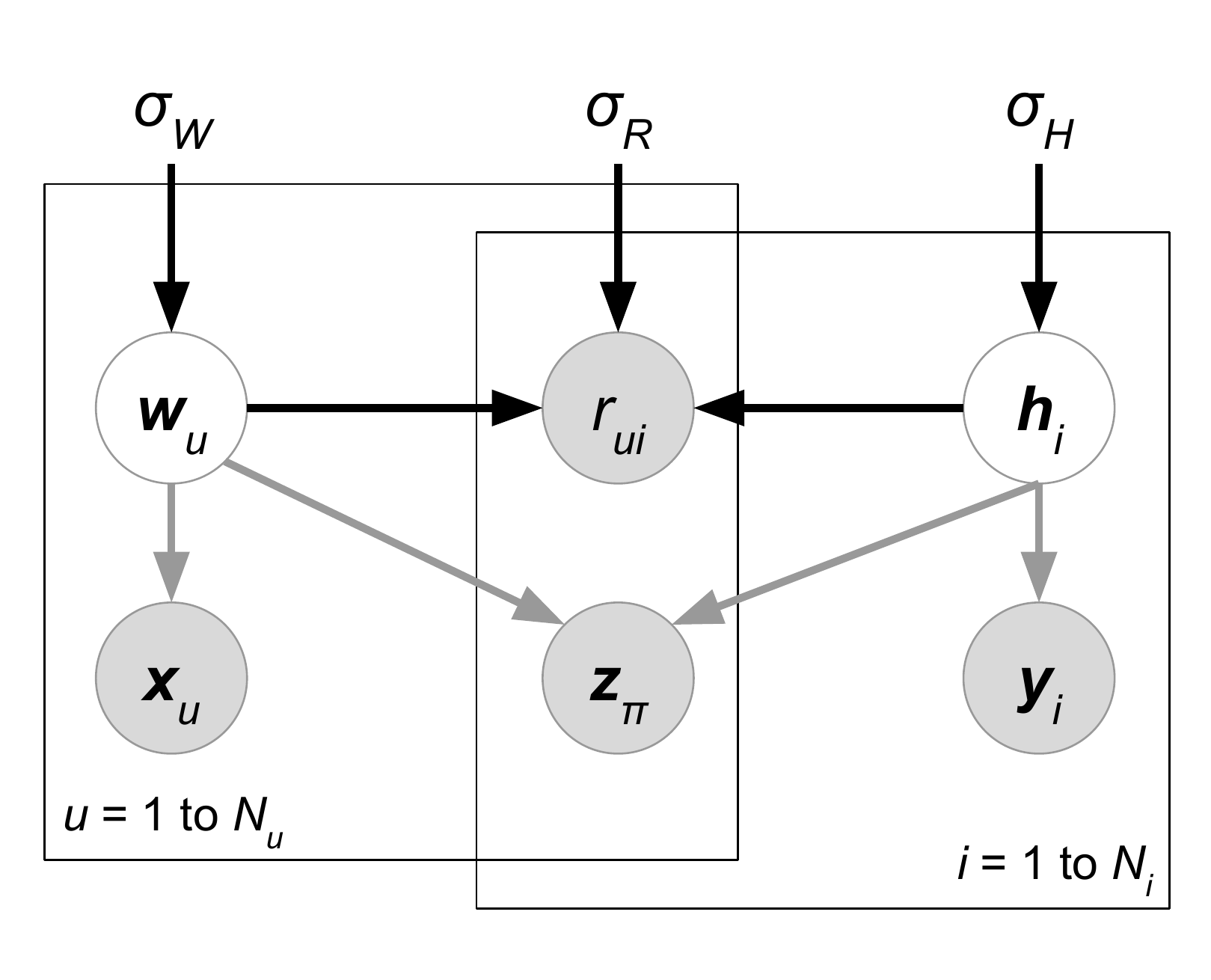}
    \caption{Graphical interpretation of generative probabilistic matrix factorization whose attributes $\bm{X}, \bm{Y}, \bm{Z}$ together with ratings are generated or predicted by latent factors. User and item-relevant attributes $\bm{X}, \bm{Y}$ could be respectively generated by corresponding latent factors $\bm{W}, \bm{H}$. Rating-relevant attributes $\bm{Z}$ is likely to result from both $\bm{W}$ and $\bm{H}$. For models of this class, some of the gray arrows are removed to represent their additional independence assumptions about attribute generation.}
    \label{figure:generative_matrix_factorization}
\end{figure}

\subsection{Generative Matrix Factorization (Figure \ref{figure:generative_matrix_factorization})}
\label{section:generative_matrix_factorization}

In Probabilistic Matrix Factorization (PMF), ratings are generated by the interactions of user or item latent factors.
However, the PMF latent factors are not limited to rating generation.
We can also make attributes be generated by the latent factors.
Mathematically, by Bayes' rule, we maximize a posteriori as follows:
\begin{align}
    \argmax_{\bm{W}, \bm{H}} \underbrace{p \left ( \bm{W}, \bm{H} \mid \bm{R}, \bm{X} \right )}_{\text{Posterior}} & = \frac{p \left ( \bm{R}, \bm{X} \mid \bm{W}, \bm{H} \right ) p \left ( \bm{W}, \bm{H} \right )}{p \left ( \bm{R}, \bm{X} \right ) } \nonumber \\
    & = p \left ( \bm{R}, \bm{X} \mid \bm{W}, \bm{H} \right ) p \left ( \bm{W}, \bm{H} \right ) \nonumber \\
    & =  \underbrace{p \left ( \bm{R} \mid \bm{W}, \bm{H} \right ) p \left ( \bm{X} \mid \bm{W}, \bm{H} \right )}_{\text{Likelihood}} \underbrace{p \left ( \bm{W} \right ) p \left ( \bm{H} \right )}_{\text{Prior}} .
    \label{equation:pmf_bayes_rule_attribute_likelihood}
\end{align}
where $p ( 
\bm{R}, \bm{X})$ does not affect the posterior maximization.
We again assume independence $\bm{R} \bot \bm{X}$ given latent factors $\bm{W}, \bm{H}$ in (\ref{equation:pmf_bayes_rule_attribute_likelihood}), which is commonly adopted in related work.
Furthermore, $\bm{X}$ may share either latent factors $\bm{W}$ (i.e., $p ( \bm{X} \mid \bm{W} )$) or $\bm{H}$ (i.e., $p ( \bm{X} \mid \bm{H} )$) with $\bm{R}$, but not both due to more generalization strength of matrix factorization.

The following relevant works are classified in this category. For explanation simplicity, all the probabilities follows normal distributions, i.e, $p ( x ) = \mathcal{N} ( x \mid \mu, \sigma^{2} )$ (i.e., squared loss objective) with mean $\mu$ and variance $\sigma^{2}$ (or mean vector $\bm{\mu}$ and covariance matrix $\bm{\Sigma}$ for multivariate normal distributions).
However the example models are never restricted in normal distributions.

\begin{figure}
    \centering
    \begin{subfigure}{0.5\linewidth}
        \centering
        \includegraphics[width=0.8\linewidth]{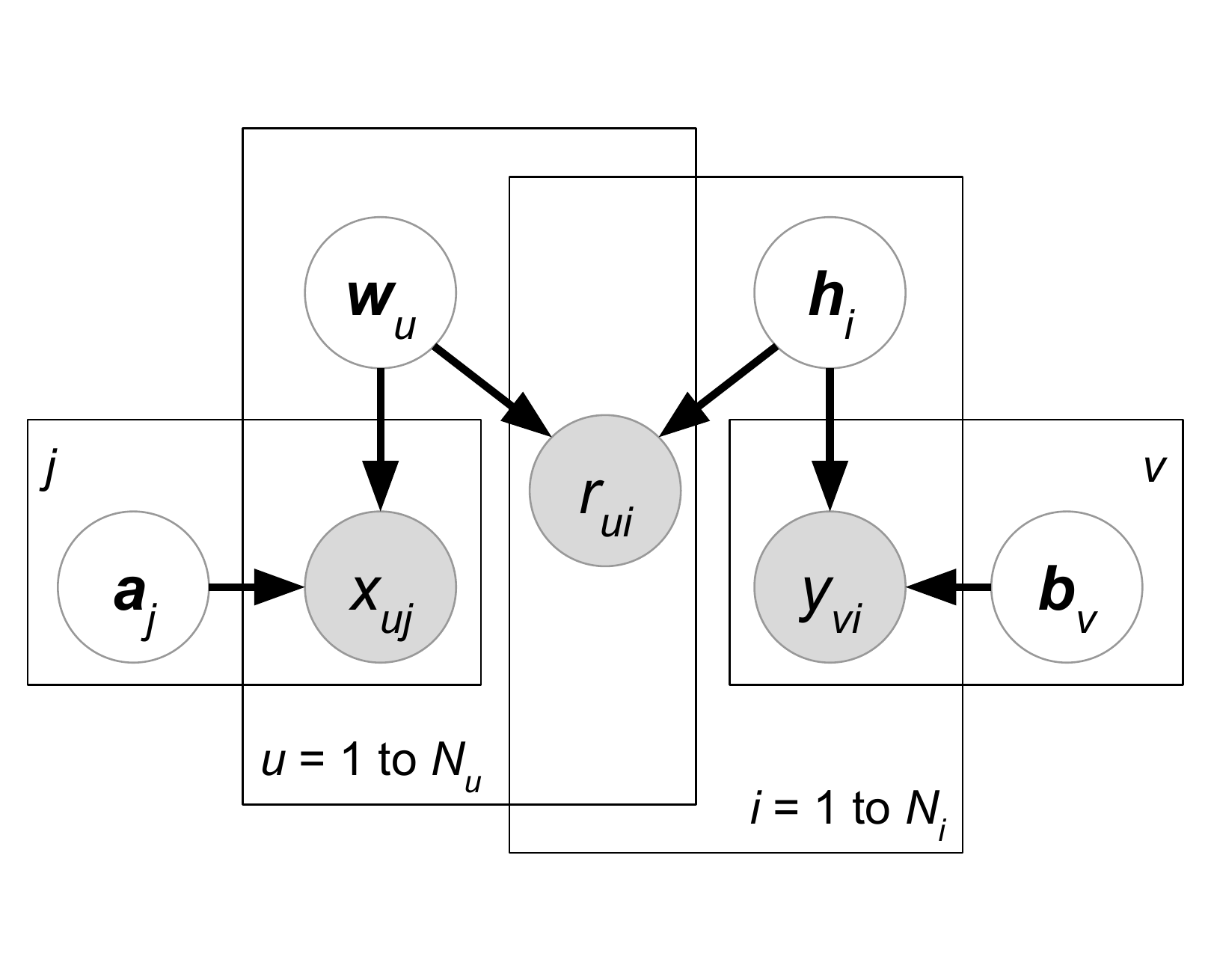}
        \caption{CMF}
    \end{subfigure}%
    \begin{subfigure}{0.5\linewidth}
        \centering
        \includegraphics[width=0.8\linewidth]{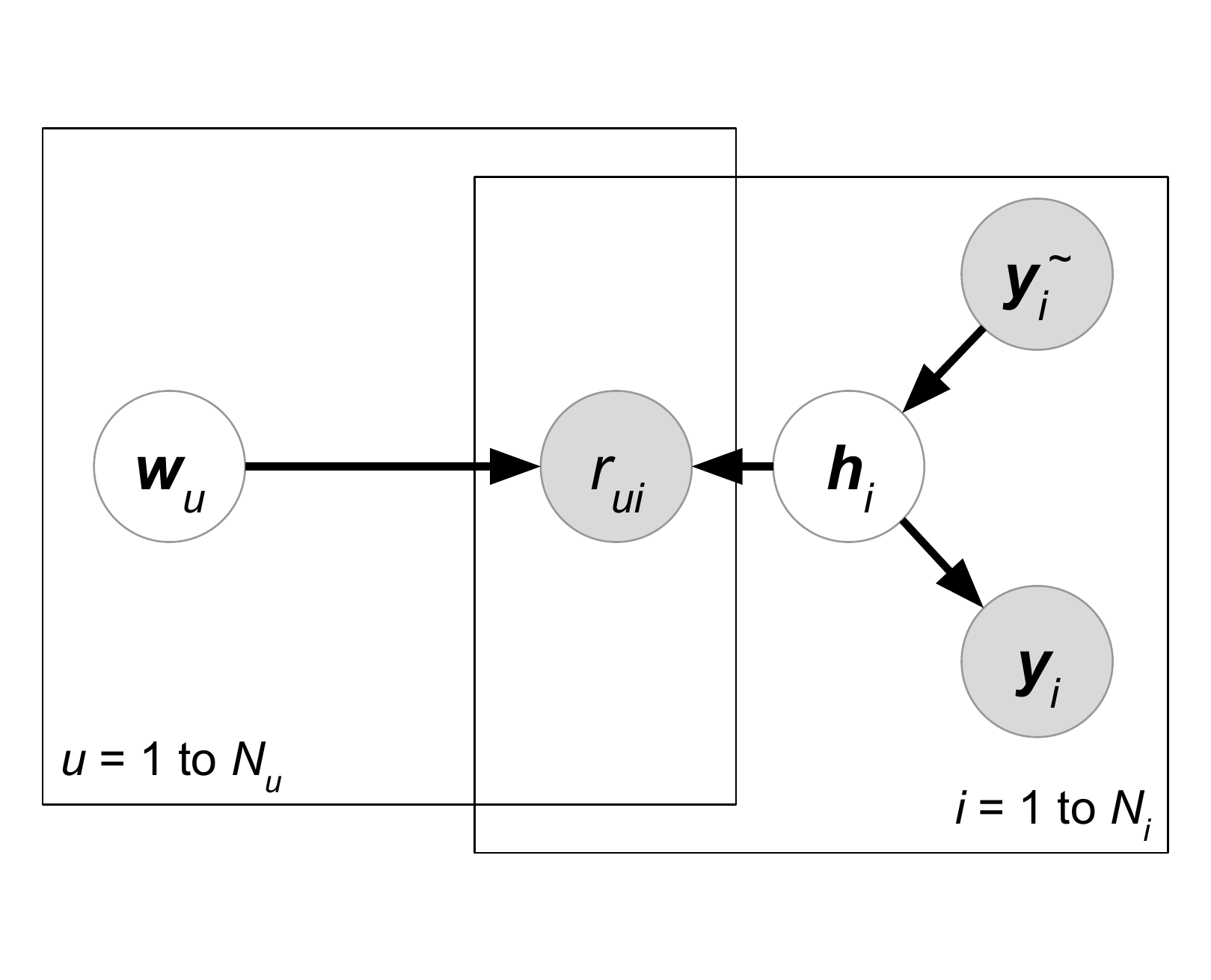}
        \caption{CDL}
    \end{subfigure}
    \caption{Graphical interpretation of the example models whose attributes are generated from latent factors. We eliminate all the hyperparameters for presentation simplicity.}
    \label{figure:generative_matrix_factorization_example}
\end{figure}

There are two different branches is this direction.
On one hand, earlier works use the matrix factorization technique again, to generate attributes from user or item latent factors.
It can be seen as a linear mapping between latent factors and attributes.
On the other hand, with the help of deep neural networks, recent works combine matrix factorization and deep autoencoders to realize non-linear mappings for attribute generation. We will introduce them in the following sections.

\subsubsection{Attributes in Multiple Matrice Factorization}

Similar to PMF $\bm{R} \approx \bm{W}^{\top} \bm{H}$ for rating distributions, attributes distributions are modeled using another matrix factorization form.
Given user attribute matrix $\bm{X}$, item attribute matrix $\bm{Y}$ and rating attribute matrix $\bm{Z}$, they can be factorized as $\bm{X} \approx \bm{A}^{\top} \bm{W}, \bm{Y} \approx \bm{B}^{\top} \bm{H}$ of low rank.
Specifically, its objective function is written as:
\begin{align}
    \argmax_{\bm{W}, \bm{H}, \bm{A}, \bm{B}} & \underbrace{ \prod_{(u, i) \mid r_{ui} \in \delta ( \bm{R} )} \mathcal{N} \left ( r_{ui} \mid \mu_{R} = \bm{w}_{u}^{\top} \bm{h}_{i}, \sigma_{R}^{2} \right ) 
    \prod_{(j, u)} \mathcal{N} \left ( x_{ju} \mid \bm{a}_{j}^{\top} \bm{w}_{u}, \sigma_{X}^{2} \right )
    \prod_{(v, i)} \mathcal{N} \left ( y_{vi} \mid \bm{b}_{v}^{\top} \bm{h}_{i}, \sigma_{Y}^{2} \right ) }_{\text{Likelihood}} \nonumber \\
    & \underbrace{ \prod_{(u, i) \mid r_{ui} \in \delta ( \bm{R} )} \mathcal{N} \left ( \bm{z}_{ui} \mid \bm{w}_{u}^{\top} \bm{C} \bm{h}_{i}, \sigma_{Z}^{2} \right ) }_{\text{Likelihood}} \underbrace{
    p ( \bm{W} ) p ( \bm{H} ) }_{\text{Prior}}
    ,
    \label{equation:multiple_matrix_factorization_form}
\end{align}
where $\delta ( \bm{R} )$ denote the non-missing entries of matrix $\bm{R}$.
The insight of (\ref{equation:multiple_matrix_factorization_form}) is to share the latent factors $\bm{W}, \bm{H}$ in multiple factorization tasks.
$\bm{W}$ is shared with user attributes, while $\bm{H}$ is shared with item attributes. 
$\bm{Z}$ requires the sharing of both $\bm{W}$ and $\bm{H}$ due to user and item-specific rating attributes.
Therefore the side information of both $\bm{X}, \bm{Y}$ and $\bm{Z}$ can indirectly transfer to rating prediction.
Auxiliary matrices $\bm{A}, \bm{B}$ and $\bm{C}$ learns the mappings between latent factors and attributes.
With respect to the mathematical form of matrix factorization, the expectation of feature values is linearly correlated with its corresponding latent factors.

\begin{itemize}
    \item \textbf{Collective Matrix Factorization (CMF)} \citep{Singh08CMF} Here we introduce a common model in this sub-category. The CMF framework relies on the combination of multiple matrix factorization objective functions.
    CMF first builds the MF for rating matrix $\bm{R}$.
    Then user and item-relevant attribute matrices $\bm{X}, \bm{Y}$ are appended to the matrix factorization objectives.
    Overall we have:
    \begin{align}
        \argmax_{\bm{W}, \bm{H}, \bm{A}, \bm{B}} & \underbrace{ p ( \bm{R} \mid \bm{W}, \bm{H} ) p ( \bm{X} \mid \bm{W}, \bm{A} ) p ( \bm{Y} \mid \bm{H}, \bm{B} ) }_{\text{Likelihood}} \underbrace{ p( \bm{W} ) p( \bm{H} ) p( \bm{A} ) p( \bm{B} ) }_{\text{Prior}} \nonumber \\
        = \argmax_{\bm{W}, \bm{H}, \bm{A}, \bm{B}} & \underbrace{ \prod_{(u, i) \mid r_{ui} \in \delta ( \bm{R} )} \mathcal{N} \left ( r_{ui} \mid \bm{w}_{u}^{\top} \bm{h}_{i}, \sigma_{R}^{2} \right ) }_{\text{Matrix factorization of } R} \underbrace{ \prod_{(j, u) } \mathcal{N} \left ( x_{ju} \mid \bm{a}_{j}^{\top} \bm{w}_{u}, \sigma_{X}^{2} \right ) }_{\text{Matrix factorization of } X}
        \underbrace{ \prod_{(v, i)} \mathcal{N} \left ( y_{vi} \mid \bm{b}_{v}^{\top} \bm{h}_{i}, \sigma_{Y}^{2} \right ) }_{\text{Matrix factorization of } Y} \nonumber \\
        & \underbrace{ \prod_{u} \mathcal{N} \left ( \bm{w}_{u} \mid \bm{0}, \bm{\Sigma}_{W} \right ) \prod_{i} \mathcal{N} \left ( \bm{h}_{i} \mid \bm{0}, \bm{\Sigma}_{H} \right ) \prod_{j} \mathcal{N} \left ( \bm{a}_{j} \mid \bm{0}, \bm{\Sigma}_{A} \right ) \prod_{v} \mathcal{N} \left ( \bm{b}_{v} \mid \bm{0}, \bm{\Sigma}_{B} \right ) }_{\text{Regularization}}
        \label{equation:cmf_posterior}
    \end{align}
    where $\delta ( \bm{R} ), \delta ( \bm{X} ), \delta ( \bm{Y} )$ denote the non-missing entries of matrix $\bm{R}, \bm{X}, \bm{Y}$ that are generated by latent factor matrices $\bm{W}, \bm{H}, \bm{A}, \bm{B}$ of zero-mean normal priors (i.e., $l2$ regularization).
    In (\ref{equation:cmf_posterior}), $\bm{W}, \bm{H}$ are shared by at least two matrix factorization objectives.
    Attribute information in $\bm{X}, \bm{Y}$ is transferred to rating prediction $\bm{R}$ through sharing the same latent factors.
    Note that CMF is not limited to three matrix factorization objectives (\ref{equation:cmf_posterior}).
\end{itemize}

\subsubsection{Attributes in Deep Neural Networks}

In deep neural networks, an autoencoder is usually used to learn latent representation of observed data.
Specifically the model tries to construct a encoder $\mathcal{E}$ and a decoder $\mathcal{D}$, where the encoder learns to map from a possibly modified attributes $\bm{\tilde{X}}$ to low-dimensional latent factors, and the decoder recover from latent factors to the original attributes $\bm{X}$.
Moreover, activation functions in autoencoders can reflect non-linear mappings between latent factors and attributes, which may capture the characteristics of attributes more accurately.

To implement an autoencoder, at first we generate another attribute matrix $\bm{\tilde{X}}$ from $\bm{X}$.
$\bm{\tilde{X}}$ could be the same as $\bm{X}$, or different due to corruption, e.g., adding random noise.
Autoencoders aim to predict the original $\bm{X}$ using latent factors that are inferred from generated $\bm{\tilde{X}}$.
Here attributes serve not only as the generation results $\bm{X}$, but also as the prior knowledge $\bm{\tilde{X}}$ of latent factors.
Let us review Bayes' Rule to figure out where autoencoders appears for generative matrix factorization:
\begin{align}
    \argmax_{\bm{W}, \bm{H}} p \left ( \bm{W}, \bm{H} \mid \bm{R}, \bm{X}, \bm{\tilde{X}} \right )
    & = \argmax_{\bm{W}, \bm{H}} \frac{ p \left ( \bm{R}, \bm{X} \mid \bm{W}, \bm{H}, \bm{\tilde{X}} \right ) p \left ( \bm{W}, \bm{H} \mid \bm{\tilde{X}} \right ) }{ p \left ( \bm{R}, \bm{X} \mid \bm{\tilde{X}} \right ) } \nonumber \\
    & = \argmax_{\bm{W}, \bm{H}} p \left ( \bm{R}, \bm{X} \mid \bm{W}, \bm{H}, \bm{\tilde{X}} \right ) p \left ( \bm{W}, \bm{H} \mid \bm{\tilde{X}} \right ) \nonumber \\
    & = \argmax_{\bm{W}, \bm{H}} \underbrace{ \underbrace{ p \left ( \bm{R} \mid \bm{W}, \bm{H}, \bm{\tilde{X}} \right ) }_{\text{Matrix factorization}} \underbrace{p \left ( \bm{X} \mid \bm{W}, \bm{H}, \bm{\tilde{X}} \right ) }_{\text{Decoder} \mathcal{D}} }_{\text{Likelihood}} \underbrace{ \underbrace{ p \left ( \bm{W} \mid \bm{\tilde{X}} \right ) p \left ( \bm{H} \mid \bm{\tilde{X}} \right ) }_{\text{Encoder} \mathcal{E}} }_{\text{Prior with assumption } \bm{W} \bot \bm{H} \mid \bm{\tilde{X}}} .
    \label{equation:autoencoder_bayes_rule}
\end{align}
$p ( \bm{R}, \bm{Y} \mid \bm{\tilde{Y}} )$ is eliminated due to irrelevance in maximization of (\ref{equation:autoencoder_bayes_rule}).
By sharing latent factors $\bm{W}, \bm{H}$ between autoencoders and matrix factorization, attribute information can affect the learning of rating prediction.
Modeling $\mathcal{D}$ with normal distributions, we can conclude that the expectation of attributes $\bm{X}$ is non-linearly mapped from from latent factors $\bm{W}, \bm{H}$.
Although latent factors have priors from attributes, we categorize relevant works into generative matrix factorization, since we explicitly model attribute distributions in the decoder part of autoencoders.

\begin{itemize}
    \item \textbf{Collaborative Deep Learning (CDL)} \citep{Wang15CDL}. The model presents a combination method of collaborative filtering and Stacked Denoising Auto-Encoder (SDAE).
    Since the model claim to exploit item attributes $\bm{Y}$ only, in the following introduction we define $\bm{Y} = \bm{X}, \bm{\tilde{Y}} = \bm{\tilde{X}}$ in (\ref{equation:autoencoder_bayes_rule}).
    
    In SDAE, input attributes $\bm{\tilde{Y}}$ is not equivalent to $\bm{Y}$ due to adding random noise to $\bm{\tilde{Y}}$.
    CDL implicitly adds several independence assumptions $( \bm{R} \bot \bm{\tilde{Y}} \mid \bm{W}, \bm{H} ), ( \bm{Y} \bot \bm{W}  \mid \bm{H}, \bm{\tilde{Y}} ), ( \bm{W} \bot \bm{\tilde{Y}} )$ to formulate its model.
    Then using identical notations in CMF introduction, normal distributions $\mathcal{N}$ are again applied to CDL:
    \begin{align}
        \argmax_{\bm{W}, \bm{H}, \bm{\theta}, \bm{\phi}} & \underbrace{ p \left ( \bm{R}, \mid \bm{W}, \bm{H} \right ) p \left ( \bm{Y} \mid \bm{H}, \bm{\tilde{Y}} \right ) }_{\text{Likelihood}} \underbrace{ p \left ( \bm{H} \mid \bm{\tilde{Y}} \right ) p \left ( \bm{W} \right ) }_{\text{Prior}} \nonumber \\
        = \argmax_{\bm{W}, \bm{H}, \bm{\theta}, \bm{\phi}} & \underbrace{ \prod_{(u, i) \mid r_{ui} \in \delta ( \bm{R} )} \mathcal{N} \left ( r_{ui} \mid \bm{w}_{u}^{\top} \bm{h}_{i}, \sigma_{R}^{2} \right ) }_{\text{Matrix factorization}} \underbrace{ \prod_{i} \mathcal{N} \left ( \bm{y}_{i} \mid \mathcal{D}_{\bm{\phi}} ( \mathcal{E}_{\bm{\theta}} ( \bm{\tilde{y}}_{i} ) ), \bm{\Sigma}_{Y} \right ) \mathcal{N} \left ( \bm{h}_{i} \mid \mathcal{E}_{\bm{\theta}} ( \bm{\tilde{y}}_{i} ), \bm{\Sigma}_{H} \right ) }_{\text{Stacked denoising auto-encoder for } \bm{Y}} \nonumber \\
        & \underbrace{ \prod_{u} \mathcal{N} \left ( w_{u} \mid \bm{0}, \bm{\Sigma}_{W} \right ) }_{\text{Regularization}}.
        \label{equation:cdl_posterior}
    \end{align}
    
    Functions $\mathcal{E}, \mathcal{D}$ indicate the encoder and the decoder of SDAE.
    The two functions could be formed by multi-layer perceptrons whose parameters are denoted by $\bm{\theta}, \bm{\phi}$.
    It is clear to see the distribution of attribute matrix $\bm{Y}$ be modeled in the decoder part.
    Last but not least, the analysis from (\ref{equation:autoencoder_bayes_rule}) to (\ref{equation:cdl_posterior}) imply that others ideas, user-relevant attributes for example, could be naturally involved in CDL, as long as we remove more independence assumptions.
\end{itemize}

\subsection{Generalized Factorization}
\label{section:generalized_factorization}

\begin{figure}[h]
    \centering
    \begin{subfigure}{0.5\linewidth}
        \centering
        \includegraphics[width=0.8\linewidth]{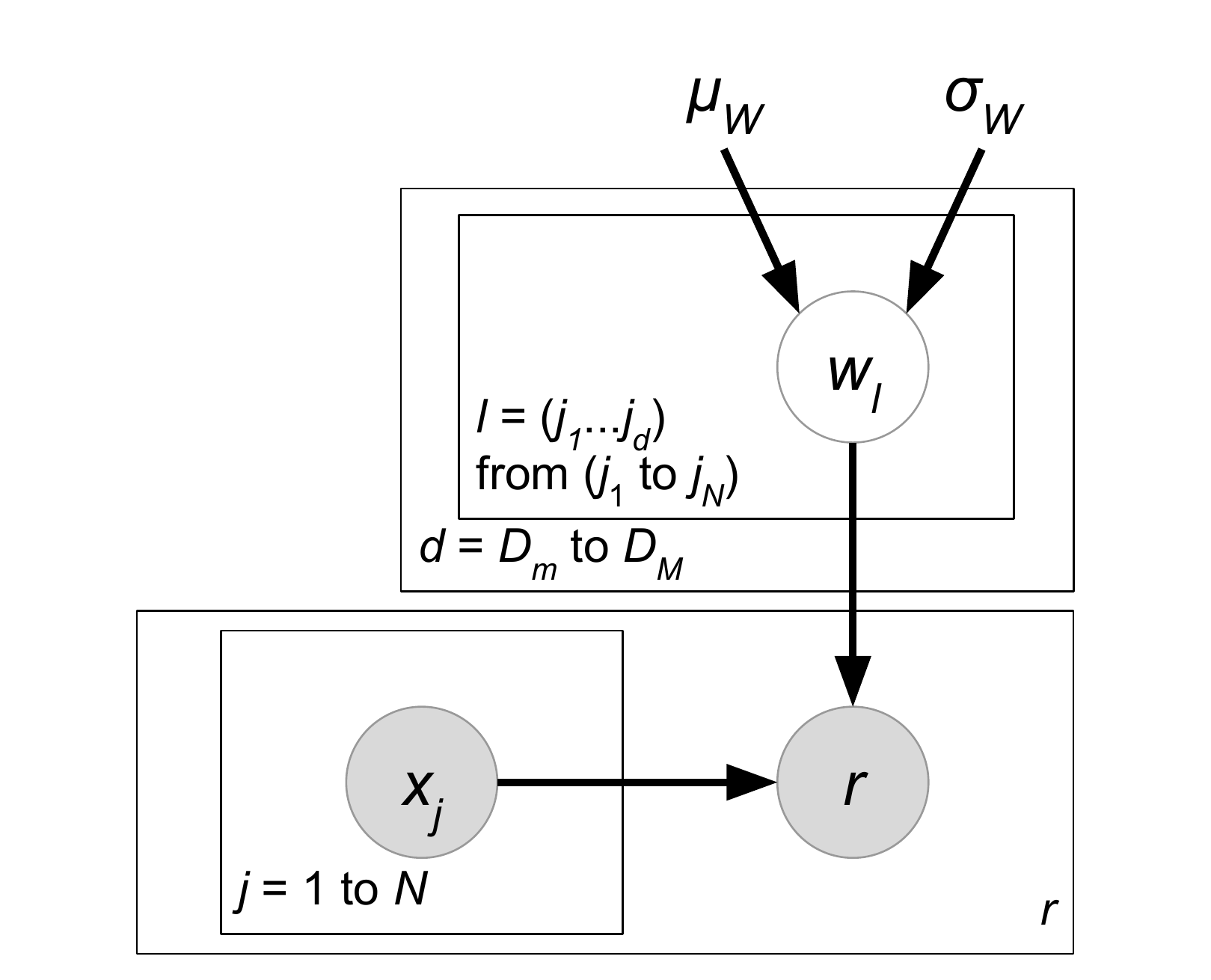}
        \caption{$w$-weighted generalization}
    \end{subfigure}%
    \begin{subfigure}{0.5\linewidth}
        \centering
        \includegraphics[width=0.8\linewidth]{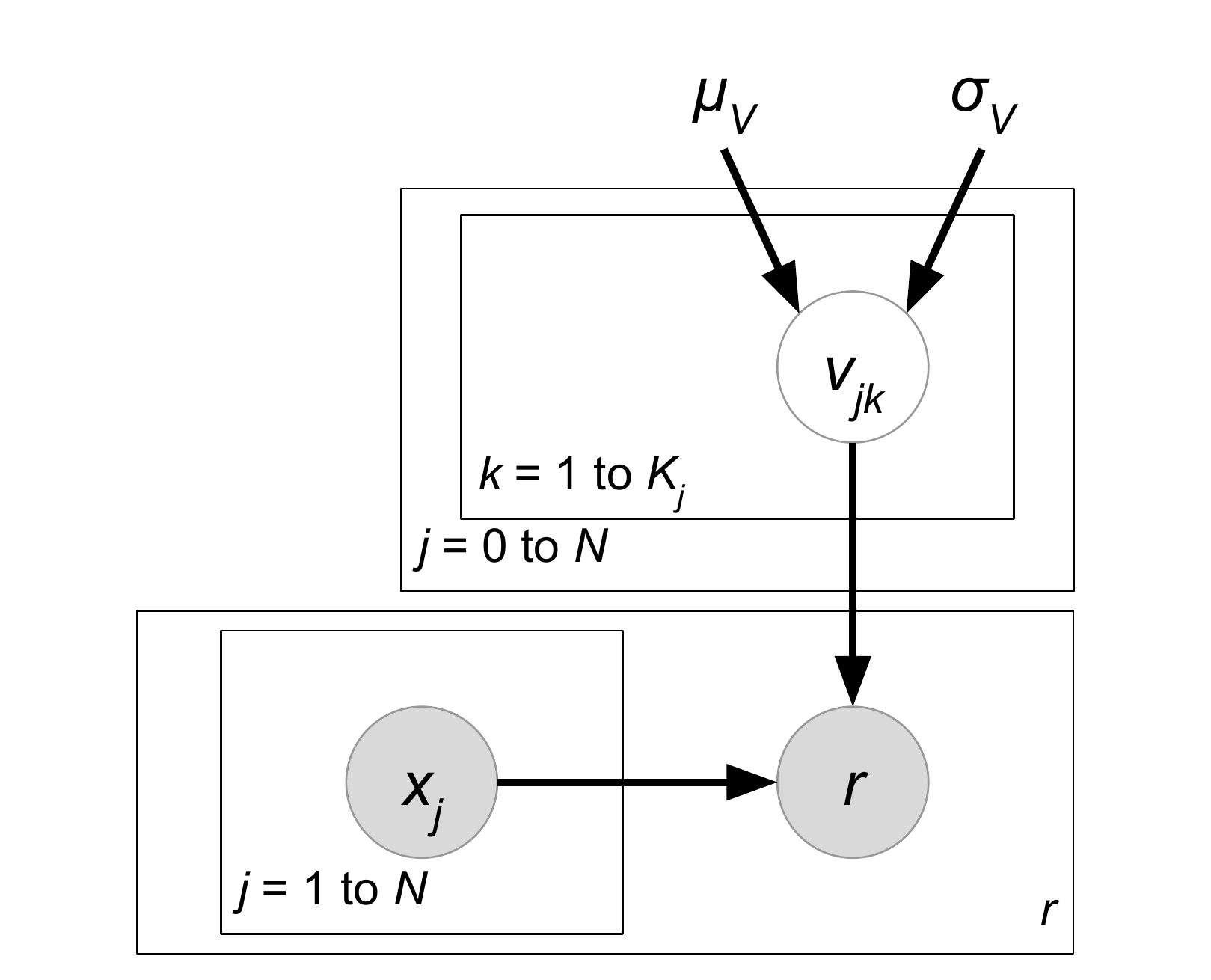}
        \caption{$v$-approximate generalization}
    \end{subfigure}
    \caption{Graphical interpretation of generalized factorization. Attributes $x$ including user or item indices are weighted with corresponding $w$ in order to fit a true rating $r$. If we have all the $w, v$'s follow normal distributions of shared hyperparameters, then there are hyperparameters $\mu_{W}, \sigma_{W}$ or $\mu_{V}, \sigma_{V}$.}
    \label{figure:generalized_factorization}
\end{figure}

Thanks to the success of matrix factorization in recommender systems, there emerge advanced works asking for generalizing the concept of matrix factorization, in order to extract more information from attributes or interactions between users and items.
The works classified in either Section \ref{section:discriminative_matrix_factorization} or Section \ref{section:generative_matrix_factorization} propose to design attribute-aware components on the basis of PMF.
They explicitly express an assumption of vanilla PMF: a latent factor matrix $\bm{W}$ to represent user preferences and another matrix $\bm{H}$ for items.
However the works classified in this section do not regard $\bm{W}$ and $\bm{H}$ as a special existence in models.
Rather, such works propose a expanded latent factor space shared by users, items and attributes.
Here neither users nor items are special entities in a recommender system.
They are simply considered as categorical attributes.
Taking rating $r_{ui}$ for example, it implies that we have a one-hot user encoding vector where all the entries are $0$ except for the $u$-th entry; similarly, we also have a one-hot item encoding vector of the $i$-th entry being $1$.
Thus external attributes $\bm{X}$ can be simply involved in the matrix-factorization-based models, because now users and items are also attributes whose interactions commonly predict or rank ratings.

We first propose the most generalized version of interpretation:
Given a rating $r$ and its corresponding attribute vector $\bm{x} \in \mathbb{R}^{N}$, then we make rating estimate:
\begin{align}
    \argmax_{w} \prod_{r \in \delta(\bm{R})} \mathcal{N} \left (
        r \mid \mu_{R} = \sum_{d = D_{m}}^{D_{M}} \sum_{j_{1} = 1}^{N} \sum_{j_{2} = j_{1} + 1}^{N} \ldots \sum_{j_{d} = j_{d - 1} + 1}^{N} w_{j_{1} j_{2} \ldots j_{d}} \left ( x_{j_{1}} x_{j_{2}} \ldots x_{j_{d}} \right )
        , \sigma_{R}^{2}
    \right )
    ,
    \label{equation:generalized_factorization_form}
\end{align}
where $\delta(\bm{R})$ indicates the set of observed ratings in training data.
Variable $d \in \{ 0 \} \cup \mathbb{N}$ determines the $d$th-order multiplication interaction between attributes $x_{j}$.
As $d = 0$, we introduce an extra bias weight $w_{0} \in \mathbb{R}$ in (\ref{equation:generalized_factorization_form}).
The large number of parameters $w \in \mathbb{R}$ is very likely to overfit training ratings due to the dimensionality curse.
To alleviate overfitting problems, the ideas in matrix factorization are applied here.
For higher values of $d$, it is assumed that each $w$ is a function of low-dimensional latent factors:
\begin{align}
    w_{j_{1} j_{2} \ldots j_{d}} = f_{d} \left ( \bm{v}_{j_{1}}, \bm{v}_{j_{2}}, \ldots , \bm{v}_{j_{d}} \right )
    ,
    \label{equation:low_dimension_approaximate_weight}
\end{align}
where $\bm{v}_{j} \in \mathbb{R}^{K_{j}}$ implies the $K$-dimensional ($K_{j} \ll N \text{ } \forall j$) latent factor or representation vector for each element $x_{j}$ of $\bm{x}$.
Function $f_{d}$ maps these $d$ vectors to a real-valued weight.
Then our learning parameters become $\bm{v}$.
The overall number of parameters ($D_{m} \leq d \leq D_{M}$) decreases from $\sum_{d = D_{m}}^{D_{M}} \frac{n !}{d ! (n - d) !} = O ( 2^{N} )$ to $\sum_{j = 1}^{N} K_{j} = O ( N K )$ where $K = \max_{1 \leq j \leq N} K_{j}$.
Next we prove that matrix factorization is a special case of (\ref{equation:generalized_factorization_form}).
Let $D_{m} = D_{M} = 2$ and $\bm{x}$ be the concatenation of one-hot encoding vectors of users as well as items.
Also we define $f_{2} ( \bm{v}, \bm{y} ) = \bm{v}^{\top} \bm{y}$.
Then for rating $r_{ui}$ of user $u$ to item $i$, we have:
\begin{align}
    \argmax_{\bm{v}} \prod_{r_{ui} \in \delta ( \bm{R} )} \mathcal{N} \left (
    \hat{r}_{ui} \mid \mu_{R} = \sum_{j_{1} = 1}^{N} \sum_{j_{2} = j_{1} + 1}^{N} \bm{v}_{j_{1}}^{\top} \bm{v}_{j_{2}} \left ( x_{j_{1}} x_{j_{2}} \right ) = \bm{v}_{u}^{\top} \bm{v}_{N_{u} + i} , \sigma_{R}^{2}
    \right ) 
    ,
    \label{equation:matrix_factorization_special_case}
\end{align}
where $N_{u}$ denotes the number of users.
(\ref{equation:matrix_factorization_special_case}) is essentially equivalent to matrix factorization.

In this class, the existing works either generalize or improve two early published works: Tensor Factorization (TF) and Factorization Machine (FM).
Both models can be viewed as the special case of (\ref{equation:generalized_factorization_form}).
We introduce TF and FM in the sections below.



\begin{figure}
    \centering
    \begin{subfigure}{0.5\linewidth}
        \centering
        \includegraphics[width=0.8\linewidth]{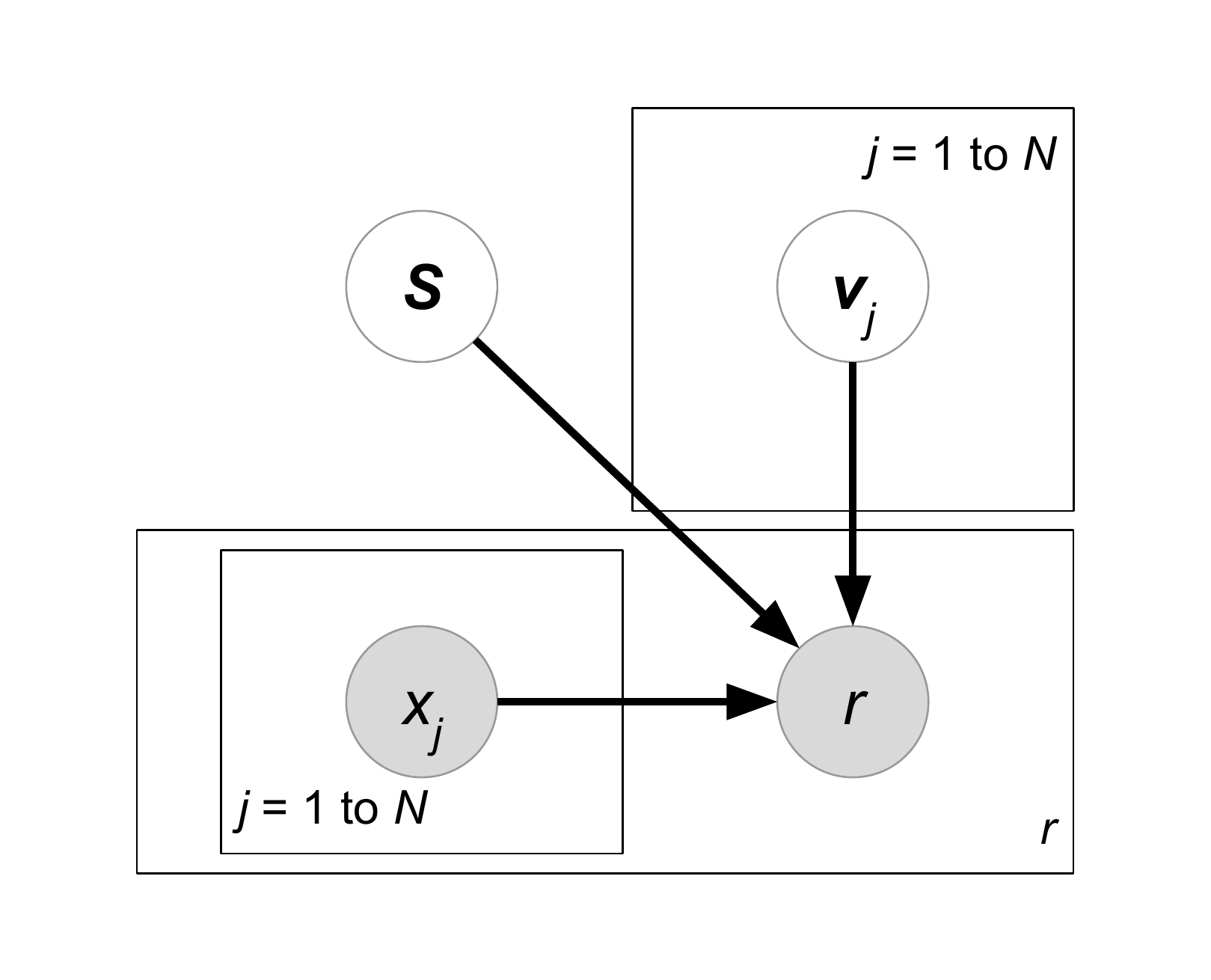}
        \caption{TF}
    \end{subfigure}%
    \begin{subfigure}{0.5\linewidth}
        \centering
        \includegraphics[width=0.8\linewidth]{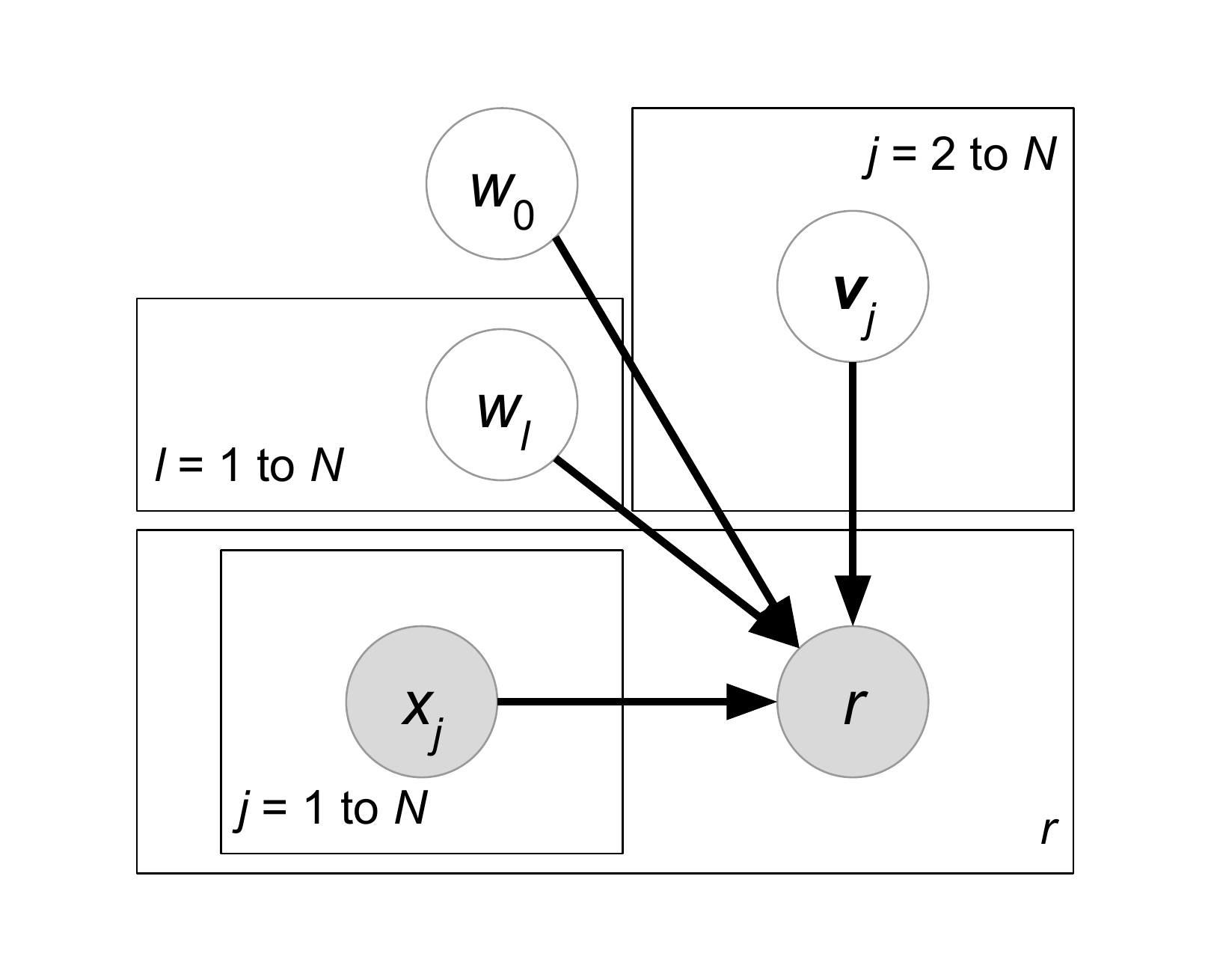}
        \caption{FM}
    \end{subfigure}
    \caption{Graphical interpretation of the example models whose attributes are put into a generalized framework of matrix factorization. All the corresponding hyperparameters are not shown in these figures.}
    \label{figure:matrix_factorization_generalization_example}
\end{figure}

\subsubsection{TF-extended Models}

Tensor Factorization (TF) \citep{Karatzoglou10TF} requires the input features to be categorical.
Attribute vector $\bm{x} \in \{ 0, 1 \} ^{N}$ is the concatenation of $D$ one-hot encoding vectors.
$( D - 2 )$ categorical rating-relevant attributes form their own binary one-hot representations.
The additional two one-hot vectors respectively represent ID's of users and items.
As a special case of (\ref{equation:generalized_factorization_form}), TF fixes $D_{m} = D_{M} = D$ to build a single $D$-order interactions between attributes.
Since weight function $f_{D}$ in (\ref{equation:low_dimension_approaximate_weight}) allows individual dimensions $K_{j}$ for each latent factor vector $\bm{v}_{j}$, TF defines a tensor $\mathcal{S} \in \mathbb{R}^{K_{1} \times K_{2} \times \ldots \times K_{D}}$ to exploit tensor product of all latent factor vectors.
In sum, (\ref{equation:generalized_factorization_form}) is simplified as the following:
\begin{align}
    \mu_{R} & = \sum_{j_{1} = 1}^{N} \sum_{j_{2} = j_{1} + 1}^{N} \ldots \sum_{j_{D} = j_{D - 1} + 1}^{N} f_{D} \left ( \bm{v}_{j_{1}}, \bm{v}_{j_{2}},  \ldots, \bm{v}_{j_{D}} \right ) \left ( x_{j_{1}} x_{j_{2}} \ldots x_{j_{D}} \right ) \nonumber \\
    & = f_{D} \left ( \bm{v}_{l_{1}}, \bm{v}_{l_{1}},  \ldots, \bm{v}_{l_{D}} \right ) \text{ as } x_{l_{1}} = x_{l_{2}} = \ldots = x_{l_{D}} = 1, \text{ other } x = 0 \nonumber \\
    & = \left < \mathcal{S}, \bm{v}_{l_{1}}, \bm{v}_{l_{2}}, \ldots, \bm{v}_{l_{D}} \right > \nonumber \\
    & = \sum_{k_{1} = 1}^{K_{1}} \sum_{k_{2} = 1}^{K_{2}} \ldots \sum_{k_{D} = 1}^{K_{D}} s_{k_{1} k_{2} \ldots k_{D}} v_{l_{1} k_{1}} v_{l_{2} k_{2}} \ldots v_{l_{D} k_{D}}
    \label{equation:tensor_factorization_reduction}
\end{align}
where function $f ( \cdot ) = < \cdot >$ denotes the tensor product.
Note that attribute vectors $\bm{x}$ in TF must consist of exact $C$ $1$'s due to one-hot encoding.
Therefore there exists only match $j_{1} = l_{1}, j_{2} = l_{2}, \ldots, j_{D} = l_{D}$ where all the attributes in these positions are set to $1$.

\subsubsection{FM-extended Models}
Factorization Machine (FM) \citep{Rendle11FM} allows numerical attributes $\bm{x} \in \mathbb{R}^{N}$ as input, including one-hot representations of users and items.
Although higher order interactions between attributes could be formulated, FM focuses on at most second-order interactions.
To derive FM from (\ref{equation:generalized_factorization_form}), let $0 = D_{m} \leq d \leq D_{M} = 2$ and $w_{j_{1} j_{2}} = f_{2} ( \bm{v}_{j_{1}}, \bm{v}_{j_{2}} ) = \bm{v}_{j_{1}}^{\top} \bm{v}_{j_{2}}$ in (\ref{equation:low_dimension_approaximate_weight}) be applied for the second-order interaction.
Then we begin to simplify (\ref{equation:generalized_factorization_form}):
\begin{align}
    \mu_{R} & = \underbrace{\vphantom{\sum_{l = 1}^{N} w_{l} x_{l}} w_{0}}_{d = 0} + \underbrace{\sum_{l = 1}^{N} w_{l} x_{l}}_{d = 1} + \underbrace{\sum_{j_{1} = 1}^{N} \sum_{j_{2} = j_{1} + 1}^{N} w_{j_{1} j_{2}} \left ( x_{j_{1}} x_{j_{2}} \right )}_{d = 2} \nonumber \\
    & = w_{0} + \sum_{l = 1}^{N} w_{l} x_{l} + \sum_{j_{1} = 1}^{N} \sum_{j_{2} = j_{1} + 1}^{N} \bm{v}_{j_{1}}^{\top} \bm{v}_{j_{2}} \left ( x_{j_{1}} x_{j_{2}} \right )
    \label{equation:factorization_machine_reduction}
\end{align}
which is exactly the formulation of FM.
Note that FM implicitly requires all the latent factor vectors $\bm{v}$ of the same dimension $K$; however the requirement could be released from the viewpoint of our general form (\ref{equation:generalized_factorization_form}).
Models in this category mainly differs in two aspects. First, linear mapping can be replaced by deep neural networks, which allows non-linear mapping of attributes. Second, FM only extracts first-order, second-order interactions. Further works such as \cite{Cao16MVM} extracts higher-order interactions between attributes.

\subsection{Heterogeneous Graphs}
\label{section:heteregeneous_graph}

We notice several relevant works that perform low-rank factorization or representation learning in heterogeneous graphs, such as \citep{Lee11GFREC, Yu14PHeteRec, Zheng16DSR, Palumbo17entity2rec, Pham16HeteRS, Jiang18HeteLearn, Nandanwar18DivHeteRec}.
The interactions of users and items can be represented by a heterogeneous graph of two node types.
An edge is unweighted for implicit feedback, while weighted for explicit opinions.
External attributes are typically leveraged by assigning them extra nodes in the heterogeneous graph.
Heterogeneous graph structure is more suitable for categorical attributes, since each candidate value of  attributes can be naturally assigned a node.

In heterogeneous graphs, recommendation can be viewed as a \emph{link prediction} problem.
Predicting a future rating corresponds to forecasting whether an edge will be built between user and item nodes.
The existing works commonly adopt a two-stage algorithm to learn the model.
At first, we perform a random-walk or a meta-path algorithms to gather the similarities between users and items from a heterogeneous graph.
The similarity information can be kept as multiple similarity matrices or network embedding vectors.
Then a matrix factorization model or other supervised machine learning algorithms are applied to extract discriminative features from the gathered similarity information, which is used for future rating prediction.
Another kind of methods is to first define the environment where ranking or similarity algorithms are applied.
The environment refers to either determining the heterogeneous graph structures, or learning the transition probabilities between nodes from observed heterogeneous graphs.
Having the environment, we can apply an existing algorithms (Rooted PageRank for example) or a proposed method to gain the relative ranking scores for each item.
In other words, the main difference between two kinds of methods is to put the similarity calculation into the first stage or the second stage.
Both kind of methods as abovementioned can be unified as a constrained likelihood maximization:
\begin{align}
    \argmax_{\theta} \underbrace{ p \left ( s, \bm{R} \mid \theta, \bm{X} \right ) }_{\text{Likelihood}} \text{ such that } \underbrace{ s \left ( u, i \right ) = \sum_{ w \in  \mathbb{P}_{u, i} \mid \bm{X}} f_{\theta} \left ( w, r_{ui} \right ) \ \forall (u, i), r_{ui} \in \delta ( \bm{R ) } }_{\text{Constraint considering attributes}}
    ,
    \label{equation:heterogeneous_graph_model}
\end{align}
where a parameterized function $f_{\theta}$ is specifically defined to estimate a similarity score $s ( u, i )$ of item $i$, given user $u$ as a query.
The calculation of a similarity score comes from the set $\mathbb{P}_{u, i}$ of random walks or paths $w$ from node $u$ to $i$ in the heterogeneous graph.
The generation of $\mathbb{P}_{u, i}$ considers the attribute node set $\bm{X}$.
Either or both of the likelihood and the constraint may involve the information of observed ratings $\delta ( \bm{R} )$ of rating matrix $\bm{R}$ for likelihood maximization or similarity calculation.
In our observation, the current heterogeneous-graph-based models do not directly solve the constrained optimization problem (\ref{equation:heterogeneous_graph_model}).
Commonly they exploit a two-stage solutions that either solves the likelihood maximization or satisfies the similarity constraint at first.
Then the output are cast into the other part of (\ref{equation:heterogeneous_graph_model}).
With different definitions of $f_{\theta}$ and $p$, the two-stage process may run only once or iteratively until convergence.
The definition of $s(u, i)$ in surveyed papers includes PageRank \cite{Lee11GFREC, Jiang18HeteLearn}, PathSim \cite{Yu14PHeteRec} and so on.
The likelihood function $p$ guides the similarity-related parameters $\theta$ to fit the distribution objective of observed similarities $s$ or ratings $\bm{R}$.
The objective may be given attributes $\bm{X}$ as learning auxiliary.
Minor works like \cite{Lee11GFREC} do not optimize the likelihood; instead, they directly compute the similarity constraint with pre-defined $\theta$ from a specifically designed heterogeneous graph.

We explain why random walk or path based algorithms in heterogeneous graphs are regarded as collaborative filtering methods.
For ease of explanations, first consider the case of no auxiliary attributes.
We have users and items as nodes in a graph structure, where edge weights denote the ratings of users toward items.
If both users $u$ and $v$ rate the same item $i$, then $i$ becomes a shortcut for a path from $u$ to $v$.
Therefore, starting from user node $u$, another user $v$ at low shortest path distances from $u$ could have similar rating behaviors as $u$.
Then we can recommend items at low distances from $u$, based on the shortcut through $v$.
It is just the spirit of collaborative filtering, which exploits the similar rating behaviors of other users for future recommendation to target users.
If attribute nodes are taken into consideration in heterogeneous graphs, they also become the shortcuts for paths between users and items.

\begin{itemize}
    \item \textbf{HeteRec} \cite{Yu14PHeteRec} .
    The model first assumes an attribute-aware heterogeneous graph which are formed by attributes and ratings.
    Then we obtain $M$ non-negative PathSim \cite{Sun11PMP} similarity matrices $\bm{S}^{(1)}, \bm{S}^{(2)}, \ldots, \bm{S}^{(m)}, \ldots, \bm{S}^{(M)}$.
    Given low-rank non-negative factorization of each $\bm{S}^{(m)} = \bm{U}^{(m) \top} \bm{V}^{(m)}$, a rating estimate $\hat{r}$ is defined as follows:
    \begin{align}
        \hat{r} = \sum_{m = 1}^{M} \theta_{m} \bm{u}^{(m) \top} \bm{v}^{(m)} .
        \label{section:heterec_rating_estimate}
    \end{align}

    \item \textbf{Graph-based Flexible Recommendation (GFREC)} \cite{Lee11GFREC} .
    This approach applies personalized PageRank, an unsupervised random walk based algorithm, to perform random walks in a bipartite heterogeneous graph for recommendation.
    Instead of independently defining a single node for each categorical attribute values, GFREC makes a node imply both an attribute value and its associated user or item.
    For example, given a user $u$ and its corresponding attribute value $x$, we can put a node named $(u, x)$ in the heterogeneous network.
    In GFREC bipartite heterogeneous graph, two disjoint sets respectively refer to users and items.
    GFREC shows that personalized PageRank can compute visiting probabilities of each node in this bipartite heterogeneous graph.
    Finally the probabilities are used to rank items to be recommended.
\end{itemize}

\subsection{Model Differences}
\label{section:model_differences}

In our previous classification, there are still a number of works in each category. 
Although Models in the same category share similar mathematical form in terms of the design of objective function, but can vary in certain design aspect. One most important difference is the task they focus on. Some models emphasize on predicting future ratings. Therefore, they usually dedicated to minimize \emph{Root Mean Square Error (RMSE)} to have a more accurate prediction on scores. Some other models care about top-N items that a user may like. Hence, they adopt pairwise ranking to predict the preference of items on a given user.
A second difference is based on the types of attributes that are exploited. For example, \cite{Yang11FIP} takes a social network as its input feature matrices. 
A third difference is that each model claimed its source of attributes. Some models claim to accept only user attributes while others might be more general for different types of attributes.

\section{Empirical Comparison}
\label{section:experiment}
In this section, we evaluate the effectiveness of each model by examining their performance on several datasets. We focus on the \emph{rating prediction} task since the majority of models have their objectives designed for this task. We also compare the performance of each competitor under different conditions: with/without user-relevant attributes, item-relevant attributes or rating-relevant attributes. Hyperparameters for each model are tuned based on grid search.

\subsection{Experiment Setup}
\label{section:experiment_setup}

\subsubsection{Model}
\label{section:experiment_setup_model}
We consider several popular models for comparison: Tensor Factorization (TF) \cite{Karatzoglou10TF}, Collective Matrix Factorization (CMF) \cite{Singh08CMF}, Regression-based Latent Factor Model (RLFM) \cite{Agarwal09RLFM}, Friendship-Interest Propagation (FIP) \cite{Yang11FIP}, Factorization Machine (FM) \cite{Rendle11FM} Neural Factorization Machine (NFM) \cite{He17NFM}, Neural Collaborative Filtering (NCF) \cite{He2017NCF} (the simple version where attributes are one-hot encoding vectors of users and items) and NCF+ (where attributes are one-hot encoding vectors appended with those from datasets). We also select Matrix Factorization (MF) \cite{Chin16LibMF} as baseline model that do not include any attribute.
The attribute types that each model accepts are concluded in Table \ref{table:model_attribute}.

\begin{table}[H]
    \centering
    \caption{Attribute types that claimed to be used for each model.}
    \begin{tabular}{c c c c}
        \toprule
        Model & User-relevant attributes & Item-relevant attributes & Rating-relevant attributes \\ \hline
        TF & $\checkmark$ & $\checkmark$ & $\checkmark$  \\
        CMF & $\checkmark$ & $\checkmark$ & \\
        RLFM & $\checkmark$ & $\checkmark$ & $\checkmark$ \\
        FIP & $\checkmark$ & $\checkmark$ & \\
        FM & $\checkmark$ & $\checkmark$ & $\checkmark$ \\
        NCF & $\checkmark$ & $\checkmark$ & \\
        NFM & $\checkmark$ & $\checkmark$ & $\checkmark$ \\
        MF & & & \\
        \bottomrule
    \end{tabular}
    \label{table:model_attribute}
\end{table}

\begin{itemize}
    \item \textbf{Tensor Factorization (TF)} \\
    TF is an $D$-dimensional extension of MF. We denote the tensor containing the ratings by $\mathcal{R} \in \mathbb{R}^{N_{1} \times N_{2} \times ... \times N_{D}}$. The tensor $\mathcal{R}$ can be factorized into $D$ matrices $\bm{V}_{j} \in \mathbb{R}^{K_{j} \times N_{j}}$ and one central tensor $\mathcal{S} \in \mathbb{R}^{K_{1} \times K_{2} \times \ldots \times K_{D}}$ where $K_{1}, K_{2}, \ldots, K_{D}$ is the dimension of latent factors.
    In this case, the predicted rating for $r_{j_{1} j_{2} \ldots j_{D}}$ is $\hat{r}_{j_{1} j_{2} \ldots j_{D}} = \mathcal{S} \times_{\bm{V}_{1}} \bm{V}_{1} \times_{\bm{V}_{2}} \bm{V}_{2} \times \ldots \times_{\bm{V}_{D}} \bm{V}_{D}$.
    Note that the subscript of the tensor-matrix multiplication operator $\times_{\bm{V}}$ shows the direction on which the tensor multiplies the matrix.
    The loss function for this model is
    \begin{align}
       \argmin_{\bm{S}, \bm{V}} L = \sum_{j_{1}, j_{2}, \ldots, j_{D} \mid r_{j_{1} j_{2} \ldots j_{D}} \in \delta ( \mathcal{R} )} \left ( \hat{r}_{j_{1} j_{2} \ldots j_{D}} - r_{j_{1} j_{2} \ldots j_{D}} \right ) ^{2} + \sum_{j = 1}^{D} \Omega \left ( \bm{V}_{j} \right ) + \Omega ( \mathcal{S} ) ,
    \end{align}
    where $\delta ( \mathcal{R} )$ is the set of non-missing entries in $\mathcal{R}$, and $\Omega ( \bm{V} ) = \frac{\lambda_{V}}{2} \left \Vert \bm{V} \right \Vert _{F}^2$ is the regularization term of squared Frobenius norm. We can update the latent factors using SGD.
    One major concern of this model is that its complexity and storage requirement grow exponentially with the number of dimensions of the rating tensor $\mathcal{R}$. 
    
    \item \textbf{Collective Matrix Factorization (CMF)} \\
    CMF is a model incorporating side information by factorizing multiple matrices simultaneously. In an $D$-entities schema, $\bm{X}^{(ij)} \in \mathbb{R}^{N_{i} \times N_{j}}$ represents the relation between entity $i$ and $j$ if the relation exists i.e. $E_{i} \sim E_{j}$. CMF factorizes these matrices into $\bm{U}^{(1)} \in \mathbb{R}^{K \times N_{1}}, \bm{U}^{(2)}, \ldots, \bm{U}^{(D)} \in \mathbb{R}^{K \times N_{D}}$ such that $\bm{X}^{(ij)} \approx f^{(ij)}(\bm{U}^{(i) \top} \bm{U}^{(j)}) $. 
    For a dataset with user and item-relevant attributes, there are four entities ($E_{1}$: user id, $E_{2}$: item id, $E_{3}$: user features and $E_{4}$: item features) and three relations ($\bm{X}^{(12)}$: ratings matrix, $\bm{X}^{(13)}$, $\bm{X}^{(24)}$: feature matrix). In our experiment, f is identity function for rating matrix and is sigmoid function for feature matrix.
    Let $E=\{(i,j): E_i \sim E_j \cap i<j \}$ denote the set of all existing relations pairs, $\bm{U}$ denote the set of latent factors, $\bm{W}$ denote the set of weight matrices, and $D_{F}( \bm{Y} || \bm{X}, \bm{W} ) = \sum_{ij} w_{ij} ( F ( y_{ij} ) + F^* ( x_{ij} ) - y_{ij} x_{ij} )$ measure the weighted divergence of two matrices $\bm{Y}$ and $\bm{X}$. The loss function for this model is
    \begin{align}
        \argmin_{\bm{U}, \bm{W}} L = \sum_{ij \in E} \alpha^{(ij)} \left ( D_{F^{(ij)}}(\bm{U}^{(i) \top} \bm{U}^{(j)}) || \bm{X}^{(ij)}, \bm{W}^{(ij)} ) + D_{G^{(i)}}(0 || \bm{U}^{(i)}) + D_{G^{(j)}}(0 || \bm{U}^{(j)}) \right )
    \end{align}
    where  $F^{(ij)}$ defines the loss for a reconstruction, and $G^{(i)}$ defines the loss for a regularizer. We can update $\bm{U}$ by Newton-Raphson step.

    \item \textbf{Regression-based Latent Factor Model (RLFM)} \\
    Let $r_{ui}$ denote the rating given by user $u$ to item $i$.
    $\bm{z}_{\pi ( u, i )} \in \mathbb{R}^{K_{Z}}$, $\bm{x}_{u} \in \mathbb{R}^{K_{X}}$ and $\bm{y}_{i} \in \mathbb{R}^{K_{Y}}$ denote attribute vectors for rating $\pi (u, i )$ (i.e., index associated to user $u$ and item $i$), user $u$ and item $i$, respectively. This model learns the latent factors ($\alpha_{u} \in \mathbb{R}, \bm{w}_{u} \in \mathbb{R}^{K}$) to user $u$, ($\beta_{i} \in \mathbb{R}, \bm{h}_{i} \in \mathbb{R}^{K}$) to item $i$ and ($\bm{b} \in \mathbb{R}^{K_{Z}}$) to rating $r_{ij}$, such that the rating is estimated by:
    \begin{align}
        \hat{r}_{ij} = \bm{z}_{\pi (i, j )}^{\top} \bm{b} + \alpha_{u} + \beta_{i} + w_{u}^{\top} h_{i}
    \end{align}
    This model assumes $\alpha_{u}$, $\beta_{i}$, $\bm{w}_{u}$ and $\bm{h}_{i}$ follow Gaussian distribution given attributes $\bm{x}_{u}$ and $\bm{y}_{i}$, so the model can be fitted by Monte Carlo EM algorithm.
    
    \item \textbf{Friendship-Interest Propagation (FIP)} \\
    FIP combines learned latent factors $( \bm{W}$, $\bm{H} )$ and given attribute matrix $( \bm{X}, \bm{Y} )$ to fit user profiles and item properties. Let $U$ be the set of users, $I$ be the set of items. For each training example $(u, i, r) \in O$, it indicates that user $u \in U$ gives item $i \in I$ a rating $r$. The objective function is as follows:
    \begin{align}
        \argmin_{\bm{W}, \bm{H}, \bm{C}} \sum_{(u,i,r) \in O} L ( r, \bm{w}_{u}^{\top} \bm{h}_{i} + \bm{x}_{u}^{T} \bm{C} \bm{y}_{i} )
        + \lambda_{C} \Omega (\bm{C})
        + \lambda_{W} (\Omega (\bm{W}) + \Omega (\bm{w}_{u} - \bm{A} \bm{x}_{u})) \notag \\
        + \lambda_{H} (\Omega (\bm{H}) + \Omega (\bm{h}_{i} - \bm{B} \bm{y}_{i}))
        + \lambda_{A} \Omega (\bm{A}) + \lambda_{B} \Omega (\bm{B})
    \end{align}
    where $L( r, \hat{r} )$ is a loss function, $\bm{C}$ is a correlation matrix, $\bm{A}$ and $\bm{B}$ are the correlation matrice between attribute and latent factors, $\Omega( \cdot )$ is a regularization term and all the $\lambda$ with subscripts are hyperparameters. If both user and item attributes are not given, the model is then reduced to matrix factorization. Since it is often the case that a dataset contains either user or item attribute, in the experiments, if user (or item) attribute is not given, we assume it is a vector of ones with the same dimension as item (or user).
    
    \item \textbf{Factorization Machine (FM)} \\
    FM reduces the original recommendation problem into a traditional classification (or regression) problem. For example, for each observation $( u, i, r ) \in O$, it can be transformed into a attribute vector $\bm{x}$ (which can be formed by representing user $u$ and item $i$ as two one-hot encoding vectors and concatenate them together) and a target rating $r$. The goal then is to fit the target value by utilizing the attribute vector. The objective function can be addressed as follows:
    \begin{align}
        \argmin_{\bm{w}, \bm{V}} \sum_{( u, i, r ) \in O} L \left ( r, w_{0} + \sum_{i=1}^{N} w_{i} x_{i} + \sum _{i=1}^{N} \sum_{j = i + 1}^{D} \left ( \sum_{k = 1}^{K} v_{ik} v_{jk} \right ) x_{i} x_{j} \right ) + \lambda_{w} \Omega(\bm{w}) + \lambda_{V} \Omega(\bm{V})
    \end{align}
    where $\bm{w}$ is the weight vector ($w_{i}$ is its $i$-th element) and $\bm{V} \in \mathbb{R}^{K \times N}$ is the latent factor matrix.
    This is called \emph{factorization machine of degree 2} (or \emph{two-way factorization machine}).
    An \emph{N-way factorization machine} can be expressed as follows:
    \begin{align}
        \argmin_{\bm{w}, \bm{V}} & \sum_{( u, i, r ) \in O} L \left ( r, w_{0} + \sum_{i = 1}^{N} w_{i} x_{i} + \sum_{l = 2}^{N} \sum_{i_{1} = 1}^{N} \sum_{i_{2} = i_{1} + 1}^{N} \cdots \sum_{i_{l} = i_{l - 1} + 1}^{N} \left ( \sum_{k = 1}^{K} \prod_{j = 1}^{l} v_{i_{j} k} \right ) \prod_{j = 1}^{l} x_{i_{j}} \right ) \nonumber \\
        & + \lambda_{w} \Omega(\bm{w}) + \lambda_{V} \Omega(\bm{V}) .
    \end{align}
    In our experiments, only two-way factorization machine is used as our baseline model, since it is the most frequent configuration in the experiments of previous works.
    
    \item \textbf{Neural Collaborative Filtering (NCF)} \\
    
    \begin{figure}[H]
        \centering
        \includegraphics[width=0.5\linewidth]{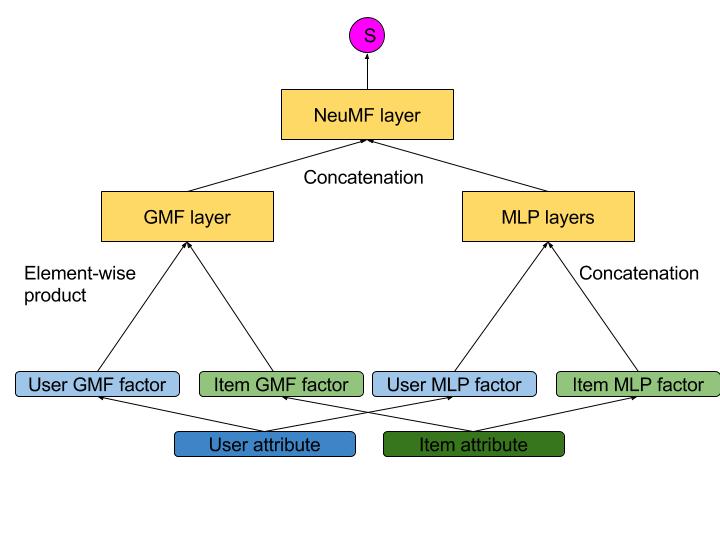}
        \caption{Model structure of NCF}
        \label{figure:NCF_model_structure}
    \end{figure}
    
    NCF consists of two parts: generalized matrix factorization (GMF) and multi-layer perceptron (MLP). GMF layer computes element-wise product of user and item latent factors. MLP layers is a neural network which takes the concatenation of user and item latent factors as inputs and outputs a vector. The results of GMF and MLP are then concatenated as a vector and served as the input of NeuMF layer, which is a one-layer perceptron and outputs the predicted rating. Normally, user/item attribute is a one-hot encoding vector which represents the user/item. However, if external attributes are provided, they can be easily modified.
    
    \item \textbf{Neural Factorization Machine (NFM)} \\
    NFM is a generalization of two-way FM. While FM extracts linear interaction between attributes, NFM is enable to extract non-linear interactions with the help of non-linear activation function in deep neural network. The objective of NFM can be seen as the following:
    
    \begin{align}
        \argmin_{\bm{w}, \bm{V}, \bm{f}} \sum_{( u, i, r ) \in O} L \left ( r, w_{0} + \sum_{i=1}^{N} w_{i} x_{i} + \bm{f}(\sum_{i=1}^{N} \sum_{j=i+1}^{N} x_{i} \bm{v}_{i} \odot x_{j} \bm{v}_{j}) \right ) \nonumber \\
        + \lambda_{w} \Omega(\bm{w}) + \lambda_{V} \Omega(\bm{V}) + \lambda_{f} \Omega(\bm{f})
    \end{align}
    where $\odot$ is element-wise product of vectors and $\bm{f}$ is the neural network. The neural network takes second-order interactions of attribute vectors in FM as input. In fact, FM can be reduced from NFM where $\bm{f}$ is a vector of ones.

\end{itemize}

\subsubsection{Dataset}
\label{section:experiment_setup_dataset}

We choose the data that are available online and widely used to evaluation to test the performance of models. Here we briefly introduce these datasets, and staststics can be seen in Table \ref{experiment:data_basic}. 
or each dataset, if train set and test set are provided by the host, we split our train set and test set accordingly. If not provided, in order to simulate real-world recommendation tasks where future ratings of users are the main concern, train set and test set are split by timestamp where train set represents the data on hand and test set represents future ratings.

\begin{itemize}
    \item \textbf{MovieLens-1M, 10M, 20M} \cite{Harper15ML1M} \\
    MovieLens datasets contain ratings that users give to different movies. 1M, 10M and 20M represents three MovieLens sizes in terms of the number of ratings. They also include some user information, such as genre, age and occupation, and item information, for example the category a movie belongs to and the year when the movie was produced. Training set and test set are divided by the time that the rating was generated. The latest 10\% ratings serve as test set while the others are served as train set.
    
    \item \textbf{Netflix} \footnote{http://www.netflixprize.com/} \\
    Netflix Prize is a competition which dedicated to developing a better movie recommendation system. The data that the host provides contain lots of rating instances. It also includes side information about the movies. Test set is extracted from the probe set, which the host has provided, and the others form training set. However, since training set is so big that most models cannot finish training in an acceptable period of time, it is randomly sampled to one-tenth of the original size in all of our experiments.
    
    \item \textbf{Yahoo Music} \footnote{https://webscope.sandbox.yahoo.com/} \\
    Yahoo provides two music datasets (denoted by Yahoo Music 1 and 2 in our experiments) for researchers to study how users rate music products. Music products include tracks and albums. Information such as genre or artist of a product is provided. The data was also used in KDD cup 2011. Among the items being rated in the original competition (albums, tracks), we extract tracks as targets to be rated. Training set and test set are split in the same way provided by the host.
    
    \item \textbf{Yelp} \footnote{https://www.yelp.com/dataset} \\
    Yelp Dataset Challenge is a contest that allows participants to come up with a research topic themselves based on the given Yelp dataset. The dataset is about how user rates a business. It includes lots of user information and item information in various types. Reviews that users give to items are also presented. Training set and test set are split in the same way as we did in MovieLens datasets.
\end{itemize}

\begin{table}[H]
    \centering
    \caption{Basic statistics of datasets. We define $\text{Density} = \frac{\text{\#(training ratings)}}{\text{\#(users)} \times \text{\#(items)}}$. }
    \begin{tabular}{c c c c c c}
        \toprule
        Dataset & Users & Items & Training ratings
        & Test ratings & Density \\ \hline
        MovieLens-1M & 6040 & 3883 & 900188 & 100021 & $3.84 \times 10^{-2}$ \\
        MovieLens-10M & 69878 & 10681 & 9000048 & 1000006 & $1.21 \times 10^{-2}$ \\
        MovieLens-20M & 138493 & 10378 & 17819935 & 1979993 & $1.24 \times 10^{-2}$ \\
        Netflix & 475708 & 17770 & 9907271 & 1408394 & $1.17 \times 10^{-3}$ \\
        Yahoo Music 1 & 129100 & 4772 & 702947 & 6858 & $1.14 \times 10^{-3}$ \\
        Yahoo Music 2 & 50751 & 3852 & 367556 & 7249 & $1.88 \times 10^{-3}$ \\
        Yelp & 1029432 & 135086 & 3635310 & 406952 & $2.61 \times 10^{-5}$ \\
        \bottomrule
    \end{tabular}
    \label{experiment:data_basic}
\end{table}

\begin{table}[H]
    \centering
    \caption{Basic statistics of cold-start setting.}
    \begin{tabular}{c c c c}
        \toprule
        Dataset & Users & Items & Cold-start test ratings \\ \hline
        MovieLens-1M & 6 & 750 & 1040 \\
        MovieLens-10M & 6 & 801 & 1196 \\
        MovieLens-20M & 16 & 633 & 1017 \\
        Netflix & 364 & 602 & 1002 \\
        Yahoo Music 1 & 984 & 446 & 1001 \\
        Yahoo Music 2 & 973 & 637 & 1001 \\
        Yelp & 592 & 930 & 1003 \\
        \bottomrule
    \end{tabular}
    \label{experiment:data_cold_start}
\end{table}


\subsubsection{Attribute extraction}
\label{section:experiment_setup_attribute}
Most models accept real value attributes as their input. For categorical attributes, since the value merely represents which category the user/item belongs to, which means there is no physical meaning of the value. Therefore, each category is treated as a new dimension of attribute. For each dimension, if user (or item) is in this category then the value is 1, otherwise 0 (i.e., one-hot encoding). However, categorical attributes are not transformed for TF due to its high sapce complexity. Since this method significantly increases the dimension of attributes (if the original attribute contains $d$ categories, the dimension of transformed attributes would be $d$), we find that most of the experimented baseline models cannot finish training in hours for some large-scale datasets.
Hence we determine to retain only top 100 representative transformed attributes that have the most value of 1. Users (or items) not belong to these top 100 categories are discarded. In MovieLens-20M, hundreds of extra attributes are provided. To reduce attribute dimension, the first 100 extra attributes in the original source of file are extracted.
For Yelp dataset, since its attirbute values have a huge range of value, $\log(1+x)$ is applied if the original attribute value x is positive and $-\log(-x)$ for negative (which is the value for longitude or latitude of a restaurant). For TF, the new attribute value is further rounded to the nearest integer.

\begin{table}[H]
    \centering
    \caption{Attribute statistics of datasets. $0$ means no such type of attributes in this dataset.}
    \begin{tabular}{c c c c}
        \toprule
        Dataset & User attributes & Item attributes
        & Rating attributes \\
        \midrule
        MovieLens-1M & 29 & 99 & 0 \\
        MovieLens-10M & 0 & 112 & 0 \\
        MovieLens-20M & 0 & 220 & 0 \\
        Netflix & 0 & 95 & 0 \\
        Yahoo Music 1 & 0 & 300 & 0 \\
        Yahoo Music 2 & 0 & 300 & 0 \\
        Yelp & 18 & 234 & 3 \\
        \bottomrule
    \end{tabular}
    \label{experiment:data_attribute}
\end{table}

\begin{table}[H]
    \centering
    \caption{Percentage of new users/items (users/items in testing data but not in training data).}
    \begin{tabular}{c c c}
        \toprule
        Dataset & \% of new users & \% of new items \\
        \midrule
        MovieLens-1M & 2.4 & 0.8 \\
        MovieLens-10M & 65.5 & 10.8 \\
        MovieLens-20M & 73.1 & 8.5 \\
        Netflix & 4.8 & 0 \\
        Yahoo music 1 & 61.1 & 0 \\
        Yahoo music 2 & 46.3 & 0 \\
        Yelp & 49.8 & 3.0 \\
        \bottomrule
    \end{tabular}
    \label{experiment:percentage_new_user_item}
\end{table}

\begin{table}[H]
    \centering
    \caption{Percentage of new users/items in cold-start setting.}
    \begin{tabular}{c c c}
        \toprule
        Dataset & \% of new users & \% of new items \\
        \midrule
        MovieLens-1M & 100.0 & 0 \\
        MovieLens-10M & 100.0 & 5.6 \\
        MovieLens-20M & 100.0 & 6.3 \\
        Netflix & 100.0 & 0 \\
        Yahoo music 1 & 100.0 & 0 \\
        Yahoo music 2 & 100.0 & 0 \\
        Yelp & 100.0 & 5.1 \\
        \bottomrule
    \end{tabular}
    \label{experiment:percentage_new_user_item_cold}
\end{table}

\subsubsection{Evaluation Metric}

Also adopted by the experiments in these baseline models, \emph{Root Mean Square Error (RMSE)} (defined in (\ref{equation:rmse_definition})) is selected as the evaluation metric in our experiments.
By our observation, RMSE is the most widely used evaluation metric for rating prediction, since most of model-based collaborative filtering methods try to minimize MSE (RMSE without root) as their objectives, including all of our experimented models.
In our opinions, it is fair to test all the baseline models using the evaluation metric they all try to optimize.

\subsubsection{Cold-start Setting}

Cold-start is a special case that many recommend systems are designed to deal with. In practical use, it is difficult to recommend items to a user especially when the user has few or even no past rating records. Since it is an important issue to deal with in the real world, we want to compare different models under this condition. Instead of extracting a new train set designed for cold-start setting (for example, a set formed by randomly reducing the size of the original train set until number of ratings for each user is less than a specific amount), we simulate the cold-start situation by evaluating the performance of a new test set. The new test set is formed by repeatedly extracting all test instances of a user from the original test set where the user has few ratings in train set. The extracting procedure halts when the size of the new test set reaches a threshold. The threshold is set to 1000 in our experiment setting. The other ratings that are not extracted form another set, called "without cold-start" in the following, to compare the result with cold-start. Compared with extracting a new train set, this evaluation metric saves the time to train a new dataset while preserving cold-start property. We list the number of cold-start statistics for each dataset in Table \ref{experiment:data_cold_start}.

\subsection{Performance Comparisons}
\label{section:experiment_result}


We run seven benchmark models on seven attribute-appended rating datasets.
All the empirical comparisons, evaluated with RMSE, are reported from Table \ref{table:rmse_movie_lens_1m} to \ref{table:rmse_yelp}.
Observing the experimental results, we prepare to answer the following four hypotheses that are often asked in attribute-aware recommender system researches:

\subsubsection{Which types of model design could extract the most recommendation-aided information from attributes?}

Section \ref{section:common_integration} introduces different types of common model designs of existing attribute-aware recommender systems.
Among the seven benchmark models, CMF belongs to generative matrix factorization, RLFM and FIP are of discriminative matrix factorization, as well as TF and FM generalizes the vanilla matrix factorization design.
In other words, by this baseline model comparison, we can roughly judge which types of model designs are more robust across different rating applications, and which types could have more improvement on future researches.

\subsubsection{Which types of attributes are the most discriminative for recommendation?}

It is intuitive that a recommender system shall perform better if it gains more additional attribute information.
However, the practical effects or interactions between user-relevant, item-relevant and rating-relevant attributes are not discussed in the previous survey works.
Especially we are curious which type of attributes is the most beneficial for item recommendation.
It is worth our wide experiments to justify the issue.



\subsubsection{Can a recommender system more accurately predict a cold-start user's preference with the help of additional attributes?}

Cold-start recommendation researchers claim to consider attributes which could indirectly reveal the preference of cold-start users.
Therefore we would like to conduct experiments to observe the change in the recommendation quality for cold-start users.

\subsubsection{Can cold-start users gain more performance enhancement than warm-start users?}

Despite more ratings given, warm-start users could obtain better recommendation if they also reveal their attributes to models.
We would like to understand whether attributes can bring more recommendation information given a user has sufficient past ratings.

\subsubsection{If attributes are given, would attribute-based recommendation models have better performance than non-attribute-based models?}

This might be the most important issue we want to discuss. Will attribute-based models have a better performance than basic models (such as MF) with the help of attributes? Or basic models which do not need any attributes could already have decent performance?

\subsection{Rating prediction performance}
\label{section:experiment_result_performance}

\begin{table}[H]
    \caption{Notations referring to attribute type combinations used in an experiment case.}
    \begin{tabular}{c c c c}
        \toprule
        Type & User attributes & Item attributes & Rating attributes \\
        \midrule
        (1) & \checkmark & & \\
        (2) & & \checkmark & \\
        (3) & \checkmark & \checkmark & \\
        \hline
        (4) & & & \checkmark \\
        (5) & \checkmark & & \checkmark \\
        (6) & & \checkmark & \checkmark \\
        (7) & \checkmark & \checkmark & \checkmark \\
        \bottomrule
    \end{tabular}
\end{table}

In the tables below, the star symbol (*) means the running time of the model on the dataset or the memory requirement is too large (over 24 hours or 64 GB memory). It usually happens when TF runs on data with a large number of features. The dash symbol (-) means that the model does not support the attribute type combinations. The results of MF and NCF are trained on ratings only. If baseline model outperforms all competitors, we mark both the baseline model and the competitor with best performance.

\begin{table}[H]
    \caption{RMSE on MovieLens-1M}
    \begin{tabular}{c c c c c c c c c}
        \toprule
        Rating & Attribute & TF & CMF & RLFM & FIP & FM & NFM & NCF+ \\
        \midrule
        \multirow{3}{*}{\makecell{All \\ MF: 0.9002 \\ NCF: 0.9082}} & (1) & 0.9315 & 0.9071 & 0.8815 & 0.9407 & \textbf{0.8793} & 0.9051 & 0.9041 \\
        & (2) & * & 0.9096 & 0.8849 & 0.9631 & \textbf{0.8824} & 0.9222 & 0.8999 \\
        & (3) &  * & 0.9088 & 0.8824 & 0.9396 & \textbf{0.8798} & 0.9162 & 0.9054 \\ \hline
        \multirow{3}{*}{\makecell{No cold-start \\ MF: 0.8986 \\ NCF: 0.9077}} & (1) & 0.9308 & 0.9059 & 0.8804 & 0.9389 & \textbf{0.8782} & 0.9046 & 0.9033 \\
        & (2) & * & 0.9086 & 0.8840 & 0.9609 & \textbf{0.8816} & 0.9218 & 0.8993 \\
        & (3) & * & 0.9075 & 0.8813 & 0.9385 & \textbf{0.8788} & 0.9156 & 0.9047 \\ \hline
        \multirow{3}{*}{\makecell{Cold-start \\ MF: 1.0419 \\ NCF: 0.9507}} & (1) & 0.9993 & 1.0126 & 0.9840 & 1.1004 & 0.9792 & \textbf{0.9481} & 0.9776 \\
        & (2) & * & 1.0036 & 0.9672 & 1.1540 & \textbf{0.9533} & 0.9622 & 0.9552 \\
        & (3) & * & 1.0273 & 0.9848 & 1.0424 & 0.9679 & 0.9691 & \textbf{0.9666} \\
        \bottomrule
    \end{tabular}
    \label{table:rmse_movie_lens_1m}
\end{table}



\begin{table}[H]
    \caption{RMSE on MovieLens-10M with attribute type (2). TF is not included due to excess amount of memory requirement. }
    \begin{tabular}{c c c c c c c c c}
        \toprule
        Rating & MF & NCF & CMF & RLFM & FIP & FM & NFM & NCF+ \\
        \midrule
        \multirow{1}{*}{All} & 0.9820 & 0.9161 & 0.9763 & 0.9111 & 1.1085 & \textbf{0.9103} & 0.9132 & 0.9129 \\ \hline
        \multirow{1}{*}{No cold-start} & 0.9821 & 0.9163 & 0.9765 & 0.9113 & 1.1086 & \textbf{0.9105} & 0.9134 & 0.9131 \\ \hline
        \multirow{1}{*}{Cold-start} & 0.8962 & 0.7724 & 0.7971 & 0.7685 & 1.0174 & \textbf{0.7600} & 0.7651 & 0.7714 \\
        \bottomrule
    \end{tabular}
\end{table}



\begin{table}[H]
    \caption{RMSE on MovieLens-20M with attribute type (2). TF is not included due to excess amount of memory requirement}
    \begin{tabular}{c c c c c c c c c}
        \toprule
        Rating & MF & NCF & CMF & RLFM & FIP & FM & NFM & NCF+ \\ 
        \midrule
        \multirow{1}{*}{All} & 0.9923 & 0.9402 & 0.9954 & \textbf{0.9227} & 1.1128 & 0.9297 & 0.9260 & 0.9240 \\ \hline
        \multirow{1}{*}{No cold-start} & 0.9923 & 0.9402 & 0.9954 & \textbf{0.9227} & 1.1128 & 0.9297 & 0.9260 & 0.9240 \\ \hline
        \multirow{1}{*}{Cold-start} & 0.9320 & 0.8832 & 0.9283 & \textbf{0.8402} & 1.0550 & 0.8434 & 0.8438 & 0.8679 \\
        \bottomrule
    \end{tabular}
\end{table}



\begin{table}[H]
    \caption{RMSE on Netflix with attribute type (2)}
    \begin{tabular}{c c c c c c c c c c}
        \toprule
        Rating & MF & NCF & TF & CMF & RLFM & FIP & FM & NFM & NCF+ \\
        \midrule
        \multirow{1}{*}{All} & 1.2033 & 1.0737 & 1.1434 & 1.0848 & 1.1325 & 1.1312 & 1.0887 & 1.0707 & \textbf{1.0705} \\ \hline
        \multirow{1}{*}{No cold-start} & 1.2033 & 1.0737 & 1.1433 & 1.0848 & 1.1325 & 1.1312 & 1.0887 & 1.0707 & \textbf{1.0705} \\ \hline
        \multirow{1}{*}{Cold-start} & 1.1940 & 1.1071 & 1.2861 & 1.1314 & 1.1614 & 1.1807 & \textbf{1.0879} & 1.0974 & 1.0980 \\
        \bottomrule
    \end{tabular}
\end{table}



\begin{table}[H]
    \caption{RMSE on Yahoo Music 1 with attribute type (2). TF is not included due to excess amount of memory requirement}
    \begin{tabular}{c c c c c c c c c}
        \toprule
        Rating & MF & NCF & CMF & RLFM & FIP & FM & NFM & NCF+ \\
        \midrule
        \multirow{1}{*}{All} & 34.9989 & 33.0522 & 34.3325 & \textbf{32.9302} & 35.8085 & 33.1422 & 33.9271 & 33.1743 \\ \hline
        \multirow{1}{*}{No cold-start} & 34.4840 & 32.5357 & 33.9653 & \textbf{32.3425} & 35.6098 & 32.5855 & 33.4257 & 32.6067 \\ \hline
        \multirow{1}{*}{Cold-start} & 37.8716 & \textbf{35.9260} & 36.4067 & \textbf{36.1779} & 36.9495 & 36.2284 & 36.7241 & 36.3181 \\
        \bottomrule
    \end{tabular}
\end{table}



\begin{table}[H]
    \caption{RMSE on Yahoo Music 2 with attribute type (2). TF is not included due to excess amount of memory requirement}
    \begin{tabular}{c c c c c c c c c c}
        \toprule
        Rating & MF & NCF & CMF & RLFM & FIP & FM & NFM & NCF+ \\
        \midrule
        \multirow{1}{*}{All} & 46.8444 & 41.2785 & 45.2139 & 45.3166 & 50.6670 & 45.4528 & 42.2594 & \textbf{40.4920} \\ \hline
        \multirow{1}{*}{No cold-start} & 45.9243 & 42.0463 & 45.2535 & 44.9473 & 50.0192 & 45.0831 & 42.6063 & \textbf{41.1551} \\ \hline
        \multirow{1}{*}{Cold-start} & 52.2223 & 36.1194 & 44.9662 & 47.5567 & 54.5368 & 47.6957 & 40.0260 & \textbf{36.0789} \\
        \bottomrule
    \end{tabular}
\end{table}



\begin{table}[H]
    \caption{RMSE on Yelp}
    \begin{tabular}{c c c c c c c c c}
        \toprule
        Rating & Attribute & TF & CMF & RLFM & FIP & FM & NFM & NCF+ \\ 
        \midrule
        \multirow{7}{*}{\makecell{All \\ MF: 1.4809 \\  NCF:1.3805}} & (1) & * & 1.3967 & 1.1434 & 1.4162 & 1.1337 & 1.1440 & \textbf{1.1280} \\
        & (2) & * & 1.3951 & 1.2672 & 1.4269 & 1.2849 & 1.2923 & \textbf{1.2586} \\
        & (3) & * & 1.3848 & 1.1029 & 1.2905 & 1.0603 & 1.0876 & \textbf{1.0372} \\
        & (4) & 1.4958 & - & 1.3114 & - & \textbf{1.3065} & 1.3386 & -\\
        & (5) & * & - & 1.1244 & - & \textbf{1.1067} & 1.1168 & - \\
        & (6) & * & - & \textbf{1.2470} & - & 1.2566 & 1.2693 & - \\
        & (7) & * & - & 1.0852 & - & \textbf{1.0372} & 1.0755 & - \\ \hline
        \multirow{7}{*}{\makecell{No cold-start \\ MF: 1.4808 \\  NCF:1.3805}} & (1) & * & 1.3967 & 1.1437 & 1.4163 & 1.1341 & 1.1444 & \textbf{1.1283} \\
        & (2) & * & 1.3950 & 1.2671 & 1.4269 & 1.2849 & 1.2923 & \textbf{1.2586} \\
        & (3) & * & 1.3847 & 1.1032 & 1.2907 & 1.0606 & 1.0879 & \textbf{1.0375} \\
        & (4) & 1.4956 & - & 1.3113 & - & \textbf{1.3064} & 1.3385 & - \\
        & (5) & * & - & 1.1247 & - & \textbf{1.1071} & 1.1171 & - \\
        & (6) & * & - & \textbf{1.2469} & - & 1.2566 & 1.2693 & - \\
        & (7) & * & - & 1.0854 & - & \textbf{1.0375} & 1.0758 & - \\ \hline
        \multirow{7}{*}{\makecell{Cold-start \\ MF: 1.5046 \\  NCF:1.3791}} & (1) & * & 1.4102 & 1.0101 & 1.3587 & \textbf{0.9585} & 0.9801 & 0.9848 \\
        & (2) & * & 1.4385 & 1.2919 & 1.4356 & 1.2835 & 1.3053 & \textbf{1.2728} \\
        & (3) & * & 1.4216 & 0.9834 & 1.2063 & 0.9209 & 0.9427 & \textbf{0.9019} \\
        & (4) & 1.5631 & - & 1.3532 & - & \textbf{1.3492} & 1.3878 & - \\
        & (5) & * & - & 1.0023 & - & \textbf{0.9512} & 0.9767 & - \\
        & (6) & * & - & 1.2846 & - & \textbf{1.2697} & 1.2870 & - \\
        & (7) & * & - & 0.9797 & - & \textbf{0.9124} & 0.9454 & - \\
        \bottomrule
    \end{tabular}
    \label{table:rmse_yelp}
\end{table}



\subsubsection{Which types of model design could extract the most recommendation-aided information from attributes?}
In general, discriminative matrix factorization models (TF, RLFM, NCF+ except FIP) and matrix factorization generalization designs (FM and NFM) perform better than generative matrix factorization design (CMF). The reason may be that in addition to reconstruct rating matrix, generative matrix factorization models have to simultaneously recover attribute matrices, which could be a lot of effort when the dimension of attributes is large. It is still challenging to design generative matrix factorization models which determined to improve RMSE.

\subsubsection{Which types of attributes are the most discriminative for recommendation?}
Since Yelp is the only dataset which contains three kinds of attributes (user, item and rating), we would focus our discussion based on the result of this datset. For RLFM, FM and NFM, the best result occurs when all of the attributes are exploited. However, if we consider three types of attribute exclusively (which are condition 1, 2 and 4 if applicable), it can be seen that user attributes are most beneficial to most models (except CMF, which shows almost no difference). This is somewhat reasonable since what influences a rating the most should be the user himself and his preference. When only rating attributes are incorporated, the results are the worst since those attributes are neither specific to users nor specific to items.


\subsubsection{Can a recommender system more accurately predict a cold-start user's preference with the help of additional attributes?}
In this section, we focus on the cold-start results of each dataset. First, we compare each competitor to the naive baseline, whcih is MF. In every dataset, all models except TF and FIP could outperform MF. Therefore, it could be infered that additional attributes indeed contribute to most recommender systems in cold-start setting. However, when comparing to NCF, which serves as the strong baseline, the effect of attributes seem not much helpful. RLFM, FM, NFM and NCF+ are the stablest models but still could not outperform NCF in all datasets. Instead, they are quite even. The reason may be that deep learning based recommender systems already equip decent ability to extract information from user and item one-hot vectors, which compensate for the effect of additional attributes to linear recommender systems. To sum it up, the help of additional attributes is beneficial when comparing to naive baseline such as MF, but is limited when comparing to NCF. A more effective way to incorporate additional attributes into recommender systems is an upcoming task to be solved.

\subsubsection{Can cold-start users gain more performance enhancement than warm-start users?}
To answer this question, we have to know the differences between baseline models and competitors in both cold start setting and non-cold start setting. We choose MF as the baseline model and the competitors are RLFM and FM, which constantly outperform MF in all datasets.Only datasets with one source of attribute are chosen to eliminate the influence of attributes as far as possible. The results are shown in the following tables.

\begin{table}[H]
    \caption{RMSE difference of MF and RLFM. MovieLens datasets are denoted as ML and Yahoo Music datasets are denoted as YM}
    \begin{tabular}{c c c c c c}
        \toprule
        Rating & ML-10M & ML-20M & Netflix & YM 1 & YM 2 \\
        \midrule
        \multirow{1}{*}{No cold-start} & 0.0708 & 0.0696 & \textbf{0.0708} & \textbf{2.1415} & 0.9770 \\ \hline
        \multirow{1}{*}{Cold-start} & \textbf{0.1277} & \textbf{0.0918} & 0.0326 & 1.6937 & \textbf{4.6656} \\
        \bottomrule
    \end{tabular}
\end{table}

\begin{table}[H]
    \caption{RMSE difference of MF and FM. MovieLens datasets are denoted as ML and Yahoo Music datasets are denoted as YM}
    \begin{tabular}{c c c c c c}
        \toprule
        Rating & ML-10M & ML-20M & Netflix & YM 1 & YM 2 \\
        \midrule
        \multirow{1}{*}{No cold-start} & 0.0716 & 0.0626 & \textbf{0.1146} & \textbf{1.8985} & 0.8412 \\ \hline
        \multirow{1}{*}{Cold-start} & \textbf{0.1362} & \textbf{0.0886} & 0.1061 & 1.6432 & \textbf{4.5266} \\
        \bottomrule
    \end{tabular}
\end{table}

For both RLFM and FM, the trend of difference is quite similar. The differences of RMSE in cold start and non-cold start setting in MovieLens and Netflix are quite subtle (less than 0.1). In Yahoo Music datasets, dataset 1 has significant improvement in non cold-start setting while the other one has improvement in cold-start setting. Since the improvements are either subtle or inconsistent in these datasets, whether cold-start users could get more improvement cannot be infered.

\subsubsection{If attributes are given, would attribute-based recommendation models have better performance than non-attribute-based models?}
The answer of this question is quite similar to that of cold-start setting. In general, most recommender systems (except TF and FIP) could outperform MF in most datasets. This implies that most MF extensions indeed benefit from additional attributes. However, the occurrence of neural network models compensate for this benefit to some extent.It might be a challenge to design neural network models which could effectively utilize additional attributes.

\section{Conclusion}
\label{section:conclusion}

Collaborative filtering has been shown a practical idea to build a recommender system.
Especially, in the case of data gathering or privacy concerns, collaborative filtering methods allow online service to  infer user preferences using the information of users' past ratings, and then successfully recommend items to target users.
Furthermore, recent ten-year researches on collaborative filtering discover that matrix factorization-based approaches commonly achieve high recommendation performance on average.
However with more accessible attributes about users, items or ratings, rating-only collaborative filtering algorithms waste the additional sources that could improve recommendation quality.
Through our wide survey (Section \ref{section:summary}), we find that there are more collaborative filtering publications taking attributes into consideration in the past ten years.
It motivates us to publish this review paper as introduction to the gradually popular domain.

The focus of our reviews lies in how the existing works build effective model-based recommender systems accepting general unstructured attribute vectors, rather than discuss rating-filtering techniques using attributes or explain attribute structures.
Our review work categorizes current works with respect to four factors: attribute source (Section \ref{section:attribute_source}), attribute type (Section \ref{section:attribute_type}), rating type (Section \ref{section:rating_type}) and recommendation goal (Section \ref{section:recommendation_goal}).
We believe that the four factors are a critical consideration for publication authors to design a novel attribute-aware recommender system.
We hope that future models can be inspired by the four factors.
On the other hand, via the probability formulation of matrix factorization, in Section \ref{section:common_integration} we systematically classify three ways of considering attributes into this currently welcome collaborative filtering method.
Modeling attributes as heterogeneous graph nodes is another minor attribute integration way.

Most of the relevant review works do not conduct any empirical evaluation for the surveyed works.
Instead, we design experiments for six attribute-aware recommendation approaches that are mostly used as baselines in other relevant papers.
Besides, seven popular benchmark datasets are adopted to examine these approaches.
Our experiments show that some of the proposed approaches can stably outperform vanilla matrix factorization due to available attributes, but several models severely suffer from time or space-efficiency problems such that they are not applicable for large real-world recommendation scenarios.
Surprisingly, the performance of certain baseline models is not beneficial from accessible attributes, maybe because their original papers emphasize the effectiveness of the item-ranking recommendation goal, which could not be correctly evaluated by RMSE.
A potentially important factor to recommendation performance lies in feature selection or dimension reduction in attributes.
It is our future work to import the additional pre-processing steps before running the baseline models.

We observe that RMSE is less applied in the experiments of the state-of-the-art recommendation works.
On one hand, recently evaluating a recommender system prefers ranking-based metric, due to the fact that users care more about the top recommended item than about the accurate rating prediction of each item.
On the other hand, labeling and gathering numerical ratings are more difficult than binary ratings, and the latter could be extracted more information if modeled as an item ranking problem (Section \ref{section:item_ranking}).
Borrowing the evaluation ideas from information retrieval, the authors of recent papers have tried precision \cite{Sedhain17LoCo}, recall \cite{Li17CVAE}, Normalized Discounted Cumulative Gain (NDCG) \cite{Yu17SQ}, Hit Rate (HR) \cite{Zhao17LDRSSI}, Mean Average Precision (MAP) \cite{Guo17CoEmbed}, and so on.
It is left as our future work to re-evaluate the classical baseline models with these ranking-based evaluation metrics.

\bibliographystyle{ACM-Reference-Format}
\bibliography{reference}

\end{document}